\documentclass[aps,pre,reprint,superscriptaddress]{revtex4-2}
\usepackage{amsmath}
\usepackage{amsthm}
\usepackage{amsfonts}
\usepackage{amssymb}
\usepackage{bm}
\usepackage{enumitem}
\usepackage{graphicx}
\usepackage{dsfont}
\graphicspath{{figures/}}
\usepackage{epstopdf}
\usepackage{hyperref}
\usepackage{xcolor}
\usepackage{physics}
\usepackage{upgreek}

\newcommand{\pderiv}[2]{\frac{\partial {#1}}{\partial {#2}}}

\newcommand{\PeR}{\text{Pe}_R}
\newcommand{\PeC}{\text{Pe}_C}
\newcommand{\pec}{\text{Pe}_C}
\newcommand{\da}{\ensuremath{\mathrm{Da}}}
\newcommand{\trJ}{\text{tr}(\tilde{\mathbf{J}})}
\newcommand{\detJ}{\text{det}(\tilde{\mathbf{J}})}
\newcommand{\bs}[1]{\boldsymbol{#1}}

\newcommand{\Da}{\ensuremath{\mathrm{Da}}}

\newcommand{\A}{\text{A}}
\newcommand{\B}{\text{B}}
\let\originalleft\left
\let\originalright\right
\renewcommand{\left}{\mathopen{}\mathclose\bgroup\originalleft}
\renewcommand{\right}{\aftergroup\egroup\originalright}

\begin{document}
\title{Dynamic Instabilities and Pattern Formation in Chemotactic Active Matter}

\author{Hongbo Zhao}
\email{hongbo@ucsd.edu}
\affiliation{Department of Physics, University of California, San Diego, CA 92093}
\affiliation{Department of Chemistry, University of California, San Diego, CA 92093}
\affiliation{Department of Chemical and Biological Engineering, Princeton University, Princeton, NJ 08544}
\affiliation{Omenn-Darling Bioengineering Institute, Princeton University, Princeton, NJ 08544}

\author{Qiwei Yu}
\email{qiweiyu@princeton.edu}
\affiliation{Lewis-Sigler Institute for Integrative Genomics, Princeton University, Princeton, NJ 08544}

\author{Andrej Ko\v{s}mrlj}
\email{andrej@princeton.edu}
\affiliation{Department of Mechanical and Aerospace Engineering, Princeton University, Princeton, NJ 08544}
\affiliation{Princeton Materials Institute, Princeton University, Princeton, NJ 08544}

\author{Sujit S. Datta}
\email{ssdatta@caltech.edu}
\affiliation{Division of Chemistry and Chemical Engineering, California Institute of Technology, Pasadena, CA 91125}
\affiliation{Department of Chemical and Biological Engineering, Princeton University, Princeton, NJ 08544}

\date{\today}

\begin{abstract}
Collectives of actively-moving particles can spontaneously segregate into dilute and dense phases through a process known as motility-induced phase separation (MIPS). This captivating phenomenon is well-studied for randomly-moving particles with no directional bias. However, many active systems perform collective chemotaxis---directed motion along a chemical gradient collectively generated by the particles themselves through consumption or production. Here, we use linear stability analysis, amplitude equations,  and numerical simulations to study how MIPS is influenced by collective chemotaxis. We find that chemotaxis can either arrest or entirely suppress MIPS, or give rise to novel dynamic instabilities such as traveling waves and spirals. We predict the stability region of the stationary and oscillatory patterns and identify four types of bifurcation that can arise: pitchfork, saddle-node, infinite period, and supercritical Hopf. We also derive analytical expressions for the amplitude of the pattern and traveling wave velocity, yielding excellent quantitative agreement with simulations. Furthermore, we generalize our model to study particles that either consume or produce chemoattractant or chemorepellent, as well as mixtures of particles with different chemotactic behaviors. By establishing quantitative principles describing the competition between MIPS and chemotaxis, our study helps deepen understanding of the rich physics underlying chemically-responsive active matter systems.
\end{abstract}

\maketitle

\section{Introduction}
\label{sec::intro}

Active matter systems---such as flocking birds~\cite{Marchetti2013,Ramaswamy2010,Vicsek1995}, swimming bacteria~\cite{Murray2007,Liu2019}, crawling cells~\cite{Alert2019,Scarpa2016}, and chemically-actuated colloids ~\cite{Elgeti2015,Palacci2013,Theurkauff2012,Palagi2018}---consume energy to generate mechanical forces, such as those that enable self-propulsion. These out-of-equilibrium systems exhibit fascinating collective behaviors, many of which have now been elucidated by theoretical models and mimicked by artificial systems~\cite{Gompper2020,Shaebani2020}. 
A paradigmatic model of such systems is to treat the constituents as active Brownian particles (ABPs)~\cite{Cates2015} that self-propel at a fixed speed while their orientation fluctuates randomly. At short distances on the order of the particle size, the particles are mutually repulsive. Nevertheless, despite having only these purely repulsive interactions, collectives of ABPs can exhibit motility-induced phase separation (MIPS)---where the particles separate into coexisting dense and dilute phases---when their motion is sufficiently directed~\cite{Redner2013a,Fily2012,Bialke2012,Cates2013,Cates2015}. This process occurs because the particles slow down in densely-packed regions, which initiates a positive feedback loop where the decrease in mobility further promotes clustering, amplifying spatial fluctuations in particle density and eventually leading to macroscopic phase separation~\cite{Tailleur2008,Cates2013,Cates2015,Stenhammar2013,Stenhammar2016,Takatori2014,Takatori2015,Takatori2016,Solon2015,Solon2018,Solon2018a,Marchetti2016}. 



In reality, however, the constituents of many active matter systems do not move purely randomly. For example, many forms of active matter are chemically responsive: the constituent particles bias their random motion in response to an external chemical gradient, a process known as chemotaxis. When the particles move up (down) the gradient, the chemical is known as a chemoattractant (chemorepellent). This gradient is often established by the constituent particles themselves through collective consumption or production of the chemical---hence, we refer to this form of biased motion as \emph{collective chemotaxis}. Manifestations of this process abound in living and active matter. For example, swimming bacteria use collective chemotaxis to gain access to more nutrients in heterogeneous terrain~\cite{Keller1971,Berg1972,Berg1975,Murray2007,Saragosti2010,Sourjik2012,Fu2018,Cremer2019,Seyrich2019,Gude2020,Bhattacharjee2021,Bai2021,Bhattacharjee2022,Moore-Ott2022,Weyer2025a}. Another biological example is that of the slime mold \textit{Dictyostelium discoideum}, whose cells move toward self-generated signaling molecules, causing them to aggregate into protective multicellular structures during starvation ~\cite{Murray2007,Haastert2004}. Chemotaxis has also been engineered in synthetic self-propelled colloids and droplets, which move using phoresis or Marangoni flow in response to self-generated chemical gradients~\cite{Howse2007,Ibele2009,Buttinoni2012,Palacci2013,Theurkauff2012,Stenhammar2016,Thutupalli2011,Zottl2016,Moran2017,Meredith2022,Liebchen2022,Michelin2013,Izri2014,Michelin2014,Maass2016,Lach2016,Hokmabad2022,Michelin2013}, mediating attraction or repulsion between the individual particles in turn~\cite{Liebchen2015,Liebchen2017,Pohl2014,Zottl2016,Hokmabad2022}. Even at the molecular scale, the catalytic activity of enzymes can direct their motion along substrate gradients~\cite{Mohajerani2018,Agudo-Canalejo2018,Jee2018,Sengupta2013,Agudo2018enhanced,Feng2020}. Given the ubiquity of collective chemotaxis in living and active matter, we ask: Does it alter the paradigm of MIPS, and if so, how? And can our findings guide ways to use chemotaxis to engineer collective behaviors, such as pattern formation and large-scale migration~\cite{Palacci2013,Aubret2018,Stark2018,Liebchen2018,Liebchen2015,Liebchen2017,Liebchen2017a,Saha2014,Saha2019,Varga2022,Varga2022pre,OByrne2020,Ziepke2022,OByrne2020,Fadda2023}, in such systems? 

Recently~\cite{Zhao2023}, we took a first step toward addressing these questions by combining established continuum models of MIPS~\cite{Solon2018,Solon2018a,Speck2014,Speck2015,Stenhammar2013,Stenhammar2014,Takatori2014,Takatori2015,Takatori2016,Tjhung2018} and collective chemotaxis~\cite{Keller1971}. We found that when ABPs consume an externally-supplied chemoattractant, collective chemotaxis can generate oscillatory instabilities that manifest as traveling waves or spirals, arrest coarsening during MIPS, or completely suppress MIPS altogether---reminiscent of phenomena observed in diverse active matter systems~\cite{You2020Marchetti,Saha2020,Saha2025,Brauns2024,Ouazan-Reboul2023,Brauns2021,Zwicker2017,Demarchi2023,Wysocki2016,Wittkowski2017,Bois2011,Radszuweit2013,Banerjee2015,Weber2018,Yin2023,Rasshofer2025,Ridgway2023,Weyer2025,Osmanovic2025,Martinez-Calvo2023,Frey2025,Romano2025,Hupe2025,VanDerKolk2023}. However, a quantitative theory for the classification of these dynamical patterns and how they emerge or vanish through different bifurcations remains lacking. Moreover, our prior work only focused on consumers of chemoattractant in two dimensions (2D)---whereas many living and active systems involve consumers or producers of chemoattractant or chemorepellent, often with multiple types of such constituents mixed together, in 2D or three dimensions (3D)~\cite{Hu2022,Ghoul2016,Hibbing2010,Ratzke2020}. Here, we build on our continuum model of chemotactic MIPS to address these knowledge gaps.


We first introduce the model in \S\ref{sec::model} and derive the linear stability analysis for the most general case, without invoking any specific constitutive laws, in \S\ref{sec::LSA}. Focusing on consumers of chemoattractant as an illustrative example, we then use linear stability analysis and 2D numerical simulations to map out the full chemotactic MIPS phase diagram in \S\ref{sec::phase_diagram}. We then study the physics of stationary and oscillatory patterns in \S\ref{sec::stationary_pattern} and \S\ref{sec::oscillatory_pattern}, respectively. By deriving amplitude equations, we go beyond the linear regime and establish analytical expressions for the pattern amplitude and traveling wave velocity, which we validate with full numerical simulations in both 2D and 3D. Finally, we broaden our study to the case of particles with different chemotactic properties in \S\ref{sec::producer}, and close with discussions in \S\ref{sec::discussion}.

\section{Model}
\label{sec::model}
\begin{figure}
  \includegraphics[width=\columnwidth]{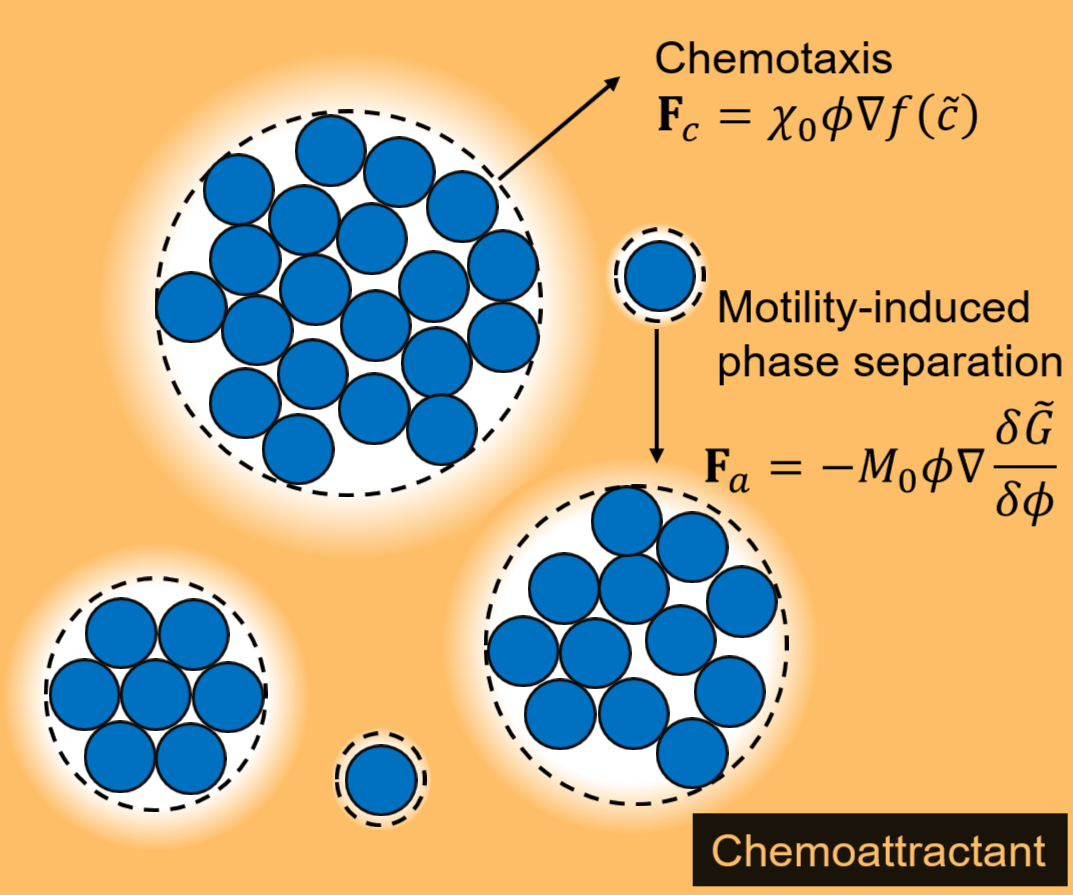}
  \caption{\textbf{Schematic illustration of chemotactic active Brownian particles (ABPs) undergoing motility-induced phase separation.} At sufficiently low reorientation P\'{e}clet number $\PeR$ and moderate particle volume fraction, non-chemotactic ABPs (blue) form dense clusters via MIPS. When particles consume a chemoattractant (yellow), they direct their motion toward higher concentrations and migrate away from depleted clusters. As a result, chemotaxic dispersal competes with MIPS clustering. Producers of chemorepellent exhibit qualitatively similar dispersal.}
  \label{fig::schematics}
\end{figure}

ABPs are particles of size $a$ that self-propel at a speed $U_0$ with a randomly fluctuating orientation that decorrelates over a time scale $\tau_R$. The particle persistence length is then $\sim U_0 \tau_R$, and the directedness of the ABP motion can therefore be described by the inverse of the reorientation P\'{e}clet number, $\PeR \equiv a/(U_0 \tau_R)$. As established previously~\cite{Stenhammar2013,Stenhammar2014,Speck2014,Cates2015,Solon2015,Marchetti2016,Bialke2013,Tailleur2008}, in the limit of large time ($t\gg\tau_R$) and length ($l \gg U_0 \tau_R$) scales, a coarse-grained continuum equation that describes the dynamics of the ABP density can be derived.
This coarse-graining procedure allows one to construct an effective ABP free energy $G$ based on their density-dependent speed, which decreases with density because of steric repulsion between particles, thereby yielding a thermodynamic description of MIPS. Alternatively, an effective free energy can also be derived from a constitutive model for the density-dependent swim pressure~\cite{Takatori2014,Takatori2015,Takatori2016}.

In this continuum description, the volume fraction $\phi$ of non-chemotactic ABPs satisfies the continuity equation $  \pderiv{\phi}{t} = -\nabla \cdot  \mathbf{F}_a,$
where $t$ is time and $\mathbf{F}_a$ is the flux due to active Brownian motion. Consistent with the free energy formulation, this flux can be written in a form that is driven by the effective chemical potential gradient (Fig.~\ref{fig::schematics}),
\begin{equation} \label{eqn::flux_ABP}
  \mathbf{F}_a = - M_0 \phi \nabla \frac{\delta \tilde{G}}{\delta \phi},
\end{equation}
where $M_0$ is the active diffusivity, $\tilde{G}$ is the free energy functional normalized by the characteristic energy scale $\sim\zeta U_0^2\tau_R$, where $\zeta$ is the drag coefficient, and the variational derivative $\delta \tilde{G}/\delta \phi$ is the chemical potential normalized by the same energy scale. When the active diffusivity is dominated by the random propulsion of ABPs, $M_0=0.5 U_0^2\tau_R$~\cite{Cates2013,Bialke2013,Takatori2014,Takatori2015,Cates2015}; the specific numeric prefactor based on the free energy model is discussed in \S\ref{sec::phase_diagram}.

Following the classical Cahn-Hilliard description of MIPS~\cite{Cates2010,Cates2015}, the free energy functional consists of a normalized bulk free energy density $\tilde{g}_h(\phi,\PeR)$ and a term that smooths sharp gradients in $\phi$ that arise from phase separation,
\begin{equation} \label{eqn::free_energy}
  \tilde{G}[\phi] = \int{\left(\tilde{g}_h(\phi,\PeR) + \frac{1}{2}\kappa |\nabla \phi|^2\right)dV},
\end{equation}
where $\sqrt{\kappa}\sim U_0\tau_R$ sets the characteristic width of the interface between the dense and dilute phases in MIPS~\cite{Stenhammar2013,Stenhammar2014,Cates2015}. The normalized chemical potential is then $\tilde{\mu} \equiv \delta \tilde{G}/\delta \phi = \tilde{\mu}_h - \kappa \nabla^2 \phi$, where the normalized bulk chemical potential is $\tilde{\mu}_h(\phi,\PeR) = \partial \tilde{g}_h/\partial \phi$. We define characteristic time and length scales $t_0 \equiv \kappa/M_0\sim \tau_R$ and $l_0\equiv\sqrt{\kappa}$, respectively, yielding the characteristic speed $v_0 \equiv M_0/\sqrt{\kappa}\sim U_0$. From the coarse-graining approach, one then obtains that for a broad range of $\PeR$ and $\phi$, $\partial_\phi \mu_h<0$---causing a net flux of particles toward high-density regions, triggering the spontaneous formation of dense clusters (Fig.~\ref{fig::schematics}), and ultimately causing these clusters to merge and coarsen into distinct macroscopic dense and dilute phases via MIPS~\cite{Takatori2014,Cates2015,Marchetti2016,Stenhammar2013}.
Other generalized models of MIPS~\cite{Tjhung2018,Solon2018} that are not derived from a free energy functional can explain additional complexities such as negative surface tension and bubble formation~\cite{Bialke2015} and will be useful to explore in future work.

In addition to moving randomly, many biological and synthetic active matter systems direct their motion via collective chemotaxis. 
Hence, in Ref.~\cite{Zhao2023}, we studied a microscopic model of chemotactic ABPs~\cite{Pohl2014,Saha2014} that experience chemotactic motion, arising from a combination of translation drift along the chemical gradient and a rotational torque that orients the direction of self-propulsion along the chemical gradient, in addition to randomly-oriented self-propulsion. Following the same coarse-graining approach as described above for non-chemotactic ABPs~\cite{Zhao2023}, one can derive a lowest-order continuum equation in the limit of long time ($t\gg \tau_R$) and length ($l\gg U_0 \tau_R$) scales. In the resulting continuum model, the active Brownian motion and chemotaxis contribute additively to the particle flux. That is, the particle volume fraction evolves in time as 
\begin{equation} \label{eqn::dphidt}
  \pderiv{\phi}{t} = -\nabla \cdot \left( \mathbf{F}_a + \mathbf{F}_c \right),
\end{equation}
where the active Brownian flux $\mathbf{F}_a$ still follows Eq.~\eqref{eqn::flux_ABP} and $\mathbf{F}_c$ is the chemotactic flux. As described by the celebrated Keller-Segel model~\cite{Keller1971}, which is widely used to describe the chemotaxis of living and synthetic active systems~\cite{keller1970initiation,keller1971traveling,Murray2007,Murray2003,Fu2018,Bhattacharjee2021,Bhattacharjee2022,Bai2021,Seyrich2019,Gude2020,Saragosti2010,Moore-Ott2022,Cremer2019,Alert2022,Agudo-Canalejo2018,Pohl2014,Masoud2013,Liebchen2022,Budrene1991,Budrene1995,Woodward1995}, 
\begin{equation}
  \label{eqn::flux_chemotax}
  \mathbf{F}_c = \chi_0 \phi \nabla f(\tilde{c}),
\end{equation}
where $\tilde{c}$ is the concentration, normalized by a fixed characteristic value, of a diffusible chemical signal that the particles sense and direct their motion in response to. The function $f(\tilde{c})$ describes the ability of the particles to sense the chemical and typically increases monotonically with $\tilde{c}$. The chemotactic coefficient $\chi_0$ describes the ability of the particles to move along the sensed chemical gradient; when $\chi_0>0$ ($<0$), the particles move up (down) the gradient (Fig.~\ref{fig::schematics}), and the chemical signal is known as a ``chemoattractant'' (``chemorepellent''). The chemotactic P\'{e}clet number $\PeC \equiv \chi_0 / M_0$ then describes the strength of directed chemotaxis relative to undirected active diffusion.

Lastly, the dynamics of the chemical signal itself are described by the reaction-diffusion equation
\begin{equation} \label{eqn::dcdt}
  \pderiv{\tilde{c}}{t} = D_c \nabla^2 \tilde{c} - R(\tilde{c},\phi),
\end{equation}
where $R(\tilde{c},\phi)$ is the net depletion rate of the chemical. For generality, we take $R = g(\tilde{c})\phi + S(\tilde{c})$, where the first term describes uptake by the particles at a rate $g(\tilde{c})$ per particle, and the second term describes the degradation (or replenishment, if $S<0$) rate of the chemical. 
We define normalized rates $\tilde{R}\equiv R/k$, $\tilde{g}\equiv g/k$, and $\tilde{S}\equiv S/k$ using the same characteristic rate constant $k$. Comparing the rates of chemical uptake and diffusion over the length scale $l_0$ then defines the Damk{\"o}hler number $\text{Da} \equiv k \kappa / D_c$. We also define the dimensionless parameter $\alpha \equiv M_0 / D_c$, which compares the ABP and chemical diffusivities.

This approach combines two established models of active matter that are typically studied separately: MIPS of randomly-moving active particles and collective chemotaxis of directed active particles, described by the first and second terms of the governing Eq.~\eqref{eqn::dphidt}, respectively. In the limit that the active particles are non-chemotactic ($\PeC=0$), Eq.~\eqref{eqn::dphidt} becomes decoupled from Eq.~\eqref{eqn::dcdt} and our model recovers conventional MIPS, as expected. 
Conversely, in the limit that the particles are strongly chemotactic, but are too dilute to cluster and exhibit MIPS, our model instead reduces to the conventional Keller-Segel model of chemotaxis, again as expected.
In this case, if the particles consume a chemoattractant or produce a chemorepellent ($\chi_0g>0$), they move along the collectively-generated gradient and disperse---as exemplified by bacteria migrating toward nutrients in traveling fronts~\cite{keller1971traveling,Fu2018,Bhattacharjee2019,Bhattacharjee2021,Bhattacharjee2022,Cremer2019,Narla2021}. Conversely, if the particles produce chemoattractant or consume chemorepellent ($\chi_0g<0$), they instead aggregate---as exemplified by the ``chemotactic condensation'' of microbes in stressful conditions~\cite{keller1970initiation,Murray2003,Budrene1991,Budrene1995,Woodward1995}. But how do these different effects of MIPS and collective chemotaxis interact? Our model of ``chemotactic active matter'' addresses this gap in knowledge. As described below, this model exhibits fascinating dynamic instabilities and pattern formation resulting from active clustering in MIPS competing with active dispersal induced by collective chemotaxis when $\chi_0 g>0$ (Fig.~\ref{fig::schematics}), or alternatively, MIPS becoming reinforced by chemotaxis when $\chi_0 g<0$.



\section{General linear stability analysis}
\label{sec::LSA}
\noindent\textbf{\emph{Linear stability framework.}} To identify how MIPS is influenced by collective chemotaxis, we first perform a linear stability analysis of our continuum model [Eqs.~\eqref{eqn::flux_ABP}, \eqref{eqn::dphidt}, \eqref{eqn::flux_chemotax}, and \eqref{eqn::dcdt}] for the most general case---without any assumptions on the functional forms of the chemical potential $\tilde{\mu}_h(\phi,\PeR)$, chemotactic sensing function $f(\tilde{c})$, and chemical depletion rate $R(\tilde{c},\phi)$. In later sections, specific functional forms relevant to experimental active systems will be discussed, and the full model will be solved numerically using these specific functional forms. The only condition we place is that there exists a homogeneous dynamic steady state with spatially uniform particle volume fraction $\phi(\mathbf{x})=\phi_0$ and spatially uniform and nonzero chemical concentration $\tilde{c}(\mathbf{x})=\tilde{c}_0>0$, from which we perform our linear stability analysis. To accomplish this steady state, the net depletion rate of the chemical must be zero everywhere: $R(\tilde{c}_0,\phi_0)=g(\tilde{c}_0)\phi_0 + S(\tilde{c}_0) = 0$. That is, the chemical is neither fully depleted nor accumulates without bound, which can be accomplished by, e.g., externally supplying the chemical when it is consumed by the particles, or by allowing the chemical to degrade when the particles produce it. 

To perform linear stability analysis, we perturb this homogeneous steady state $\phi(\mathbf{x})$ and $\tilde{c}(\mathbf{x})$ with small-amplitude fluctuations of spatial wavevector $\mathbf{q}$ and growth rate $\omega$, $\delta \phi = \delta \hat{\phi} e^{\omega t + i \mathbf{q}\cdot\mathbf{x}}$, and $\delta \tilde{c} = \delta \hat{c} e^{\omega t + i \mathbf{q}\cdot\mathbf{x}}$, respectively. Substituting this perturbed state into linearized versions of Eqs.~\eqref{eqn::dphidt} and \eqref{eqn::dcdt} yields the eigenvalue problem $(\omega \mathbf{I}-\mathbf{J}) \cdot (\delta \hat{\phi}, \delta \hat{c})^T = 0$, where $\mathbf{I}$ is the identity matrix and the Jacobian matrix is
\begin{equation} \label{eqn::Jacobian}
  \mathbf{J} = \begin{pmatrix}
    -M_0 \phi_0 q^2 (\partial_\phi \tilde{\mu}_h + \kappa q^2 ) & \chi_0 \phi_0 q^2 f' \\
    -g & -D_c q^2 - \partial_c R
  \end{pmatrix};
\end{equation}
here, $q = |\mathbf{q}|$ is the wavenumber, $f'$ is the derivative of $f(c)$ with respect to $c$, and all derivatives are taken at $\phi_0$ and $\tilde{c}_0$.
The solution to the eigenvalue problem then gives the dispersion relation $\omega(q)$ describing the rate at which fluctuations of different wavenumbers $q$ grow or shrink. We normalize these quantities by the corresponding values of the most unstable mode in non-chemotactic MIPS: $\tilde{\omega}\equiv\omega/\omega_0$ and $\tilde{q}\equiv q/q_0$, where the maximal growth rate $\omega_0 = \max_{q}{\omega(q;\chi_0=0)} = M_0 \phi_0(\partial_\phi \tilde{\mu}_h)^2 /\kappa$ and corresponding most unstable wavenumber $q_0 = \text{arg}\max_{q}{\omega(q;\chi_0=0)} = (|\partial_\phi \tilde{\mu}_h|/\kappa)^{1/2}$. Similarly, we define $\tilde{\mathbf{J}}\equiv\mathbf{J}/\omega_0$.

The normalized Jacobian $\tilde{\mathbf{J}}$ has two invariants, which are the trace $\trJ$ and the determinant $\detJ$, related to the two eigenvalues $\tilde{\omega}_\pm$ by $\trJ=\tilde{\omega}_- + \tilde{\omega}_+$ and $\detJ = \tilde{\omega}_- \tilde{\omega}_+$.
The two invariants of $\tilde{\mathbf{J}}$ are described by the three dimensionless parameters governing chemotactic active matter:
\begin{align}
  \alpha' &\equiv \frac{M_0 \phi_0 |\partial_\phi \tilde{\mu}_h|}{D_c} = \frac{\text{ABP diffusion rate}}{\text{Chemical diffusion rate}}, \\
  \text{Da}' &\equiv \frac{ \partial_c R}{D_c q_0^2} = \frac{\text{Differential reaction rate}}{\text{Chemical diffusion rate}}, \\
  \PeC' &\equiv \frac{\chi_0 \phi_0 f' }{M_0 \phi_0 |\partial_\phi \tilde{\mu}_h|} \frac{g}{\partial_c R} = \frac{\text{Chemotactic response rate}}{\text{ABP diffusion rate}},
\end{align}
which as expected are proportional to the parameters $\alpha$, $\text{Da}$ and $\PeC$ introduced in \S\ref{sec::model}. However, these parameters are weighted by additional factors that arise from the linear stability analysis: $\alpha' = \alpha \phi_0 |\partial_\phi \tilde{\mu}_h|$, $\text{Da}'=\text{Da} (\partial_c \tilde{R})/|\partial_\phi \tilde{\mu}_h|$, and $\PeC'=\PeC f'\tilde{g}/(|\partial_\phi \tilde{\mu}_h| \partial_c \tilde{R})$. Using these definitions, the two invariants of $\tilde{\mathbf{J}}$ are
\begin{align}
  \trJ &= -\tilde{q}^4 - \left(\sigma+\frac{1}{\alpha'}\right) \tilde{q}^2 - \frac{\text{Da}'}{\alpha'} \label{eqn::tr} \\
  \detJ &= \frac{\tilde{q}^2}{\alpha'} \left[ (\tilde{q}^2+\sigma)(\tilde{q}^2+\text{Da}') + \text{Da}'\PeC' \right] ,\label{eqn::det}
\end{align}
where $\sigma = \text{sgn}(\partial_\phi \tilde{\mu}_h)$; that is, $\sigma=1$ or $-1$ outside or inside the region, respectively, in which non-chemotactic ABPs exhibit spinodal decomposition (hereafter referred to as the ``MIPS spinodal region''), respectively.

Before proceeding, we comment on the sign of the dimensionless parameters. Since the diffusivities are positive ($M_0>0$ and $D_c>0$), $\alpha'>0$. We restrict our subsequent analysis to the case of $\partial_c R>0$, that is, the chemical uptake rate increases with increasing chemical concentration. In other words, we require that the chemical reaction is autoinhibitory, because if the reaction is autocatalytic ($\partial_c R<0$), the chemical concentration field is always unstable under a spatially uniform perturbation ($q=0$), when one of the eigenvalues is $-\partial_c R$. We focus on the case where $R$ increases strictly monotonically with $c$ such that a unique solution for $\tilde{c}_0$ exists for $R(\tilde{c}_0,\phi_0)=0$ given $\phi_0$. Therefore, $\Da'>0$.
Lastly, because $f'>0$, the sign of $\PeC'$ depends on $\chi_0 g$. As discussed in \S\ref{sec::model}, when $\PeC'>0$ ($\chi_0 g>0$), chemotaxis causes particles to disperse while when $\PeC'<0$ ($\chi_0 g<0$), it causes particles to aggregate. Hence, in the following sections, we term $|\chi_0 g|$ the chemotactic dispersal or aggregation rate, depending on the sign of $\chi_0 g$.\\

\begin{figure*}
  \includegraphics[width=\textwidth]{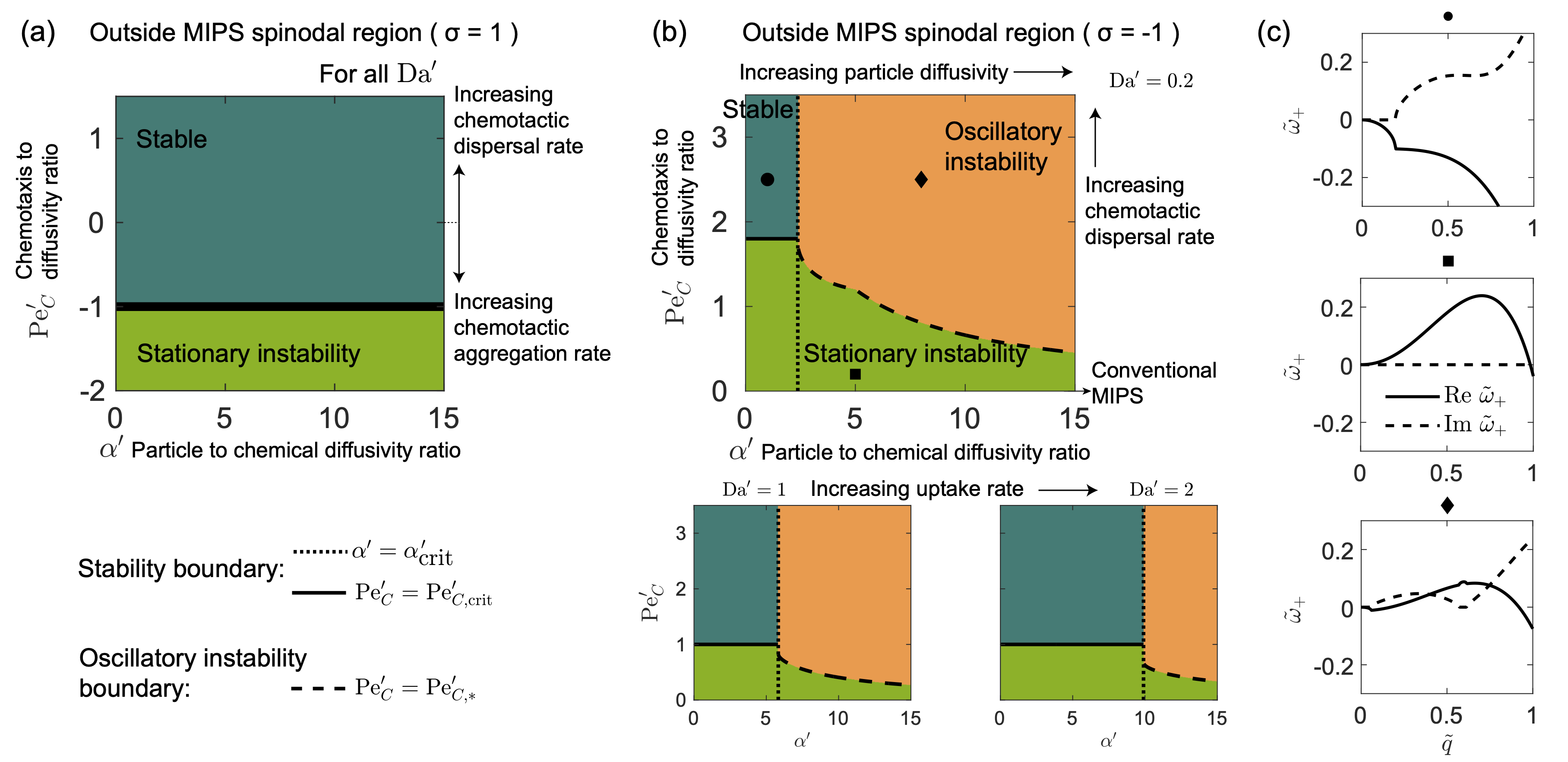}
  \caption{\textbf{Linear stability analysis reveals three distinct regimes of chemotactic MIPS depending on chemical diffusivity and chemotactic strength.} 
  (a) Outside the MIPS spinodal region ($\sigma=1$), the uniform state is stable when $\PeC'\geq-1$, independent of $\Da'$ and $\alpha'$; stationary instability occurs when $\PeC'<-1$. 
  (b) Inside the MIPS spinodal region ($\sigma=-1$), three regimes emerge: stable (dark teal, where $\alpha' \leq \alpha'_\text{crit}$ and $\PeC' \geq \text{Pe}'_{C,\text{crit}}$), stationary instability (light green), and oscillatory instability (orange, where $\alpha'>\alpha'_\text{crit}$ and $\PeC'>\text{Pe}_{{C,*}}'$). Increasing $\Da'$ (faster chemical uptake) shrinks both instability regions. 
  (c) Representative dispersion relations showing the positive branch of the nondimensionalized eigenvalue $\tilde{\omega}_+$ at $\text{Da}'=0.2$ for the three regimes marked in (b): stable (dot), stationary instability (square), and oscillatory instability (diamond).}
  \label{fig::PeC_alpha_phase_diagram}
\end{figure*}

\noindent\textbf{\emph{Stability conditions.}} Next, we use this linear stability analysis to identify the conditions under which chemotactic active matter systems are linearly stable. In particular, the stability condition for the homogeneous state $\text{Re}~\tilde{\omega}_\pm \leq 0$ is equivalent to $\trJ \leq 0$ and $\detJ \geq 0$ for all $\tilde{q}$. 
Since $\max_{\tilde{q}}{\trJ} = \frac{1}{4\alpha^{\prime2}}\left( \text{min}\{\sigma \alpha' + 1,0\} \right)^2 - \frac{\text{Da}'}{\alpha'}$ and $\min_{\tilde{q}}{\alpha'\tilde{q}^{-2}\detJ} = -\frac{1}{4} \left( \text{min}\{\sigma+\text{Da}',0\} \right)^2 + \text{Da}'(\PeC'+\sigma)$,
this stability condition is then equivalent to
\begin{equation} \label{eqn::Da_crit}
  \text{Da}' \geq \frac{ \left( \text{min}\{\sigma \alpha' + 1,0\} \right)^2 }{4\alpha'} \equiv \text{Da}'_\text{crit},
\end{equation}
and
\begin{equation} \label{eqn::PeC_crit}
  \PeC' \geq \frac{\left(\text{min}\{\text{Da}'+\sigma,0\}\right)^2}{4\text{Da}'} - \sigma \equiv \text{Pe}'_{C,\text{crit}}.
\end{equation}
At the critical condition $\Da' = \Da'_\text{crit}$, the critical wavenumber is $\tilde{q}^2_{\text{crit},1} = \text{min}\left\{0, -\frac{\sigma+\alpha'^{-1}}{2} \right\}$
and at the critical condition $\PeC' = \text{Pe}'_{C,\text{crit}}$, the critical wavenumber is $  \tilde{q}^2_{\text{crit},2} = \text{min}\left\{0, -\frac{\sigma+\text{Da}'}{2} \right\}$.
Thus, outside the MIPS spinodal region ($\sigma=1$), the stability condition reduces to $\PeC' \geq -1$, while the other condition $\text{Da}'\geq0$ is satisfied automatically. This condition represents the stability condition of the classical Keller-Segel model --- chemotactic aggregation occurs when $\chi_0 g$ is sufficiently negative. 

As shown in Fig.~\ref{fig::PeC_alpha_phase_diagram}(a), the system is linearly stable when $\PeC' \geq -1$, and exhibits a stationary instability when $\PeC' < -1$, independent of $\text{Da}'$ and $\alpha'$. By contrast, inside the MIPS spinodal region ($\sigma=-1$), phase separation is suppressed when $\PeC' \geq \text{Pe}'_{{C,\text{crit}}}>0$, i.e., chemotactic dispersal is sufficiently rapid, \textit{and} $\text{Da}'>\text{Da}'_\text{crit}$. This latter condition is equivalent to $\alpha' \leq 1+2\text{Da}'+2\sqrt{\text{Da}'(1+\text{Da}')} \equiv \alpha'_\text{crit}$, i.e., the chemical diffusivity is sufficiently fast compared to the particle diffusivity, as shown by the dark teal region in Fig.~\ref{fig::PeC_alpha_phase_diagram}(b). An example of the dispersion relation $\tilde{\omega}_+(\tilde{q})$ in this linearly stable regime ($\text{Re}~\tilde{\omega}_+<0$) is shown by the dot in Fig.~\ref{fig::PeC_alpha_phase_diagram}(c).
This region of linear stability expands with faster chemical uptake (increasing $\text{Da}'$), as shown by the different panels of Fig.~\ref{fig::PeC_alpha_phase_diagram}(b).\\ 

\noindent\textbf{\emph{Classification of instabilities.}} Having established the conditions under which chemotactic active matter is linearly stable, we next categorize the conditions under which it is unstable. We focus on two classes of instability. The first is known as a stationary instability, in which all unstable modes with $\text{Re}~\tilde{\omega}(\tilde{q})>0$ have $\text{Im}~\tilde{\omega}(\tilde{q})=0$. This instability arises at sufficiently small $\PeC'$, as shown in light green in panels (a)-(b), consistent with the fact that as $\PeC'$ decreases to 0, the model reduces to standard MIPS. An example of the corresponding dispersion relation is shown by the square in Fig.~\ref{fig::PeC_alpha_phase_diagram}(c). The second class is known as an oscillatory instability, in which there exists a wavenumber $\tilde{q}$ for which $\text{Re}~\tilde{\omega}(\tilde{q})>0$ and $\text{Im}~\tilde{\omega}(\tilde{q})\neq0$. This condition is equivalent to the condition that there exists $\tilde{q}$ for which $\trJ>0$ and $\Delta<0$, where $\Delta = \trJ^2 - 4\detJ$ is the discriminant. As noted earlier, the requirement that $\trJ>0$ is equivalent to $\text{Da}'<\text{Da}'_\text{crit}$, which is only possible when $\sigma=-1$ (inside the MIPS spinodal region), and is equivalent to $\alpha'>\alpha'_\text{crit}$.  
In this case, Eqs.~\eqref{eqn::tr} and \eqref{eqn::det} yield
\begin{equation} \label{eqn::Delta}
  \Delta(\tilde{q}) = \left[ \tilde{q}^4 + \left(\sigma-\frac{1}{\alpha'}\right) \tilde{q}^2 - \frac{\text{Da}'}{\alpha'} \right]^2 - \frac{4\text{Da}'\PeC'}{\alpha'}\tilde{q}^2,
\end{equation}
which shows that that $\Delta(\tilde{q})<0$ is possible only when $\PeC'$ is positive and sufficiently large. Altogether, this analysis indicates that an oscillatory instability arises when four criteria are met: (i) The system is in the MIPS spinodal region, (ii) chemical diffusivity is sufficiently slow relative to the particle diffusivity, i.e., $\alpha'>\alpha'_\text{crit}$, (iii) chemotaxis drives particles to disperse, i.e., $\chi_0 g>0$, and (iv) chemotaxis is sufficiently strong, i.e., $\PeC'>\text{Pe}_{{C,*}}'$ where the critical value $\text{Pe}_{{C,*}}'$ is given by Eq.~\eqref{eqn::critical_condition_oscillatory}. The emergence of oscillatory instability at sufficiently large $\PeC'$ and $\alpha'$ is shown by the orange region in Fig.~\ref{fig::PeC_alpha_phase_diagram}(b).
An example of the corresponding dispersion relation is shown by the diamond in Fig.~\ref{fig::PeC_alpha_phase_diagram}(c). These criteria can be summarized in a simple, intuitive picture: an oscillatory instability emerges due to the competition between clustering via MIPS and dispersal via collective chemotaxis, in the case where the chemical signal diffuses so slowly that its gradient lags behind the moving particles. 

\section{Phase diagram of chemotactic MIPS}
\label{sec::phase_diagram}
\begin{figure*}
  \includegraphics[width=\textwidth]{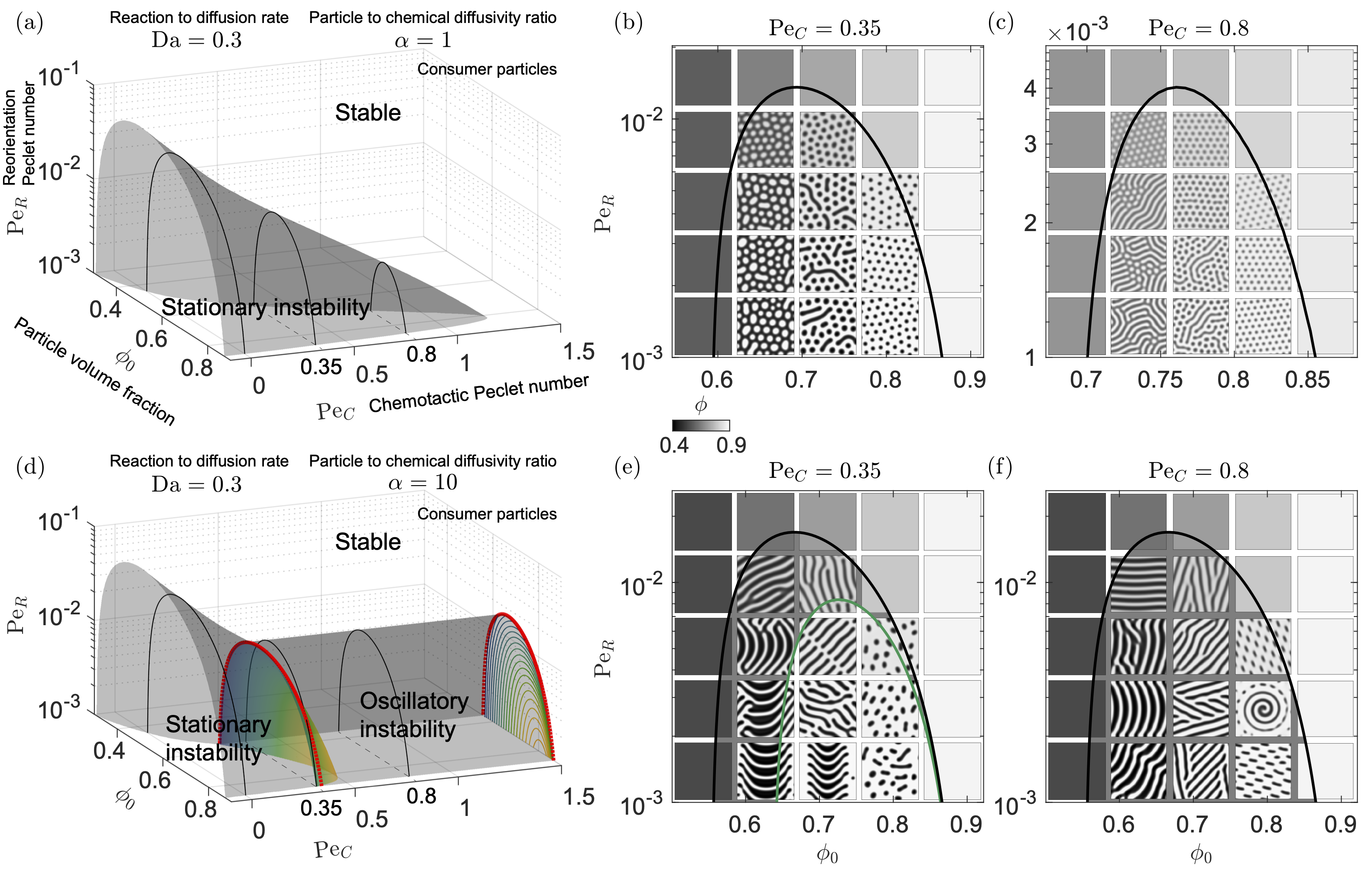}
  \caption{\textbf{Three-dimensional phase diagram reveals how chemotaxis suppresses MIPS and generates oscillatory instabilities in consumer particles.} (a,d) Phase behavior of chemotactic ABPs that consume a chemoattractant in the $\PeR-\phi_0-\PeC$ phase space at $\text{Da}=0.3$ with $\alpha=1$ (a) and $\alpha=10$ (d). Gray surface marks the stability boundary from linear stability analysis; above it, the homogeneous state is stable, while below it, phase separation occurs. At small $\alpha$ (a), chemotaxis arrests coarsening below the boundary, producing stationary dots and stripes. At large $\alpha$ (d), the colored surface delineates stationary (left) from oscillatory (right) instability; the color represents the value of $\PeC$, and the colormap ranges from the minimum to maximum values of $\PeC$ of the surface. The red dashed curve marks the transition from one regime to the other ($\PeC'=\text{Pe}'_{{C,\text{crit}}}$ and $\alpha' = \alpha'_\text{crit}$). (b-c, e-f) Cross-sections of the $\PeR-\phi_0$ phase space at $\PeC=0.35$ and $0.8$; black curves are cross-sections through the gray surface, and the green curve in (e) shows a cross-section through the colored surface in (d). We expect a stationary instability below the black curves in (b)-(c) and below the green curve in (e), and an oscillatory instability in the gray shaded region between the black and green curves in (e) and below the black curve in (f). Simulation snapshots at $t=2\times 10^4t_0$ confirm our predictions, with stationary dots/stripes at small $\alpha$ and traveling waves/dots at large $\alpha$. Simulations use $[100l_0,100l_0]$ periodic domains with homogeneous initial conditions plus small perturbations. Increasing $\PeC$ shrinks the phase separation region; increasing $\alpha$ expands the oscillatory regime.}
  \label{fig::PeC_PeR_phi_phase_diagram}
\end{figure*}

\noindent\textbf{\emph{Construction of the phase diagram.}} The phase diagram of non-chemotactic ABPs is conventionally parameterized by the average particle volume fraction $\phi_0$ and the reorientation P\'{e}clet number $\PeR$~\cite{Cates2015}. In particular, MIPS occurs at intermediate $\phi_0$ and high directedness of self-propulsion (small $\PeR$). We now build on our analysis in \S\ref{sec::LSA} to extend the canonical MIPS phase diagram to account for the effect of chemotaxis using specific illustrative models of $\tilde{\mu}_h(\phi,\PeR)$, $f(c)$, and $R(c,\phi)$. 

Refs.~\cite{Takatori2014,Takatori2015} used a mechanical model to derive the effective free energy density of non-chemotactic ABPs as a function of $\phi_0$ and $\PeR$. In particular, for ABPs in 2D or 3D, the normalized chemical potential is given by 
\begin{multline} \label{eqn::ABP_2D}
  \tilde{\mu}_h(\phi,\PeR) = \ln{\phi}+1 - 2\phi - 0.3\phi^2  \\ - \frac{4}{\pi}\PeR \phi_m \left( \ln{\left[ 1- \frac{\phi}{\phi_m}\right]} - \frac{\phi}{\phi_m - \phi} \right)~(\textrm{2D})
\end{multline}
\begin{multline} \label{eqn::ABP_3D}
  \tilde{\mu}_h(\phi,\PeR) = \ln{\phi}+1 - 2\phi - 1.5\phi^2 \\ - 3\PeR\phi_0 \left[ \ln{\left(1-\frac{\phi}{\phi_m}\right)} - \frac{\phi}{\phi_m-\phi} \right]~(\textrm{3D}),
\end{multline}
where the maximum dense-packing density in 2D or 3D is $\phi_m =0.9$ or $0.65$, respectively, and the chemical potential is normalized by $\zeta U_0^2\tau_R/2$ or $\zeta U_0^2 \tau_R/6$, respectively. We use these established relations here, first focusing our analysis on the well-studied 2D case [Eq.~\eqref{eqn::ABP_2D}], and then extending our analysis to 3D in \S\ref{sec::oscillatory_pattern} [Eq.~\eqref{eqn::ABP_3D}]. 

As an illustrative example, we first examine particles that consume the chemical signal ($g>0$); the opposite case of particles that instead produce the chemical signal is then discussed in \S\ref{sec::producer}. To satisfy the conditions specified in \S\ref{sec::LSA} that $\partial_c R>0$ and a homogeneous steady state exists, we take $g(\tilde{c})=k\tilde{c}$ with the chemical signal supplied at a constant, spatially-uniform rate $S(\tilde{c})=-S_0$. The uniform chemical concentration at steady state is then $\tilde{c}_0=S_0/(k \phi_0)$. We also take $f(\tilde{c})=\tilde{c}$ as is commonly done. 
Because the governing Eqs.~\eqref{eqn::dphidt} and \eqref{eqn::dcdt} are then linear with respect to $\delta\tilde{c}\equiv\tilde{c}-\tilde{c}_0$, scaling the supply rate $S_0$ by a constant is equivalent to scaling the chemotactic rate and hence $\PeC$ by the same constant. Therefore, without loss of generality, we fix the supply rate to be $S_0/k=1$. Supplementary Table~\ref{table::consumer_producer} summarizes all these constitutive relations as well as the corresponding relationships between the dimensionless parameters $\alpha$, $\Da$, $\PeC$, and the reduced versions $\alpha'$, $\Da'$, and $\PeC'$. 

\begin{figure}
  \includegraphics[width=\columnwidth]{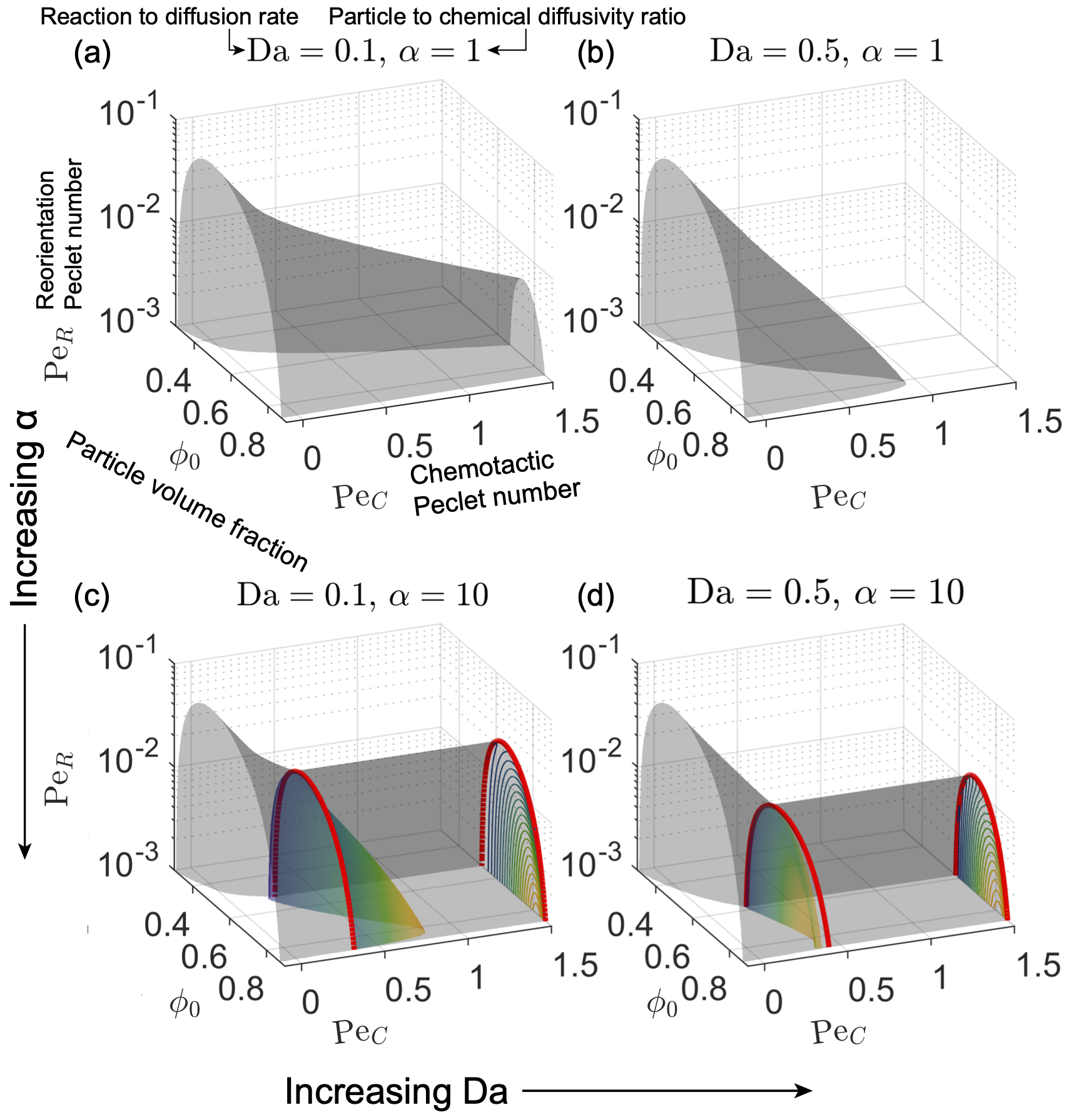}
  \caption{\textbf{Increasing chemical uptake rate ($\text{Da}$) and particle-to-chemical diffusivity ratio ($\alpha$) systematically alter the phase separation and oscillatory instability regions.} (a-b) At small $\alpha = 1$, increasing $\text{Da}$ from $0.1$ to $0.5$ shrinks the stationary instability region, demonstrating that faster chemical uptake suppresses phase separation. (c-d) At large $\alpha = 10$, increasing $\text{Da}$ similarly contracts both the phase separation region and the oscillatory instability region (colored surface). }
\label{fig::PeC_PeR_phi_phase_diagram_Da0_comparison}
\end{figure}

With these choices, we can now use the results of our linear stability analysis in \S\ref{sec::LSA} to construct the full phase diagram of chemotactic MIPS---which is parameterized by $\{\phi_0,\PeR\}$ as well as the three new dimensionless parameters $\{\alpha,\text{Da},\PeC\}$. For ease of visualization, we show the linear stability boundary by the gray surface in the three-dimensional $\{\phi_0,\PeR,\PeC\}$ phase space for $\text{Da}=0.3$ and $\alpha=1$ or $10$ in Fig.~\ref{fig::PeC_PeR_phi_phase_diagram}(a),(d). Below the stability boundary, the homogeneous state is unstable to small perturbations and the system undergoes phase separation, whereas above it, phase separation is suppressed. Cross-sections through this stability boundary at $\PeC=0.35$ and $0.8$ are shown by the black curves in panels (b)-(c) and (e)-(f) in Fig.~\ref{fig::PeC_PeR_phi_phase_diagram}, along with snapshots of numerical simulations at $t=2\times 10^4t_0$ shown by the inset panels, confirming that phase separation occurs below the stability boundary. The boundary at $\PeC=0$ corresponds to the conventional spinodal curve of non-chemotactic ABPs, below which typical MIPS occurs. We also show how the stability boundary changes with varying $\text{Da}$ in Fig.~\ref{fig::PeC_PeR_phi_phase_diagram_Da0_comparison}. Below, we highlight the key results emerging from our analysis; their mathematical proofs are given in Supplementary Information \S\ref{sec::SI_LSA}.\\




\noindent\textbf{\textit{Chemotaxis suppresses phase separation.}} In the case of $\text{Da}=0.3$, $\alpha=1$ [Fig.~\ref{fig::PeC_PeR_phi_phase_diagram}(a)-(c)], the stability boundary corresponds to $\PeC'=\text{Pe}_{C,\text{crit}}'$; the boundary given by $\alpha' = \alpha'_\text{crit}$ corresponds to $\PeR<0$ and is therefore inapplicable. Below the stability boundary shown in (a), the system thus exhibits a stationary instability. 
As indicated by the sloping nature of this boundary, the region of phase separation shrinks with increasing $\PeC$. Additionally, visual inspection of the simulation results in (b)-(c) shows that, whereas the dense and dilute phases continually coarsen over time in conventional MIPS, chemotaxis arrests this process---resulting in the formation of stationary, finite-sized dot and stripe patterns. We characterize these patterns further in \S\ref{sec::stationary_pattern}. 

Notably, these results are for consumers of a chemoattractant ($\PeC>0$), for which chemotaxis drives particle dispersal (Fig.~\ref{fig::schematics}). However, for the opposite case of consumers of a chemorepellant, chemotaxis drives aggregation, not dispersal, as discussed in \S\ref{sec::model}. Consistent with this expectation, we find the opposite behavior when $\PeC$ is increasingly negative---the region of phase separation instead \emph{expands} compared to the non-chemotactic case. \\



\noindent\textbf{\emph{Chemotaxis can generate an oscillatory instability.}} As noted in \S\ref{sec::LSA}, chemotaxis can also generate an oscillatory instability when the chemical signal diffuses so slowly that its gradient lags behind the moving particles. We explore this behavior by increasing $\alpha$ from $1$ to $10$, keeping $\text{Da}=0.3$. As shown by the gray surface in Fig.~\ref{fig::PeC_PeR_phi_phase_diagram}(d), the stability boundary now consists of two parts. The first part lying to the left of the left-most red dashed curve is independent of $\alpha$ and corresponds to $\PeC'=\text{Pe}'_{{C,\text{crit}}}$, as in panel (a). In this region, below the stability boundary, chemotaxis suppresses phase separation and generates a stationary instability, as discussed previously for $\alpha=1$. The second part of the stability boundary lies to the right of the left-most red dashed curve, and corresponds to $\alpha'=\alpha'_\text{crit}$, which is independent of $\PeC$.
In this region, below the stability boundary, phase separation always occurs, regardless of the value of $\PeC$. Moreover, when chemotactic dispersal is sufficiently fast ($\PeC' > \text{Pe}_{{C,*}}'$) --- i.e., to the right of the bowed colored surface in panel (d), above the green boundary in (e), and below the black boundary in (f) --- the system exhibits an oscillatory instability. The simulation results shown in (e)-(f) present examples of this instability, which results in the formation of traveling waves, traveling dots, and dynamic spiral patterns. Interestingly, the simulations indicate that oscillatory instability also arises below the green boundaries in (e), where the linear stability analysis predicts a stationary instability in this region instead. We revisit this point and provide a more accurate theoretical prediction for the incidence of oscillatory instability in \S\ref{sec::oscillatory_pattern}.\\

\noindent{\textbf{\emph{Faster chemical uptake suppresses phase separation.}}} How does increasing the rate of chemical uptake by the ABPs relative to its diffusion (i.e., increasing $\text{Da}$) influence phase behavior? We first examine the boundary corresponding to $\PeC'=\text{Pe}'_{C\text{crit}}$ demarcating the onset of stationary instability. When chemotactic dispersal is slow --- specifically, when $\PeC<\text{Da}\cdot\phi_0^3  k/S_0$ --- this stability boundary is independent of $\text{Da}$, as we show in \S\ref{sec::SI_LSA}. However, when chemotactic dispersal is sufficiently fast, the stability boundary shifts downward with increasing $\text{Da}$, as seen by comparing Figs.~\ref{fig::PeC_PeR_phi_phase_diagram_Da0_comparison}(a), \ref{fig::PeC_PeR_phi_phase_diagram}(a), and \ref{fig::PeC_PeR_phi_phase_diagram_Da0_comparison}(b). Next, we examine the boundary corresponding to $\alpha' = \alpha'_\text{crit}$ demarcating the onset of oscillatory instability. In this case, increasing $\text{Da}$ also shifts the stability boundary downward, as seen by comparing Figs.~\ref{fig::PeC_PeR_phi_phase_diagram_Da0_comparison}(c), \ref{fig::PeC_PeR_phi_phase_diagram}(d), and \ref{fig::PeC_PeR_phi_phase_diagram_Da0_comparison}(d). These results are also corroborated by Fig.~\ref{fig::PeC_alpha_phase_diagram}(b). Altogether, they demonstrate that faster chemical uptake suppresses phase separation, no matter the nature of the instability.

\section{Stationary patterns}
\label{sec::stationary_pattern}
\noindent\textbf{\emph{Amplitude equation formulation.}} The results in Fig.~\ref{fig::PeC_PeR_phi_phase_diagram}(b)-(c) showed the emergence of finite-sized dot and stripe patterns in the stationary instability region. While linear stability analysis can capture the onset of the instability that causes these patterns to form (\S\ref{sec::phase_diagram}), it cannot predict the morphology of the resulting patterns. To extend our analysis beyond the linear regime, we employ the amplitude equation method~\cite{Cross1993}---a systematic perturbative approach that describes pattern formation in the weakly nonlinear regime by tracking the slow evolution of pattern amplitude and phase. This method enables us to predict not only when stationary patterns emerge, but also their characteristic size, morphology, and stability, as demonstrated in other phase-separating biophysical systems~\cite{Yu2025_patternmembranes}.

To derive the amplitude equation, we expand the particle density and chemical concentration fields around the homogeneous steady state as $\phi = \phi_0 + \phi_1$ and $\tilde{c} = \tilde{c}_0 + \tilde{c}_1$, where $\phi_0$ and $\tilde{c}_0$ are the homogeneous steady state solutions introduced in \S\ref{sec::LSA}, and $\phi_1$ and $\tilde{c}_1$ are the leading-order perturbations that represent small but finite-amplitude deviations from the uniform state.
Following the standard amplitude equation formalism~\cite{Cross1993} and an established mean-field model of MIPS~\cite{Speck2014}, we approximate the normalized bulk free energy density $\tilde{g}_h$ using a fourth-order Ginzburg-Landau expansion around $\phi_0$. Taylor expanding the normalized bulk chemical potential $\tilde{\mu}_h(\phi) = \partial \tilde{g}_h/\partial \phi$ to third order in $\phi_1$ yields $\tilde{\mu}_h(\phi) \approx \tilde{\mu}_h(\phi_0) - a \phi_1 + b \phi_1^2 + \gamma \phi_1^3$ where $a \equiv - \partial \tilde{\mu}_h/\partial \phi$, $b \equiv (\partial^2 \tilde{\mu}_h/\partial \phi^2)/2$ and $\gamma \equiv (\partial^3 \tilde{\mu}_h/\partial \phi^3)/6$, with all derivatives evaluated at $\phi_0$. 
Since $\phi_1$ is small, the ABP mobility $M_0\phi_0$ and chemotactic coefficient $\chi_0\phi_0$ can be treated as constants.

With $g(\tilde{c}) = f(\tilde{c}) = \tilde{c}$ as before, the governing equations for $\phi_1$ and $\tilde{c}_1$ are then
\begin{equation}\label{eqn::perturbation}
\begin{aligned} 
  \pderiv{\phi_1}{t} &= M_0 \phi_0 \nabla^2 (-a\phi_1 + \gamma \phi_1^2 + b \phi_1^3 - \kappa \nabla^2 \phi_1) - \chi_0 \phi_0 \nabla^2 \tilde c_1, \\
  \pderiv{\tilde c_1}{t} &= D_c \nabla^2 \tilde c_1 - k (\phi_0 \tilde c_1 + \tilde{c}_0 \phi_1 + \tilde c_1\phi_1).
\end{aligned}
\end{equation}
The perturbations $\phi_1$ and $\tilde c_1$ can be  decomposed into Fourier modes, $\phi_1 = \sum_{n=1}^N A_{\phi,n} e^{i \mathbf{k}_n \cdot \mathbf{x}} +\mathrm{c.c.}$ and $\tilde c_1 = \sum_{n=1}^N A_{c,n} e^{i\mathbf{k}_n \cdot \mathbf{x}} + \mathrm{c.c.}$,
where the subscript $n=1,\ldots, N$ labels the Fourier modes, $\mathbf{k}_n$ is the wavevector of the $n$th mode (distinct from $\mathbf{q}$ used in the linear stability analysis), $A_{c,n}$ and $A_{\phi_n}$ are the complex amplitudes, and c.c. denotes the complex conjugate. 
To analyze spatially periodic patterns, it suffices to consider spatially uniform amplitudes with a single characteristic wavenumber $k_c=\abs{\bs{k}_n}$. 
Substituting the amplitude expansion into Eq.~\eqref{eqn::perturbation} then yields the amplitude equations:
\begin{widetext}
\begin{equation}\label{Eq:amplitude}
\begin{aligned}
    \dv{A_{\phi,n}}{t} &= - M_0 \phi_0k_c^2 \left[ \qty(\kappa k_c^2- a)A_{\phi,n} + \gamma \sum_{\mathbf{k}_n=\mathbf{k}_{m}+\mathbf{k}_{p}} A_{\phi,m}A_{\phi, p} + b \sum_{\mathbf{k}_n=\mathbf{k}_{m}+\mathbf{k}_{p}+\mathbf{k}_{r}} A_{\phi,m}A_{\phi,p}A_{\phi,r}\right] + \chi_0 \phi_0 k_c^2 A_{c,n},\\
    \dv{A_{c,n}}{t} &= -\qty(D_ck_c^2 + k \phi_0) A_{c,n} - k \tilde{c}_0 A_{\phi,n} - k \sum_{\mathbf{k}_n=\mathbf{k}_{m}+\mathbf{k}_{p}} A_{c,m} A_{\phi,p},
\end{aligned}
\end{equation}
\end{widetext}
which represent a dynamical system whose attractors represent patterns of different morphologies. We classify the emergence and stability of these states using bifurcation theory~\cite{Strogatz2018} below.\\

\begin{figure}
  \includegraphics[width=0.9\columnwidth]{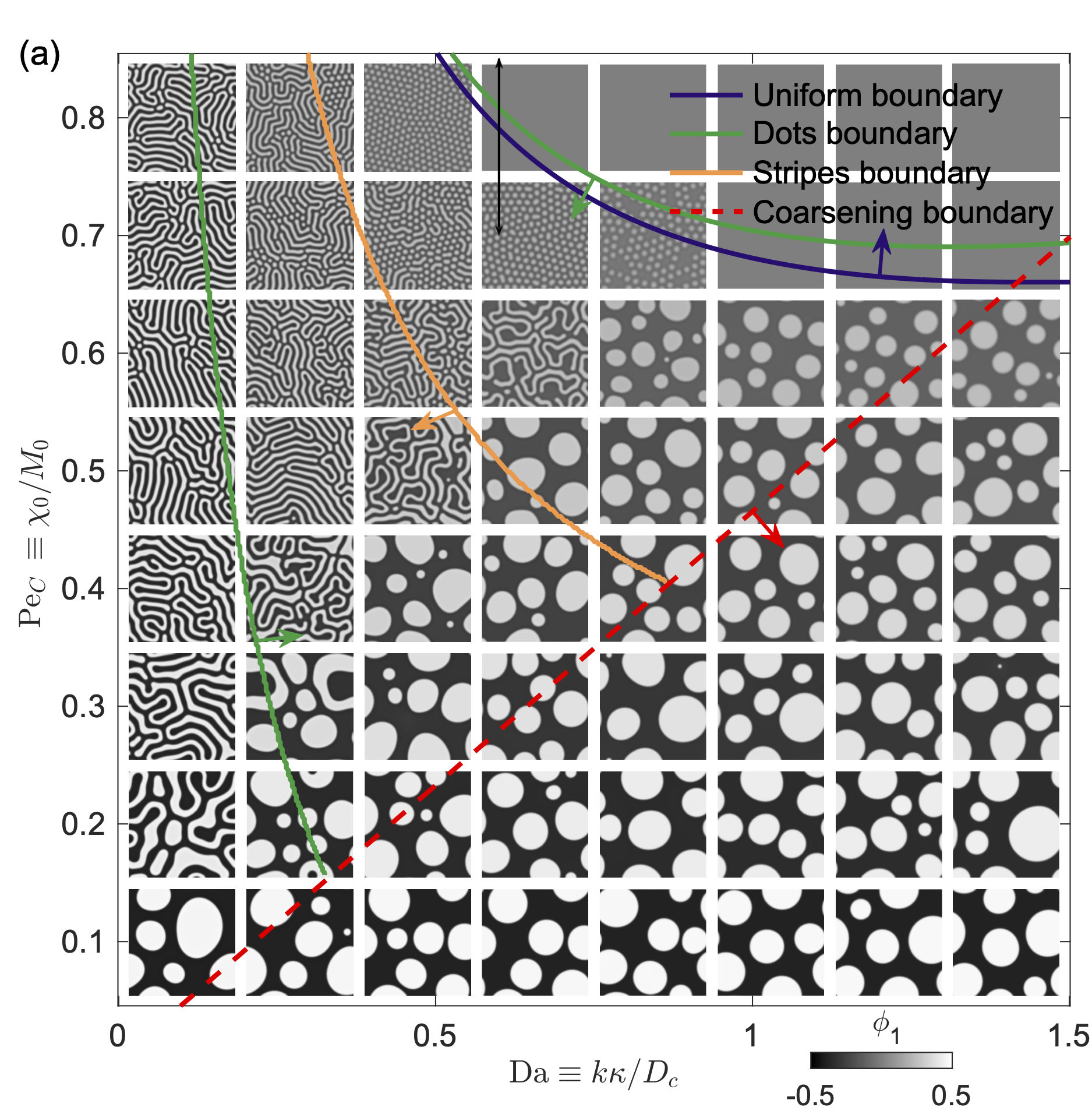}
  \includegraphics[width=0.9\columnwidth]{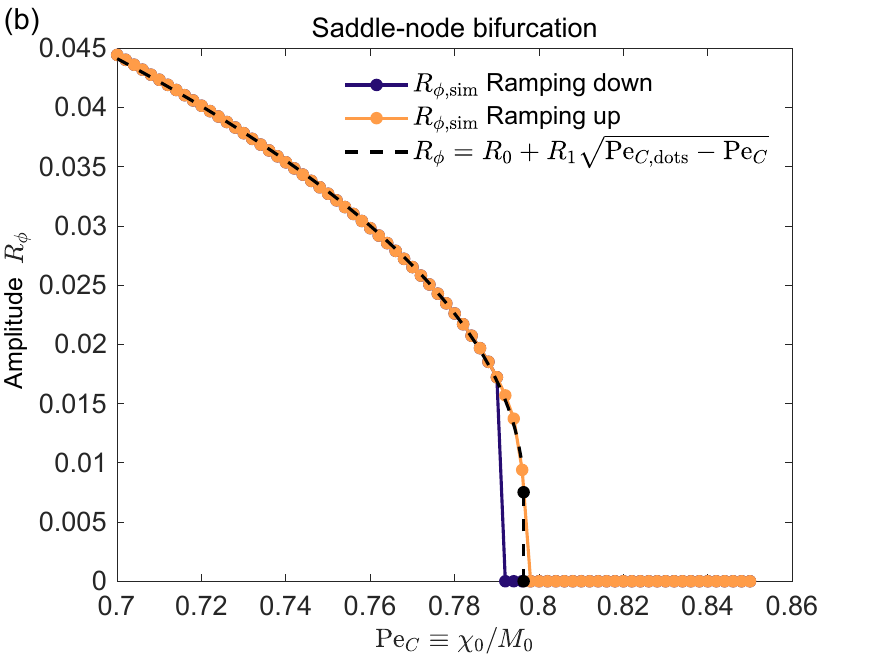}
  \caption{\textbf{Amplitude equation analysis quantitatively predicts the stability boundaries and bifurcation types of stationary patterns.} 
  (a) Phase diagram in $\PeC-\Da$ space with $\PeR=10^{-3}$, $\phi_0=0.775$, and $\alpha=2$, showing pattern morphologies from simulations of the weakly nonlinear model [Eq.~\ref{eqn::perturbation}]. 
  Stability boundaries predicted by the amplitude equation (solid curves) separate uniform (purple), stripes (yellow), and dots (green) solutions, with excellent agreement with simulations. Bistable regions exist where dots coexist with either stripes or uniform states. Below the red dashed line ($\sqrt{\Da\cdot\PeC \cdot\tilde{c}_0}=\Da\cdot\phi_0$), patterns coarsen to system size. Simulations are performed on a $[200l_0,200l_0]$ periodic domain and solved on a $512\times 512$ grid. The snapshots are taken at $t=10^5t_0$.
  These parameters are selected such that no oscillatory patterns emerge.
  (b) Hysteresis at the uniform-to-dots transition [$\Da=0.6$, double-headed vertical arrow in (a)] demonstrates a saddle-node bifurcation. We calculate the amplitude from each simulation as $R_{\phi,\text{sim}} = \sqrt{\int{(\phi(\mathbf{x})-\phi_0)^2d\mathbf{x}}/(6\int{d\mathbf{x}})}$. Ramping $\PeC$ down (blue) shows that the pattern persists below the stability threshold, while ramping up (orange) shows a delayed onset. Amplitude follows the fitted curve $R_\phi = R_0 + R_1 \sqrt{\text{Pe}_{\text{C,dots}}-\PeC}$ (dashed line), where $R_0$ and $R_1$ are fit parameters; $R_0$ is nonzero at the bifurcation, indicating saddle-node behavior.}
  \label{fig::stationary_pattern}
\end{figure}

\noindent\textbf{\emph{Stability of pattern morphologies.}} We consider three steady-state solutions of the amplitude equations --- uniform, stripes, and arrays of dots. It suffices to consider up to $N=3$ modes whose wavevectors are separated by $2\pi/3$ angles; the normalized characteristic wavenumber $\tilde{k}_c \equiv k_c l_0$. For convenience, we express the complex amplitudes in polar form $A_{\phi,n} = R_{\phi,n} e^{i\psi_n}$ and $A_{c,n} = R_{c,n} e^{i\varphi_n}$; explicit expressions for the moduli $R_{\phi,n}$, $R_{c,n}$ and phases $\psi_n$, $\varphi_n$ are given by Eqs.~\eqref{eqn::SI_complex_1}-~\eqref{eqn::SI_complex_4} in Supplementary Information \S\ref{sec::SI_amplitude}.
Our results are summarized in Fig.~\ref{fig::stationary_pattern}(a), which shows the amplitude equation predictions in the $\PeC-\Da$ phase diagram along with simulations of the weakly nonlinear model [Eq.~\eqref{eqn::perturbation}]. 

First, we consider the uniform (homogeneous) solution. Since all amplitudes are zero, the stability is completely captured by linear stability analysis. As detailed in \S\ref{sec::LSA}, we expect this state to be stable in the region above the purple curve in Fig.~\ref{fig::stationary_pattern}(a). In this regime of large $\PeC$ and $\Da$, fast chemotaxis and chemical uptake suppress phase separation.  

Next, we consider the stripes solution, which has one nonzero Fourier mode: $R_{\phi,1}=R_\phi$,  $R_{c,1}=R_c$, and $R_{\phi,2}=R_{\phi,3}=R_{c,2}=R_{c,3}=0$.
As derived in \S\ref{sec::SI_stripes}, the steady-state amplitude is
\begin{equation} \label{eqn::R_phi_stripe}
  R_\phi = \frac{1}{\sqrt{3b}}\sqrt{a-\tilde{k}_c^2-  \frac{\Da\cdot\PeC \cdot\tilde{c}_0}{\tilde{k}_c^2 + \Da\cdot\phi_0}} \leq \sqrt{\frac{a-a_c}{3b}}
\end{equation}
with $R_c = \frac{\da\cdot c_0}{\tilde{k}_c^2 + \da\cdot\phi_0} R_\phi$. Therefore, the stripe pattern exists only when $a>a_c\equiv 2\sqrt{\Da\cdot\PeC \cdot\tilde{c}_0} - \Da\cdot\phi_0$, with an amplitude that scales as $R_\phi \sim \sqrt{a-a_c}$. That is, as $a$ exceeds $a_c$, stripes emerge from the uniform state through a pitchfork bifurcation~\cite{Strogatz2018}. Moreover, the pattern amplitude is maximized when $\tilde{k}_c^2=\tilde{k}_{c,\max}^2\equiv\max\qty{\sqrt{\Da\cdot\PeC \cdot\tilde{c}_0} - \Da\cdot\phi_0,0}$. 
Note that the conditions $a>a_c$ and $k_{c,\text{max}}^2>0$ are equivalent to $\PeC'<\text{Pe}'_{{C,\text{crit}}}$ and $\text{Da}'<1$, which lie in the unstable regime of the linear stability analysis [Eqs.~\eqref{eqn::Da_crit} and \eqref{eqn::PeC_crit}].


The stability of the stripe morphology, which is determined by the Jacobian of the full amplitude equation, can be decomposed into stability in two directions: $R$, which is the amplitude, and $\Theta \equiv \theta-\psi$, which is the phase difference between $\phi$ and $c$. At steady state, $\Theta =\pi (+2n\pi)$, which means that they are in antiphase: Regions abundant with particles are scarce in the chemoattractant and vice versa. In the $\Theta$ direction, the antiphase fixed point ($\cos\Theta=-1$) is stable when $\alpha < 4/(\PeC \tilde{c}_0)$. When $\alpha> 4/(\PeC \tilde{c}_0)$, a new fixed point appears with a finite phase shift. It represents oscillatory patterns with a velocity proportional to $\sin\Theta$ [\S\ref{sec::oscillatory_pattern}]. 
The stability in the $R$ direction can be obtained by numerically computing the eigenvalues of the $4\times4$ Jacobian and finding the region with only negative eigenvalues. The combined stability region of the stripe solution is to the left of the orange line in Fig.~\ref{fig::stationary_pattern}(a). 



We then consider the dots solution, which is represented by 3 Fourier modes ($N=3$) of equal amplitudes: $R_{\phi,1}=R_{\phi,2}=R_{\phi,3}=R_\phi$, $R_{c,1}=R_{c,2}=R_{c,3}=R_c$.  
Using the same characteristic wavenumber $\tilde{k}_c^2=\tilde{k}_{c,\max}^2$ that maximizes the amplitude of $R_\phi$ (which also maximizes the amplitude for the stripes solution), as derived in \S\ref{sec::SI_dots}, the steady-state amplitude is
\begin{equation} \label{eqn::dots_amplitude}
  R_\phi = \frac{1}{15b}\qty(\abs{\gamma-\Da} + \sqrt{\qty(\gamma-\Da)^2 + 15b(a-a_c)}),
\end{equation}
again with $R_c = \frac{\da c_0}{\tilde{k}_c^2 + \da\phi_0} R_\phi$. Therefore, the dots pattern exists only when $\qty(\gamma-\Da)^2 + 15b(a-a_c)>0$.
It emerges through a saddle-node bifurcation, with a finite amplitude $R_{\phi, \mathrm{crit}}=\abs{\gamma-\Da}/(15b)$ at the critical point. 
The dots and homogeneous solution can coexist in the bistable region of $- \qty(\gamma-\Da)^2 / (15b) < a-a_c < 0$. 
Similar to the stripes case, the stability boundaries of the dots solution are obtained by numerically calculating the eigenvalues of the Jacobian [\S\ref{sec::SI_dots}]. The resulting stable region of the dots solution is in between the green lines in Fig.~\ref{fig::stationary_pattern}(a). 


To examine this saddle-node bifurcation in detail, we perform a hysteresis analysis.  In Fig. \ref{fig::stationary_pattern}(b), we zoom in on the region near the stability boundary of the uniform and dots solution and track the amplitude of the pattern $R_\phi$ as we ramp the value of $\PeC$ up and down across the phase boundary [along the vertical double-headed black arrow in Fig.~\ref{fig::stationary_pattern}(a)]. Since the uniform solution is stable when $a<a_c$, as we decrease $\PeC$ [which decreases $a_c$; see dark blue points in (b)], the amplitude remains zero, until $a=a_c$ where the uniform state is destabilized and dots emerge. 
On the other hand, since the dots solution is stable when $a>a_c-(\gamma-\da)^2/(15b)$, as we increase $\PeC$ [which increases $a_c$; see orange points in (b)], the dots solution remain stable until $a=a_c-(\gamma-\da)^2/(15b)$, which is the saddle-node bifurcation point (black point). 
Thus, simulation of the model confirms the predicted hysteresis in $a\in(a_c-(\gamma-\da)^2/(15b),a_c)$.

Expanding the amplitude [Eq.~\eqref{eqn::dots_amplitude}] with respect to $\PeC$ at the critical value at the bifurcation point $\text{Pe}_{\text{C,dots}}$, the amplitude follows $R_\phi = R_0 + R_1 \sqrt{\text{Pe}_{\text{C,dots}}-\PeC}$ for $\PeC<\text{Pe}_{\text{C,dots}}$ (black dashed line). As a key characteristic of the saddle-node bifurcation, the amplitude has a nonzero value $R_0$ at the bifurcation point. We find excellent agreement between the black dashed line and the numerical results.\\

\noindent\textbf{\emph{Pattern selection and coarsening.}} For both stripes and dots, the amplitude is maximized at characteristic wavevector $\tilde{k}_c=\tilde{k}_{c,\max}$, which establishes the preferred domain size $L_c=2\pi/k_{c,\mathrm{max}}$. When $\sqrt{\Da\cdot\PeC\cdot \tilde{c}_0} < \Da\cdot\phi_0$, however, the maximum amplitude is reached in the long wavelength limit, which means that the domains will coarsen until reaching the system size. This boundary is indicated by the red dashed line in Fig. \ref{fig::stationary_pattern}(a) ($\PeC = \Da \cdot \phi_0^2 \tilde{c}_0^{-1}$). Coarsening occurs below this line, where chemotaxis is slow, and the chemical uptake rate is fast (small $\PeC$ and large $\Da$). 

Altogether, this analysis predicts six different stable stationary patterns in 
the $\PeC-\Da$ phase diagram shown in Fig.~\ref{fig::stationary_pattern}(a): (i) uniform above the purple curve, (ii) dot-uniform coexistence between the purple curve and the rightmost green curve, (iii) dots in between the orange and the purple curve, (iv) dot-stripe coexistence in between the leftmost green and orange curves, (v) stripes to the left of the leftmost green curve, and (vi) coarsening below the red dashed line. 
Our simulations confirm all these predictions, as shown by the inset panels in Fig.~\ref{fig::stationary_pattern}(a), except for slightly above the coarsening boundary, where the observed coarsening is possibly due to a finite system size. 




Having established the stability boundaries of each pattern type, we now examine the bistability and hysteresis that arise when multiple solutions are stable. The phase diagram in Fig.~\ref{fig::stationary_pattern}(a) contains two bistable regions in between the leftmost green and orange curves and between the rightmost green and purple curves, where the dots pattern can coexist with stripes or the uniform state, respectively. A consequence of this bistability is that the pattern morphology should be hysteretic if parameters are varied in different directions. Fig.~\ref{fig::stationary_pattern}(b) already demonstrated hysteresis at the uniform-dots saddle-node bifurcation. We extend this analysis in Fig.~\ref{fig::stationary_pattern_hysteresis}, which shows the influence of ramping up and down the value of $\Da$ for each $\PeC$, using the final steady state at the previous $\Da$ as the initial condition for the next $\Da$ value. We again observe the coexistence of dots and stripes (or uniform) solutions in the region where both states are stable. As expected for a bistable system, the dominance of one state over the other depends on the direction of the ramping: for example, when exploring dots-stripes coexistence, we find that stripes dominate in the coexistence region when ramping up the value of $\Da$, while dots dominate when ramping down the value of $\Da$. We observe a similar hysteretic effect in the region where both dots and the uniform state are stable.

While the above results demonstrate excellent agreement between amplitude equation predictions and simulations of the weakly nonlinear model [Eq.~\eqref{eqn::perturbation}], we now assess the range of validity of the amplitude equation approach by comparing with numerical simulations of the full model. As shown in Fig. \ref{fig::stationary_pattern_full_model}, as we move from the uniform phase to pattern forming phase by decreasing $\Da$ and $\PeC$, the amplitude equation qualitatively captures the transition from uniform to dots to stripes as well as bistability between them. However, the agreement deteriorates far from the stability boundaries, where the weakly nonlinear approximation breaks down.
Notably, the coarsening boundary $\PeC > \Da \phi_0^2 \tilde{c}_0^{-1}$ (below the red dashed line) remains an accurate predictor of when patterns will coarsen to the system size, even in regions where other quantitative predictions deviate.

\section{Oscillatory patterns}
\label{sec::oscillatory_pattern}
\noindent\textbf{\emph{Plane wave analysis and existence conditions.}} Having analyzed stationary patterns in \S\ref{sec::stationary_pattern}, we now turn to the oscillatory patterns---traveling waves, traveling dots, and spirals---that emerge when chemical diffusivity is sufficiently slow compared to particle mobility ($\alpha'>\alpha'_\text{crit}$) and chemotactic dispersal is sufficiently strong ($\PeC'>\text{Pe}'_{{C,\text{crit}}}$). We again use the amplitude equation method to predict when these patterns emerge, their velocities, and their amplitudes, which we validate through numerical simulations.

To understand the basic properties of oscillatory patterns, we begin by analyzing the simplest case: a plane-wave solution with $N=1$ mode. We assume the wave propagates with a constant phase difference $\Theta\equiv \theta-\psi$ between the chemical and particle density fields, and with angular velocity $\omega = \dd\theta/\dd{t} = \dd{\psi}/\dd{t}$. That is, $\phi(\mathbf{x},t) - \phi_0 = R_\phi e^{i (k_c x + \omega t)}+\mathrm{c.c.}$ and $\tilde{c} - \tilde{c}_0 =  R_c e^{i (k_c x + \omega t + \Theta) }+\mathrm{c.c}$. 
Substituting this ansatz into the amplitude equations [Eq.~\eqref{Eq:amplitude}] yields the traveling wave velocity, as derived in \S\ref{sec::SI_oscillatory}:
\begin{equation} \label{eqn::traveling_wave_velocity}
  v = \frac{\omega}{k_c} = v_0 \sqrt{\Da\cdot \tilde{c}_0 \phi_0 \alpha^{-1} \cdot\PeC} \sin\Theta,
\end{equation}
where the phase difference satisfies
\begin{equation} \label{eqn::traveling_wave_phase_difference}
  \cos\Theta = - \frac{\tilde{k}_c^2 + \Da \cdot\phi_0}{\tilde{k}_c \sqrt{\Da \cdot\tilde{c}_0 \alpha \phi_0\cdot\PeC}}.
\end{equation}

Three key insights emerge from these equations. First, a finite phase shift between the particle and chemical fields ($\sin\Theta\neq 0$) is required for traveling waves to emerge; when $\Theta\to\pi$, the velocity $v$ vanishes, and patterns become stationary. 
Second, since $\cos\Theta >-1$, traveling waves can only exist when
\begin{equation} \label{eqn::traveling_wave_existence}
  \tilde{k}_c + \frac{\Da\cdot\phi_0}{\tilde{k}_c} <  \sqrt{\Da\cdot \tilde{c}_0\alpha \phi_0\cdot\PeC}.
\end{equation}
This condition requires both fast chemotaxis (large $\PeC$) and slow chemical diffusion (large $\alpha$). 
Third, the amplitude is given by 
\begin{equation} \label{eqn::R_phi_traveling_wave}
  R_\phi = \frac{1}{\sqrt{3b}}\sqrt{ a-\tilde{k}_c^2 -  \frac{\PeC \cdot\Da \cdot\tilde{c}_0}{\tilde{k}_c^2+\Da\cdot\phi_0}\cos^2\Theta},
\end{equation}
which is only stable when
\begin{equation} \label{eqn::traveling_wave_stability}
  \tilde{k}_c^2 + \frac{\Da}{\tilde{k}_c^2 \alpha} + \frac{1}{\alpha\phi_0} < a,
\end{equation}
as detailed in \S\ref{sec::SI_oscillatory}.\\

\noindent\textbf{\emph{Two bifurcation regimes.}} Two distinct bifurcation regimes emerge depending on which condition---existence [Eq.~\eqref{eqn::traveling_wave_existence}] or stability [Eq.~\eqref{eqn::traveling_wave_stability}]---is more restrictive. 
In both regimes, traveling waves are only stable when chemical diffusion is sufficiently slow compared to particle mobility: namely $\alpha>\alpha_c$ where $\alpha_c$ is a critical value that depends on $\PeC$.

The first regime corresponds to small $\PeC$, such that the existence condition [Eq.~\eqref{eqn::traveling_wave_existence}] is more restrictive. Minimizing the left-hand side of this condition with respect to $\tilde{k}_c$ gives $\alpha_c = 4/(\PeC \cdot\tilde{c}_0)$, which is achieved at the critical wavenumber  $\tilde{k}_c = \sqrt{\Da\cdot \phi_0}$. At this wavenumber, the wave velocity is maximized:
\begin{equation} \label{eqn::traveling_wave_velocity_low_PeC}
  \frac{v}{v_0} = \alpha^{-1} \sqrt{{\Da \cdot\tilde{c}_0}{ \phi_0\cdot\PeC}\qty(\alpha-\alpha_c)}.
\end{equation}
The key signature of this regime is that as when $\alpha<\alpha_c$, the stripes solution is stable, while as $\alpha$ increases past $\alpha_c$, stripes continuously transition to traveling waves. The amplitude $R_\phi$ remains finite and continuous across the transition, while the velocity grows continuously from zero as $v/v_0 \propto \alpha^{-1}\sqrt{\alpha-\alpha_c}$. This behavior is the signature of an infinite-period bifurcation at $\alpha=\alpha_c$.

The second regime corresponds to large $\PeC$, where the stability condition [Eq.~\eqref{eqn::traveling_wave_stability}] is more restrictive.
Minimizing the left-hand side of this condition yields $a > 2\sqrt{\Da/\alpha} + (\alpha\phi_0)^{-1}$, which gives $\alpha_c = a^{-2} \qty(\sqrt{\Da} + \sqrt{\Da+a\phi_0^{-1}})^2 $, above which the traveling wave solution is stable. Note that $\alpha \leq \alpha_c$ is equivalent to the stability condition of the uniform state $\alpha' \leq \alpha'_\text{crit}$. At the critical wavenumber $\tilde{k}_c = (\Da/\alpha)^{1/4}$, the velocity is
\begin{multline} \label{eqn::traveling_wave_velocity_high_PeC}
  \frac{v}{v_0} = \alpha^{-1} \sqrt{\Da \cdot\tilde{c}_0 \phi_0\cdot\PeC} \\
  \times \sqrt{\alpha - \frac{\sqrt{\alpha\cdot\Da} \phi_0}{\PeC \cdot\tilde{c}_0} \qty(1+\frac{1}{\sqrt{\Da\cdot\alpha}\phi_0})^2}
\end{multline}
and the amplitude is
\begin{equation} \label{eqn::R_phi_traveling_wave_high_Pe}
  R_\phi = \frac{1}{\sqrt{3b}}\sqrt{ a-\frac{1}{\alpha\phi_0}-2\sqrt{\frac{\Da}{\alpha}}}.
\end{equation}
The key signature of this regime is that as $\alpha$ increases past $\alpha_c$, the uniform state transitions to traveling waves. The amplitude grows continuously from zero as $R_\phi \propto \sqrt{\alpha-\alpha_c}$, while the velocity undergoes a discontinuous jump from zero to a finite value. This behavior is the signature of a supercritical Hopf bifurcation at $\alpha=\alpha_c$.

In summary, the critical $\alpha$ above which traveling waves exist and are stable is
\begin{equation} \label{eqn::alpha_c}
  \alpha_c = \max{\left\{ \frac{4}{\PeC \cdot\tilde{c}_0}, \frac{1}{a^2}\qty(\sqrt{\Da} + \sqrt{\Da+a\phi_0^{-1}})^2 \right\}}.
\end{equation}
The $\PeC$ value dividing the two regimes is given by equating these two expressions.\\

\begin{figure*}
  \includegraphics[width=\textwidth]{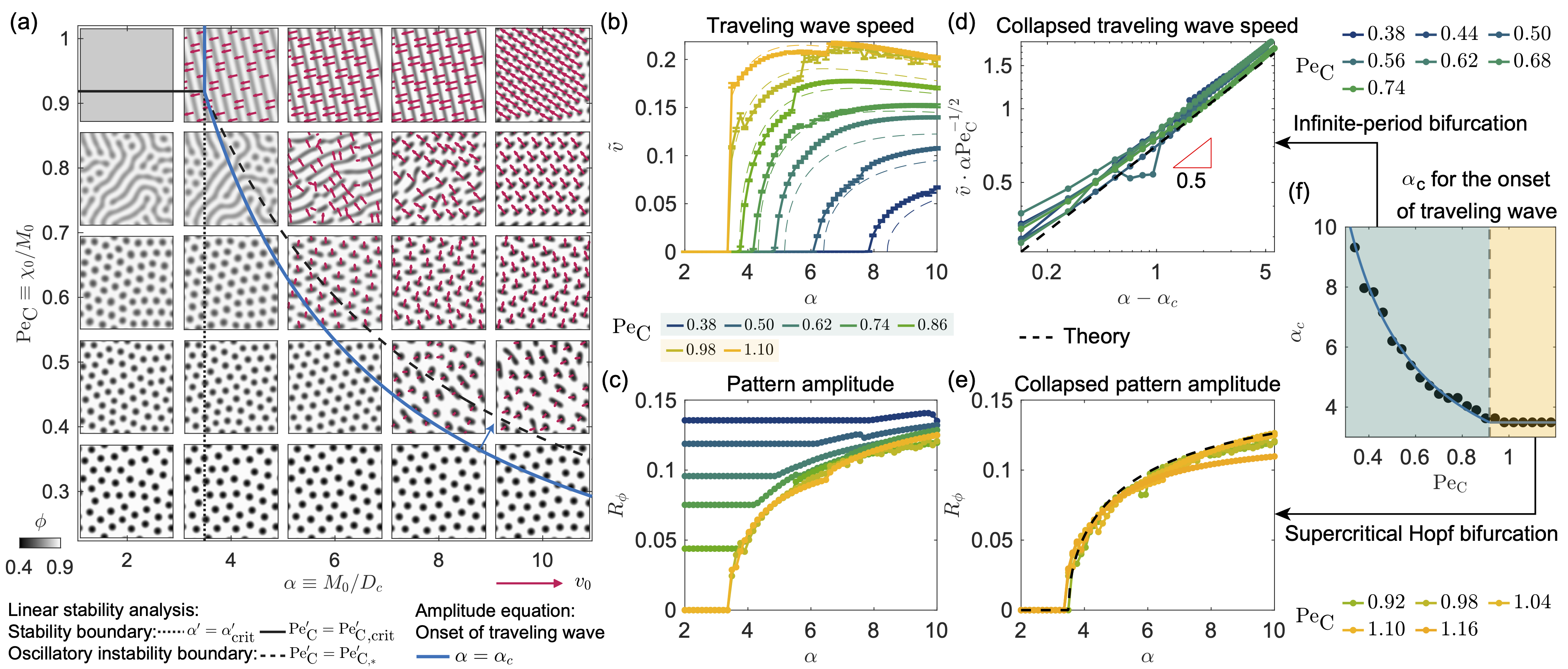}
  \caption{\textbf{Amplitude equation predictions for oscillatory patterns agree quantitatively with simulations across the full $\PeC-\alpha$ phase space.} (a) Simulations in $\PeC-\alpha$ space ($\text{Da}=0.5$, $\PeR=10^{-3}$, $\phi_0=0.8$ and $S_0/k=1$) show traveling waves and dots emerging when $\alpha$ exceeds a critical value that decreases with $\PeC$. Blue curve shows amplitude equation prediction [Eq.~\eqref{eqn::alpha_c}], which accurately captures the onset of oscillatory patterns. Black dashed curve shows the linear stability prediction [Eq.~\ref{eqn::critical_condition_oscillatory}], which overestimates the transition. Simulations at $\alpha=2$ start from the uniform state with added noise as the initial condition. The value of $\alpha$ was ramped up for each $\PeC$, i.e., simulations at $\alpha>2$ use the final state of the previous $\alpha$ as the initial condition. Red arrows indicate pattern velocities; snapshots are at $t=5\times 10^3t_0$. 
  (b) Average normalized pattern speed $\tilde{v}=v/v_0$ as a function of $\alpha$ at different $\PeC$. The speed is obtained from each simulation by averaging in space for each frame after the characteristic size of the pattern is less than 10\% different from the final state, and then averaging in time to obtain the mean and standard deviation, as shown by the error bars. Dashed curves show the amplitude equation predictions [Eqs.~\eqref{eqn::traveling_wave_velocity_low_PeC} and \eqref{eqn::traveling_wave_velocity_high_PeC}]. At small $\PeC$ (blue, green), velocity grows continuously from zero (infinite-period bifurcation); at large $\PeC$ (orange), velocity jumps discontinuously (supercritical Hopf bifurcation). (c) Pattern amplitude increases with $\alpha$ and decreases with $\PeC$, reflecting stronger phase separation when chemotaxis and chemical diffusion are slow. The amplitude is obtained from each simulation as  $R_{\phi,\text{sim}} = \sqrt{\int{(\phi(\mathbf{x})-\phi_0)^2d\mathbf{x}}/(2\int{d\mathbf{x}})}$.
  (d) Velocity scaling collapse in infinite-period regime ($\PeC < 0.917$) follows $\tilde{v} \cdot \alpha \cdot\PeC^{-1/2} = \sqrt{\Da\cdot \tilde{c}_0 \phi_0 (\alpha-\alpha_c)}$ (dashed line).
  (e) Amplitude scaling collapse in Hopf regime ($\PeC > 0.917$): $R_{\phi}$ follows Eq.~\eqref{eqn::R_phi_traveling_wave_high_Pe} (dashed line) independent of $\PeC$. 
  (f) Critical $\alpha_c$ for oscillatory onset matches analytical prediction [Eq.~\eqref{eqn::alpha_c}, blue curve] across all $\PeC$; vertical dashed line separates the two bifurcation regimes. Points are obtained from simulations by identifying the first $\alpha$ when $\tilde{v}>0.01$.
  }
  \label{fig::alpha_ramp}
\end{figure*}

\noindent\textbf{\emph{Comparison with numerical simulations.}} We now validate these predictions through numerical simulations. 
Fig.~\ref{fig::alpha_ramp}(a) shows the full model simulated across the $\PeC-\alpha$ phase space with $\text{Da}=0.5$, $\PeR=10^{-3}$, $\phi_0=0.8$.
As expected from linear stability analysis [Fig. \ref{fig::PeC_alpha_phase_diagram}(b)], to the left of the stability boundary $\alpha'=\alpha'_\text{crit}$ (dotted line at $\alpha=2$), the system reaches stationary states: either uniform (above the solid curve where $\PeC'>\text{Pe}'_{C,\text{crit}}$) or stationary stripes/dots (below it).

To study the transition to oscillatory patterns, we gradually increase $\alpha$ for each value of $\PeC$, using the final state from the previous $\alpha$ as the initial condition. Traveling stripes and dots emerge when $\alpha$ exceeds a critical value that decreases with increasing $\PeC$, consistent with our predictions. We quantify the velocity of stripes that span the entire domain using the level-set velocity ($\mathbf{u}=-\partial_t \phi \nabla \phi / |\nabla\phi|^2$, i.e., the velocity at which contours of $\phi(x,t)$ move in the direction along the gradient), denoted by the red arrows in Fig.~\ref{fig::alpha_ramp}(a). 
The speed of the pattern $v$ is the average of $|\mathbf{u}|$ in space, excluding regions with small $|\nabla\phi|$.
For traveling dots and short stripes with a small aspect ratio, the velocity $\mathbf{u}$ is obtained by tracking the center of mass of the pattern instead. The speed $v$ is the average of $|\mathbf{u}|$ of all the dots or stripes.

The amplitude equation prediction [Eq.~\eqref{eqn::alpha_c}, blue curve] accurately captures the onset of oscillatory patterns, as shown in Fig.~\ref{fig::alpha_ramp}(a). By contrast, the linear stability prediction [Eq.~\ref{eqn::critical_condition_oscillatory}] overestimates the critical $\alpha$, as shown by the black dashed curve. 
Nevertheless, it captures the qualitative trend of decreasing critical $\alpha$ with increasing $\PeC$. This discrepancy is expected: linear stability analysis applies only near the uniform state, while amplitude equations remain valid for transitions from both uniform and patterned states.\\

\noindent\textbf{\emph{Velocity and amplitude scaling.}} Next, we examine the velocity and amplitude predicted by the amplitude equations by performing numerical simulations with high resolutions of $\alpha$ and $\PeC$ values. The results are shown in Fig.~\ref{fig::alpha_ramp}(b,c) with snapshots shown in Fig.~\ref{fig::alpha_ramp_phase_diagram}. 

Fig.~\ref{fig::alpha_ramp}(b) shows the average normalized speed $\tilde{v}=v/v_0$ as a function of $\alpha$ for different values of $\PeC$, with dashed lines showing amplitude equation predictions [Eqs.~\eqref{eqn::traveling_wave_velocity_low_PeC} and \eqref{eqn::traveling_wave_velocity_high_PeC}].
The data confirm the two predicted bifurcation types: at small $\PeC$ (blue and green curves), $\tilde{v}$ increases continuously from zero (infinite-period bifurcation), while at large $\PeC$  (orange curves), $\tilde{v}$ jumps discontinuously (Hopf bifurcation). The critical $\alpha$ decreases with $\PeC$ as predicted. Additionally, at a constant $\PeC$ (constant particle diffusivity $M_0$), increasing $\alpha$ by decreasing chemical diffusivity first increases and then decreases the velocity $\tilde{v}$, vanishing as $\tilde{v} \sim \alpha^{-1/2}$ when $\alpha \to \infty$, just as predicted by the theory [Eqs.~\eqref{eqn::traveling_wave_velocity} and \eqref{eqn::traveling_wave_phase_difference}].
Some of the lines also feature a second discontinuous jump, which can be attributed to a morphological change from stripes to dots (e.g., as seen in Fig.~\ref{fig::alpha_ramp_phase_diagram}). 
Stripes are preferred at large $\PeC$ and small $\alpha$ while dots are found at small $\PeC$ and large $\alpha$.


Fig.~\ref{fig::alpha_ramp}(c) shows that the pattern amplitude $R_\phi$ increases with increasing $\alpha$ and decreasing $\PeC$, indicating stronger phase separation when chemotaxis and chemical diffusion are slow---consistent with the competition between MIPS clustering and chemotactic dispersal.
For small $\PeC$ (blue and green curves), the amplitude remains constant when the pattern is stationary ($\alpha<\alpha_c$) and increases continuously with increasing $\alpha>\alpha_c$. For large $\PeC$ (orange curves), the amplitude grows continuously from zero beyond the Hopf bifurcation point. 

To further distinguish different bifurcation types at small and large $\PeC$, we investigate the scaling of velocity and amplitude. 
As shown in Fig. \ref{fig::alpha_ramp}(d-f), we observe the predicted scaling of velocity and pattern amplitude near critical points.  
In the infinite-period regime ($\PeC < 0.917$), when $\alpha>\alpha_c$, velocities at different $\PeC$ collapse onto the universal curve 
$\tilde{v} \cdot \alpha \cdot \PeC^{-1/2} = \sqrt{\Da\cdot \tilde{c}_0 \phi_0 (\alpha-\alpha_c)}$ [Eq.~\eqref{eqn::traveling_wave_velocity_low_PeC}], as shown by the black dashed line in Fig.~\ref{fig::alpha_ramp}(d).
In the supercritical Hopf regime ($\PeC > 0.917$), when $\alpha>\alpha_c$, amplitudes at different $\PeC$ increases continuously from zero with increasing $\alpha$ and is independent of $\PeC$ [Fig.~\ref{fig::alpha_ramp}(e)], in good agreement with the analytical prediction [Eq.~\eqref{eqn::R_phi_traveling_wave_high_Pe}]. The traveling wave speed jumps from 0 to a finite value when $\alpha=\alpha_c$ as shown in Fig.~\ref{fig::alpha_ramp}(b). Fig.~\ref{fig::alpha_ramp}(f) shows excellent agreement between the simulated and theoretically predicted [Eq.~\eqref{eqn::alpha_c}] critical $\alpha_c$ across the full $\PeC$ range, with the vertical dashed line marking the regime boundary.

\begin{figure}
  \centering
  \includegraphics[width=\columnwidth]{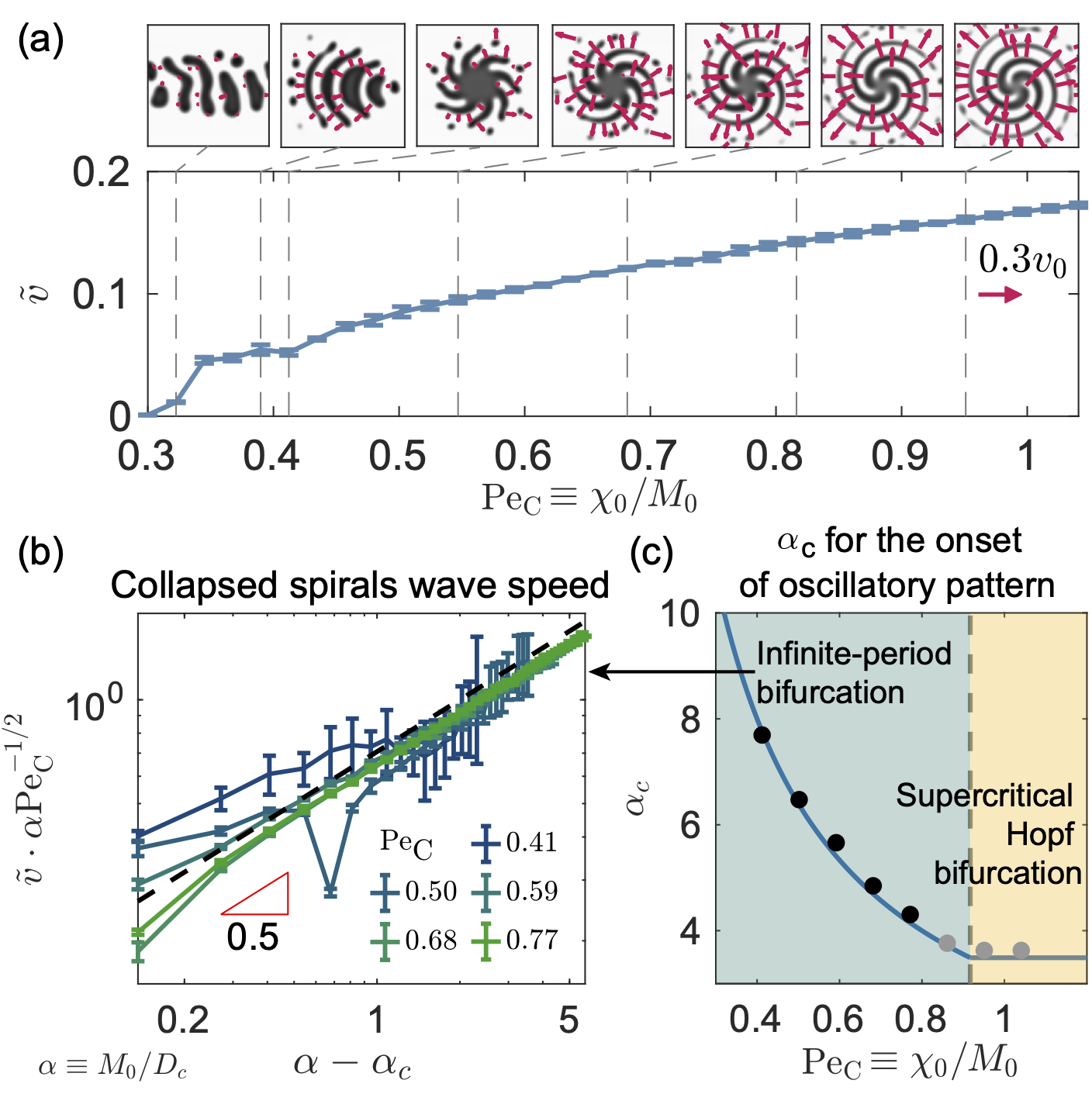}
  \caption{\textbf{Spiral waves obey the same universal scaling laws as traveling waves despite their distinct topology.} (a)~A family of spiral waves as a function of $\PeC$ at $\alpha=10$ and $\text{Da}=0.5$, $\PeR=10^{-3}$, $\phi_0=0.8$, and $S_0/k=1$.   Snapshots are taken at $t=5\times 10^3t_0$; dashed vertical lines mark the corresponding $\PeC$ in the plot below, which shows the dependence of the normalized average speed $\tilde{v}$ on $\PeC$. Red arrows indicate the level set speed of the patterns with the same scale bar next to the snapshots. Velocities smaller than $0.005u_0$ are not shown. Increasing $\PeC$ increases spiral velocity and decreases topological charge (number of arms); below $\PeC\approx0.4$, spirals fragment into ripples before becoming stationary. 
  (b)~The spiral wave speed collapses onto the same curve as traveling stripes: $\tilde{v} \cdot \alpha \cdot\PeC^{-1/2} = \sqrt{\Da\cdot \tilde{c}_0 \phi_0 (\alpha-\alpha_c)}$ (dashed line), demonstrating universality of the infinite-period bifurcation. $\PeC$ values correspond to the black dots in (c).
  (c)~The critical $\alpha_c$ for the onset of spiral waves (black dots) as a function of $\PeC$ matches the onset of traveling stripes (gray dots) and the analytical prediction from Eq.~\eqref{eqn::alpha_c} (blue curve).
  }
  \label{fig::spiral}
\end{figure}

Altogether, these results show that the amplitude equation analysis of the traveling wave solution based on the weakly nonlinear model provides a strikingly good agreement with the simulations over all ranges of $\alpha$ and $\PeC$, despite the fact that the majority of the observed patterns are traveling dots, and the full nonlinear model is solved in the simulations. We also find excellent agreement between the results of the amplitude equations and simulations in 1D (where the dot patterns do not exist) [\S\ref{sec::SI_oscillatory_patterns_1D}]. 

\noindent\textbf{\emph{Spiral waves obey the same scaling laws.}} Beyond traveling stripes and dots, the oscillatory regime also supports spiral wave solutions---rotating patterns with topological defects at their cores~\cite{Rana2024}. By slowly ramping up $\alpha$ in Fig.~\ref{fig::alpha_ramp}, we deliberately avoided these to focus on traveling dots and stripes.
However, spirals can be observed by starting from a suitable initial condition. 
By starting from a spiral wave solution and systematically ramping up and down $\PeC$, we obtain a family of spiral wave solutions shown in Fig.~\ref{fig::spiral}(a). 
As $\PeC$ increases, spiral velocity increases while the topological charge (number of arms) decreases. Remarkably, the system can sustain spirals with large topological charges, in contrast to examples in other reaction-diffusion systems where only spirals with charge $\pm1$ are stable~\cite{Hagan1982}.
At small $\PeC$, the oscillatory solution becomes unstable, and so does the spiral.
Below $\PeC\approx0.4$, spirals fragment into ripples before transitioning to a stationary pattern. 

Using these solutions as initial conditions, we then ramp down $\alpha$, and obtain spiral wave solutions at various $\alpha$ and $\PeC$. Fig.~\ref{fig::spiral}(b) shows that in the infinite-period regime, the spiral wave speed collapses onto the same universal curve as traveling waves, $\tilde{v} \cdot \alpha \cdot\PeC^{-1/2} = \sqrt{\Da \cdot\tilde{c}_0 \phi_0 (\alpha-\alpha_c)}$, as predicted by the amplitude equation. At small $\PeC$, the spiral wave state remains stable as $\alpha$ decreases all the way until the pattern becomes stationary [Fig.~\ref{fig::spiral_phase_diagram}]. The critical $\alpha_c$ for the onset of spiral waves [black dots in Fig.~\ref{fig::spiral}(c)] agrees well with Eq.~\eqref{eqn::alpha_c} (blue curve) as well as the critical $\alpha_c$ for the onset of traveling stripes (gray dots).
This remarkable agreement, despite spirals having fundamentally different topology, demonstrates the robustness of the amplitude equation predictions, as detailed further in \S\ref{sec::spiral_waves}.\\

\noindent\textbf{\emph{Transient dynamics before steady state.}} The above analysis focused on periodic steady states---ordered arrays of traveling stripes or dots moving coherently in a hexagonal lattice. However, before reaching steady state, the system undergoes rich transient dynamics with disordered oscillatory patterns. These transients are practically important because real systems, whether cells or synthetic colloids, experience continuous perturbations and may never reach true steady state. Hence, Fig.~\ref{fig::morphology} illustrates the time evolution of four representative transient scenarios starting from homogeneous initial conditions:
\begin{figure*}
  \includegraphics[width=\textwidth]{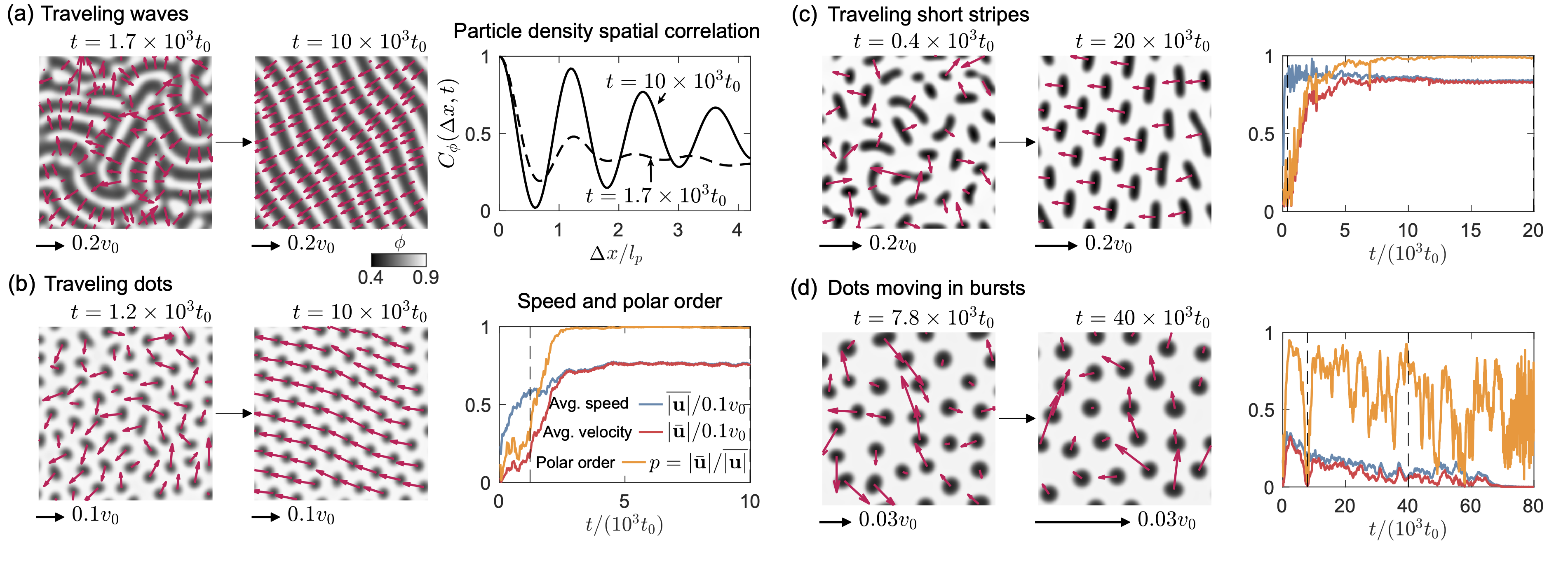}
  \caption{\textbf{Transient dynamics before reaching steady state reveal diverse oscillatory morphologies and synchronization pathways.} Snapshots show particle density with red arrows indicating pattern velocity, defined by the level set velocity for stripes and the velocity of the center of mass for dots. All simulations start from uniform initial conditions with small perturbations.
    (a) Traveling waves spanning the domain initially form with random orientations in different regions, then align over time to propagate coherently. Spatial correlation $C_\phi(\Delta x,\Delta y=0,t)$ in the horizontal direction (right) develops peaks at the predicted wavelength $l_p = 2\pi/q_{\text{max}}$ where $q_\text{max} = \text{argmax}\ {\text{Re}~\omega_+(q)}$ is the most unstable wavenumber. Parameters: $\PeR=1.9\times 10^{-3}$, $\phi_0=0.70$, $\Da=0.2$, $\alpha=4$, $\PeC=1$. 
    (b) Traveling dots initially move in random directions with low polar order $p \equiv |\bar{\mathbf{u}}| / \overline{|\mathbf{u}|}$, then synchronize into a hexagonal traveling crystal as speed and $p$ increase. Plot to the right shows the time evolution of averaged speed $\overline{|\mathbf{u}|}$, magnitude of average velocity $|\bar{\mathbf{u}}|$, and polar order $p \equiv |\bar{\mathbf{u}}| / \overline{|\mathbf{u}|}$. The average is taken over all the patterns. Vertical dashed lines in the plots correspond to the time points of the two snapshots. Parameters: $\PeR=1.2\times 10^{-3}$, $\phi_0=0.81$, $\Da=0.2$, $\alpha=4$, $\PeC=1$.
    (c) Short traveling stripes gradually align, with polar order approaching unity as stripes propagate coherently along their minor axes. Parameters: $\PeR=1.4\times 10^{-3}$, $\phi_0=0.79$, $\Da=0.56$, $\alpha=15$, $\PeC=0.44$. (d) Dots exhibit intermittent bursts along transient pathways without achieving sustained order; polar order fluctuates and speed decays until dots become stationary. Parameters: $\PeR=2.4\times 10^{-3}$, $\phi_0=0.80$, $\Da=0.5$, $\alpha=10$, $\PeC=0.34$.}
  \label{fig::morphology}
\end{figure*}

\emph{Traveling waves with ordering.} Fig.~\ref{fig::morphology} (a) and Supplementary Movie S1 shows a representative example of traveling waves that span the entire domain. Initially, traveling waves form with random orientations in different spatial domains. The spatial correlation $C_\phi(\Delta \mathbf{x},t) \equiv \langle \Delta\phi(\mathbf{x},t),\Delta\phi(\mathbf{x}+\Delta \mathbf{x},t) \rangle / \langle \Delta\phi(\mathbf{x},t),\Delta\phi(\mathbf{x},t) \rangle $ shows limited spatial extent. Over time, domains align and all waves propagate in the same direction, with the correlation extending in space, with a second peak approaching $l_p= 2\pi/(q_{\text{max}})$, where $q_\text{max} = \text{argmax}\ {\text{Re}~\omega_+(q)}$ is the most unstable wavelength from linear stability analysis, confirming theoretical predictions.


\emph{Traveling dots with synchronization.} Fig.~\ref{fig::morphology}(b) and Supplementary Movie S2 shows an example of traveling dots. Initially, dots that form as a result of spinodal decomposition move slowly, as seen from the evolution of the average speed $|\bar{\mathbf{u}}|$. They move in random directions while avoiding each other, undergoing fusion, fission, and dissolution. As a result, the polar order parameter $p \equiv |\bar{\mathbf{u}}| / \overline{|\mathbf{u}|}$~\cite{Menzel2013}, where $\overline{|\mathbf{u}|}$ is the magnitude of the velocity averaged over all dots, is low. 
Eventually, dots synchronize: the speed increases to a steady state value, while the polar order parameter also increases, until eventually, the dots form a traveling hexagonal crystal.

\emph{Traveling stripes with alignment.} Fig.~\ref{fig::morphology}(c) and Supplementary Movie S3 shows an example of short stripes that propagate along their minor axis. Initially, stripes move randomly, with longer stripes expanding concentrically and occasionally fragmenting. Over time, $p$ approaches 1 as the stripes align and propagate coherently. 


\emph{Intermittent bursts.} In Fig.~\ref{fig::morphology}(d) and Supplementary Movie S4, dots also move and do not coarsen. However, in contrast with previous cases, polar order is never achieved. Instead, dots exhibit intermittent bursts of collective motion along transient local ``pathways'' but never achieve sustained order. Therefore, $p$ fluctuates in time while the speed decays until the dots become stationary after $t/t_0\approx 7 \times 10^4$.

These diverse morphologies span a spectrum: 
stripes extending across the system (large $\PeC$ and small $\alpha$), to ellipsoids and circles (small $\PeC$ and large $\alpha$), as seen throughout Figs.~\ref{fig::morphology} (a)-(c), ~\ref{fig::alpha_ramp}, and ~\ref{fig::alpha_ramp_phase_diagram}.\\

\begin{figure*}
  \includegraphics[width=0.8\textwidth]{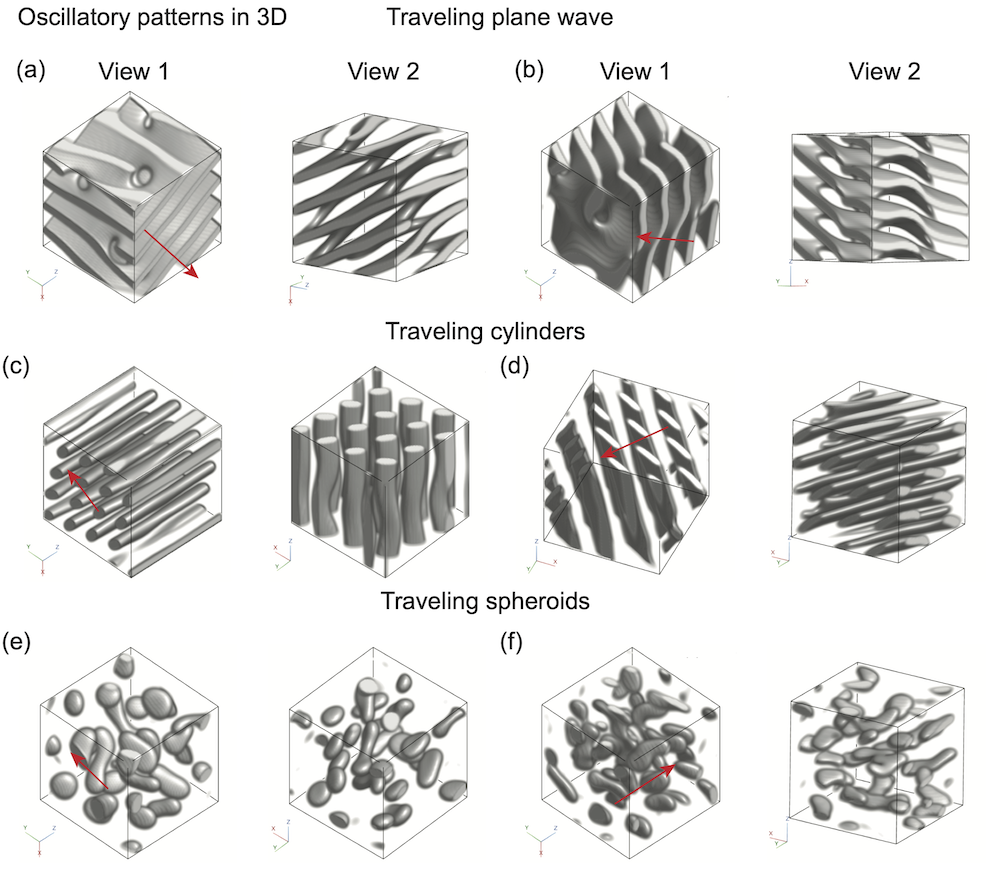}
  \caption{\textbf{Oscillatory patterns in 3D exhibit rich morphologies depending on volume fraction and chemotactic strength.} Simulations in a $[50l_0,50l_0,50l_0]$ periodic domain at $\PeR=10^{-3}$, $\text{Da}=1$, and $\alpha=10$. Each row shows two viewing angles; snapshots are taken at $t=10^4t_0$. For clarity, gray regions show dilute phase ($\phi<0.65$) and transparent regions show dense phase. (a-b) Traveling plane waves at $\phi_0=0.55$: layered structures propagate with interconnected topology and embedded voids. $\PeC=0.40$ in (a) and $0.71$ in (b). (c-d) Traveling cylinders at $\phi_0=0.58$: elliptical cylinders with waviness along their axes, resembling 2D dots extruded in 3D. $\PeC=0.42$ in (c) and $0.80$ in (d). (e-f) Traveling spheroids at $\phi_0=0.61$: ellipsoidal droplets with aspect ratio depending on $\PeC$. $\PeC=0.37$ in (e) and $0.59$ in (f). The red arrow indicates the direction in which the patterns move. See Supplemental Movies 5-10. }  
  \label{fig::3D}
\end{figure*}

\noindent\textbf{\emph{Oscillatory patterns in 3D.}} While we have focused on ABPs in 2D space, as is conventionally done, recent work has begun to explore both MIPS~\cite{Stenhammar2014,Wysocki2014,Turci2021} and chemotaxis~\cite{Bhattacharjee2019,Bhattacharjee2021,Bhattacharjee2022,Martinez-Calvo2022} in 3D. Thus, we use the established 3D free energy model [Eq.~\eqref{eqn::ABP_3D}], which reproduces MIPS in 3D~\cite{Takatori2015}, to numerically simulate oscillatory traveling wave patterns that result from the competition between chemotaxis and MIPS. We focus on a $[50l_0,50l_0,50l_0]$ periodic domain with $\PeR=10^{-3}$, $\text{Da}=1$, and $\alpha=10$. 
Fig.~\ref{fig::3D} and the corresponding Supplementary Movies 5-8 shows three classes of 3D traveling patterns (opaque regions represent $\phi<0.65$):

(a)-(b) \emph{Traveling plane waves} ($\phi_0=0.55$, $\PeC=0.40$ and $0.71$). Waves propagate as layered structures. Interestingly, different layers can interconnect while maintaining holes, creating complex 3D topology.

(c)-(d) \emph{Traveling cylinders} ($\phi_0=0.58$, $\PeC=0.42$ and $0.80$). Patterns resemble 2D dots extruded into the third dimension, forming elliptical cylinders with waviness along their axes.

(e)-(f) \emph{Traveling spheroids} ($\phi_0=0.61$, $\PeC=0.37$ and $0.59$). Ellipsoidal droplets travel through space. The aspect ratio depends on $\PeC$, with traveling speed increasing with $\PeC$.\\

\noindent\textbf{\emph{In summary.}} Altogether, these results demonstrate that the amplitude equation approach successfully predicts not only when oscillatory patterns emerge, but also their velocities, amplitudes, and bifurcation types. 
The excellent quantitative agreement---despite analyzing traveling dots using plane wave solutions and despite the full nonlinear dynamics---demonstrates the power and robustness of weakly nonlinear theory for understanding chemotactic MIPS.

\begin{figure*}
  \includegraphics[width=\textwidth]{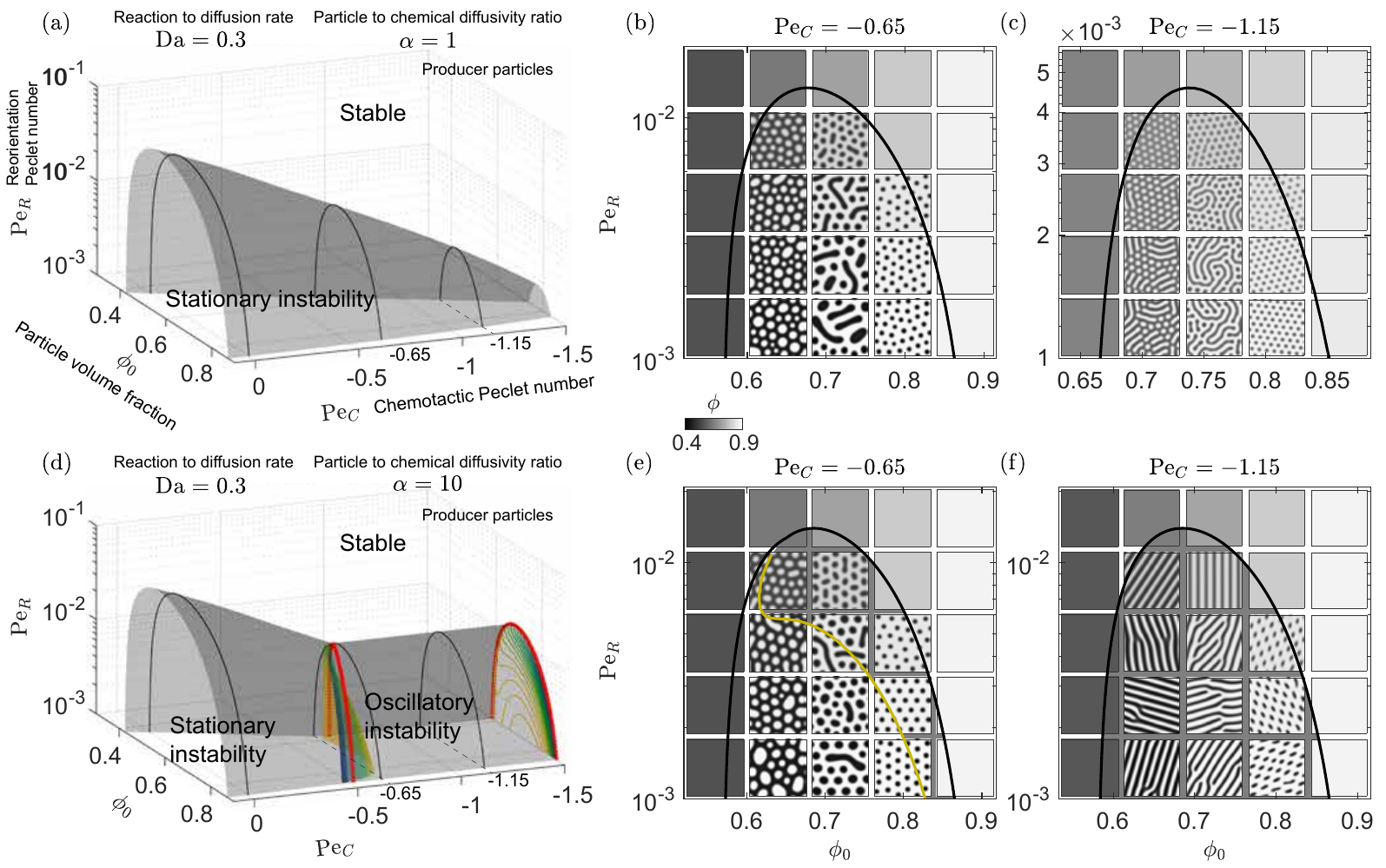}
  \caption{ \textbf{Producer particles exhibit similar behavior as consumers but with opposite sign of $\PeC$.} Details are as in Fig. \ref{fig::PeC_PeR_phi_phase_diagram}, but for chemotactic ABPs that produce a chemorepellent, and with the yellow replacing the green curve in (e).
  At small $\alpha$ (a), chemorepellent producers exhibit stationary dots and stripes below the stability boundary. At large $\alpha$ (d), oscillatory instability emerges when dispersal is sufficiently strong. (b-c, e-f) Cross-sections of the $\PeR-\phi_0$ phase space at $\PeC=-0.65$ and $-1.15$ confirm patterns analogous to the consumer case. Simulation snapshots at $t=2\times 10^4t_0$ in $[100l_0,100l_0]$ periodic domains confirm our predictions. 
  }
  \label{fig::PeC_PeR_phi_phase_diagram_producer}
\end{figure*}

\begin{figure}
  \includegraphics[width=\columnwidth]{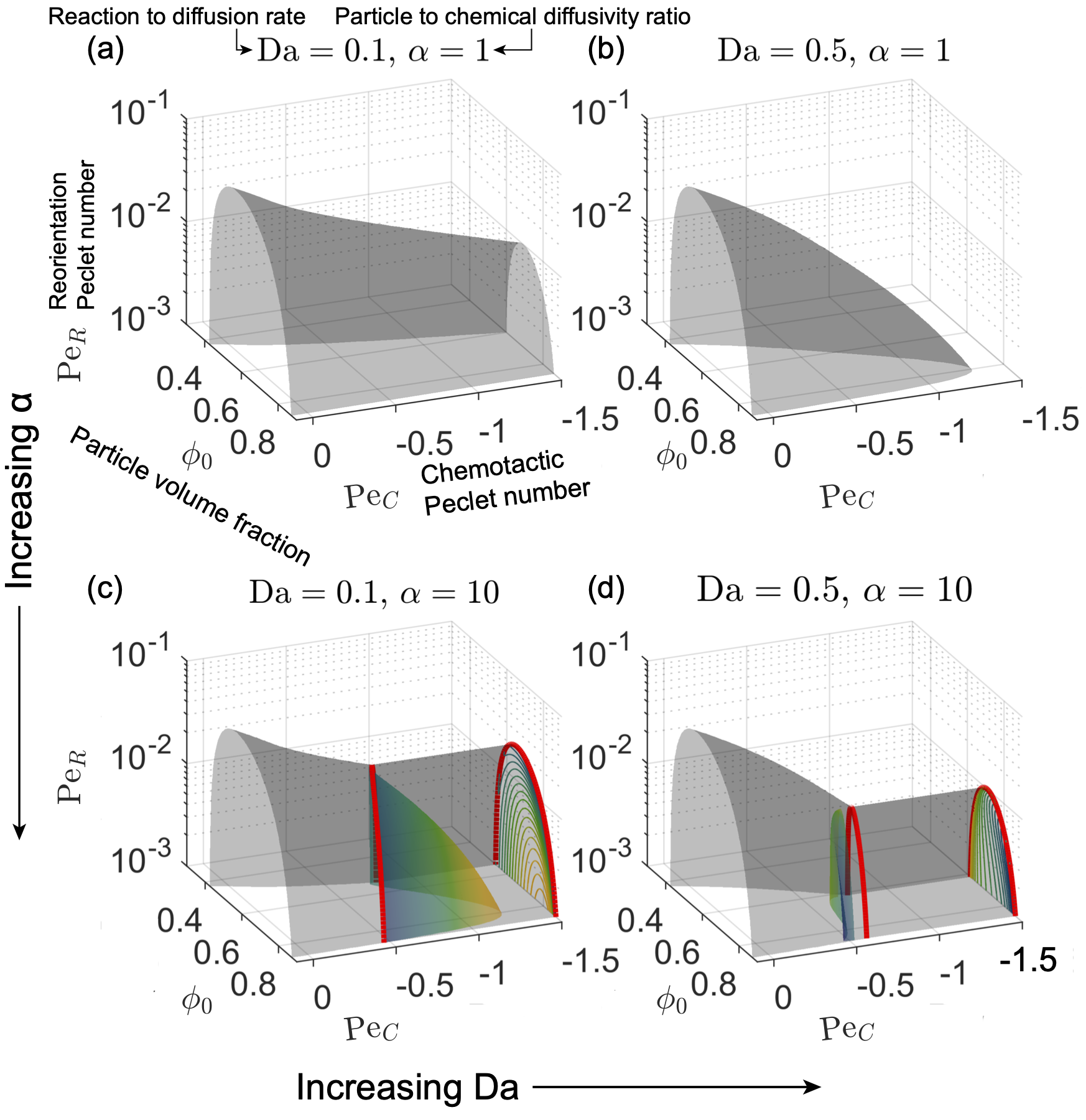}
  \caption{\textbf{Increasing $\text{Da}$ and $\alpha$ alter producer particle phase diagrams similar to consumers.} (a-d) Phase behavior at varying $\text{Da}$ and $\alpha$. Faster chemical production (larger $\text{Da}$) shrinks both phase separation and oscillatory regions; slower chemical diffusion (larger $\alpha$) expands them. These trends are identical to consumer particles [Fig.~\ref{fig::PeC_PeR_phi_phase_diagram_Da0_comparison}], confirming that the competition between MIPS and chemotactic dispersal (regardless of whether via consumption or production) governs pattern formation.}
  \label{fig::PeC_PeR_phi_phase_diagram_Da0_comparison_producer}
\end{figure}

\section{Active Brownian particles with different chemotactic properties}
\label{sec::producer}
In \S~\ref{sec::phase_diagram}-\ref{sec::oscillatory_pattern}, we studied the illustrative example of ABPs that consume a chemoattractant, for which chemotaxis drives particle dispersal and competes with MIPS clustering, resulting in the formation of complex patterns. However, active matter systems can exhibit diverse chemotactic behaviors: particles can produce rather than consume chemicals, and can respond to chemorepellents rather than chemoattractants. Moreover, many scenarios involve mixtures of particles with different chemotactic properties. Here, we extend our framework to encompass this broader landscape of chemotactic active matter.\\

\noindent\textbf{\emph{Producer particles.}} We first consider particles that produce rather than consume the chemical signal. We use the same free energy functional [Eq.~\eqref{eqn::ABP_2D}] as for consumers, and the same chemotactic sensing function $f(\tilde{c})=\tilde{c}$.
An illustrative model satisfying the requirements that $\partial_c R>0$ and that there exists a homogeneous steady state is that the particles produce the chemical at a constant rate $g(\tilde{c})=-g_0$ and that the chemical degrades via first-order kinetics $S(\tilde{c})=k\tilde{c}$. At steady state, $\tilde{c}_0=g_0\phi_0/k$.
The governing equations [Eqs.~\eqref{eqn::dphidt} and \eqref{eqn::dcdt}] are again linear in $\delta\tilde{c}=\tilde{c}-\tilde{c}_0$, so scaling the production rate $g_0$ is equivalent to scaling $\PeC$ proportionally. Therefore, without loss of generality, we fix $g_0/k=1$. 

Following the analysis in \S\ref{sec::phase_diagram}, we generate the chemotactic MIPS phase diagram parameterized by $\{\phi,\PeR,\PeC\}$ for producer particles, again at $\text{Da}=0.3$ and $\alpha=1$ or $10$. The results are shown in Fig.~\ref{fig::PeC_PeR_phi_phase_diagram_producer}(a)-(c) and (d)-(f), respectively. We also investigate the influence of varying $\text{Da}$ in Fig.~\ref{fig::PeC_PeR_phi_phase_diagram_Da0_comparison_producer}. While the stability and stationary/oscillatory instability boundaries are shifted compared to those of the consumer particles, taken altogether, our results show that producer particles exhibit qualitatively the same types of behavior as consumer particles, but with $\PeC$ having the opposite sign. This sign reversal reflects the fundamental relationship between chemical production/consumption and chemotactic response: producers ($g<0$) of a chemorepellent ($\chi_0<0$) direct their motion away from the chemical they produce, causing them to disperse and suppressing MIPS, similar to consumers of a chemoattractant. By contrast, producers of a chemoattractant ($\chi_0>0$) direct their motion toward the chemical they produce, driving aggregation and enhancing MIPS, similar to consumers of a chemorepellent. 

Our simulations confirm these predictions. At small $\alpha$ [Fig.~\ref{fig::PeC_PeR_phi_phase_diagram_producer}(a)-(c)], when chemotactic dispersal is sufficiently strong (chemorepellent producers with $\PeC<0$), stationary dots and stripes emerge below the stability boundary, similar to the patterns in Fig.~\ref{fig::PeC_PeR_phi_phase_diagram}(a)-(c). These patterns arise because chemotaxis arrests coarsening, just as for consumers of a chemoattractant. For large $\alpha$, when chemical diffusivity is sufficiently slow ($\alpha' > \alpha'_\text{crit}$, below the dashed red curve in Fig.~\ref{fig::PeC_PeR_phi_phase_diagram_producer}(d)), phase separation always occurs. In the region where chemotactic dispersal is strong ($\PeC$ sufficiently negative, right of the colored surface), oscillatory instability emerges again. The simulations in panels (e)-(f) confirm traveling waves and dots in this regime, qualitatively matching the oscillatory patterns observed for consumer particles [Fig.~\ref{fig::PeC_PeR_phi_phase_diagram}(e)-(f)].

Conversely, when $\PeC>0$ (chemoattractant producers), chemotaxis enhances aggregation, expanding the phase separation region compared to non-chemotactic MIPS---as seen by comparing the instability regions at positive versus negative $\PeC$ in Fig.~\ref{fig::PeC_PeR_phi_phase_diagram_producer}(a) and (d). This behavior is similar to that of consumers of a chemorepellent (\S\ref{sec::phase_diagram}).

The dependencies on $\alpha$ and $\text{Da}$ established in \S\ref{sec::phase_diagram} for consumer particles also carry over to producer particles. As shown in Fig.~\ref{fig::PeC_PeR_phi_phase_diagram_Da0_comparison_producer}, increasing $\Da$ (faster chemical uptake/production relative to diffusion) shrinks both the phase separation region and the oscillatory instability region [compare panels (a)-(b) and (c)-(d)]
Conversely, increasing $\alpha$ (slower chemical diffusion relative to particle mobility) expands these regions. The mathematical proof of these dependencies in \S\ref{sec::SI_LSA} applies generally because it relies only on the structure of the dimensionless parameters $\alpha'$, $\text{Da}'$, and $\PeC'$, not on the specific signs of $g$ and $\chi_0$.\\


\noindent\textbf{\emph{Mixtures of chemotactic particles.}} Real ecosystems typically contain diverse populations---different bacterial species~\cite{Hu2022,Ghoul2016,Martinez-Calvo2025,Mitri2016,Hallatschek2007}, or mixtures of synthetic colloids with different surface chemistries~\cite{Agudo-Canalejo2019, Sengupta2011, Meredith2022}---that may compete or cooperate through chemical signaling. Such mixtures can exhibit nonreciprocal interactions: particle A may attract B while B repels A, for instance. How does MIPS, which arises from particle motility, interact with these chemically-mediated interactions in mixtures of chemotactic particles? We address this question by extending our continuum model to mixtures of particles with different chemotactic properties. For simplicity, we consider particles that share the same size, self-propulsion speed, and reorientation time---so if they were non-chemotactic, they would be indistinguishable and exhibit conventional single-component MIPS. The particles differ only in their chemical properties, i.e., in their chemical production/consumption rates and chemotactic responses.

Since the particles have identical active Brownian motion characteristics, the same effective free energy [Eq.~\eqref{eqn::free_energy}] applies, augmented by the entropy of mixing: $\tilde{G}[\phi_1,\ldots,\phi_N] = \tilde{G}[\phi] + \int{\phi \sum_i{\zeta_i \ln{\zeta_i}} \dd{V}}$,
where $\phi_i$ is the volume fraction of particles of type-$i$, $\phi=\sum_i{\phi_i}$ is the total volume fraction, and $\zeta_i = \phi_i/\phi$ is the fraction of type-$i$ particles. The second term is the conditional entropy for mixing, where $\phi \sum_i{\zeta_i \ln{\zeta_i}} = \sum_i{\phi_i\ln{\phi_i}} - \phi\ln{\phi}$. The normalized chemical potential of type-$i$ particles is then $  \tilde{\mu}_i = \frac{\delta G}{\delta \phi} + \ln{\zeta_i}
$.
Using the 2D free energy model [Eq.~\eqref{eqn::ABP_2D}]: $\tilde{\mu}_i = \ln{\phi_i}+1 - 2\phi - 0.3\phi^2   - \frac{4}{\pi}\cdot\PeR\cdot \phi_m \left[ \ln{\left( 1- \frac{\phi}{\phi_m}\right)} - \frac{\phi}{\phi_m - \phi} \right] - \kappa \nabla^2 \phi.$
Each particle type obeys the continuity equation [Eq.~\eqref{eqn::dphidt}] with fluxes due to active Brownian motion and chemotaxis [Eqs.~\eqref{eqn::flux_ABP} and \eqref{eqn::flux_chemotax}, respectively]: $  \pderiv{\phi_i}{t} = \nabla\cdot \left[ M_0 \phi_i \nabla \tilde{\mu}_i - \chi_{0,i} \phi \nabla f_i(\tilde{c})  \right],$
where $\chi_{0,i}$ and $f_i(\tilde{c})$ are the chemotactic coefficient and chemotactic sensing function of type-$i$ particles, respectively. The chemical concentration then evolves in time according to $\pderiv{\tilde{c}}{t} = D_c \nabla^2 \tilde{c} - \sum_i{R_i(c,\phi_i)}$, 
where $R_i(c,\phi_i)$ is the net consumption rate of type-$i$ particles.

\begin{figure}
  \includegraphics[width=\columnwidth]{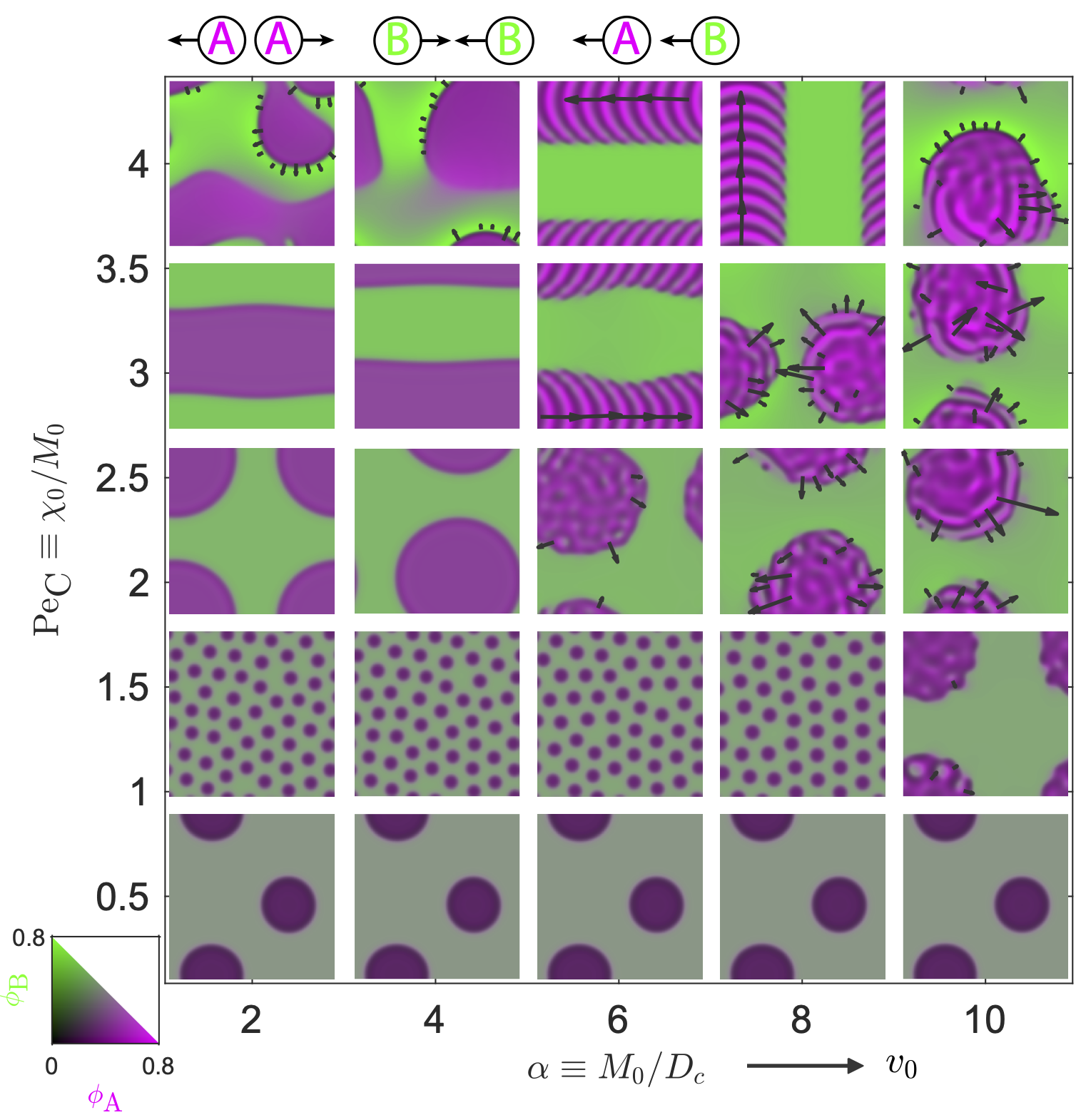}
  \caption{\textbf{Nonreciprocal chemotactic interactions drive spatial segregation and selective oscillations in mixtures.} Phase diagram in $\PeC-\alpha$ space shows mixtures of chemoattractant consumers (A, green) and chemorepellent consumers (B, magenta). Both types consume the same chemical but have opposite responses: $\chi_{0,\A}=\chi_0>0$ and $\chi_{0,\B}=-\chi_0<0$. At $\PeC=0$ (non-chemotactic), A and B mix uniformly in MIPS clusters. As $\PeC$ increases, chemotaxis drives segregation: A particles move away from B-rich regions while B particles are attracted to A-rich regions. At large $\alpha$, oscillatory patterns emerge selectively in A-rich phases (which satisfy conditions for oscillatory instability) while B-rich phases remain stationary. Simulations are performed in a $[100l_0,100l_0]$ periodic domain. Snapshots are taken at $t=2\times 10^4 t_0$. The initial condition is a uniform mixture of A and B particles where $\phi_\A(\mathbf{x}) = \phi_{\text{A}0}=0.3$ and $\phi_\B(\mathbf{x})=\phi_{\text{B}0}=0.3$ with small perturbations and a uniform chemical concentration at $c_0=S/\qty[k(\phi_{\text{A}0}+\phi_{\text{B}0})]$. Other fixed parameters are $\PeR=10^{-3}$, $\Da=0.5$, $S/k=1$. See Supplementary Movie S11.}
  \label{fig::CMIPS2}
\end{figure}

As an illustrative example demonstrating effective nonreciprocal interactions mediated by chemotaxis, we consider two particle types ($i={\A,\B}$) that both consume the same chemical ($R_A=R_B= k \tilde{c} \phi_i - S$) but have opposite chemotactic responses: particle A moves toward the chemical ($\chi_{0,\A}=\chi_0>0$, chemoattractant consumer) while particle B moves away ($\chi_{0,\B}=-\chi_0<0$, chemorepellent consumer). Both types have the same sensing function $f_i(\tilde{c})=\tilde{c}$. The dimensionless parameters $\alpha \equiv M_0/D_c$, $\Da \equiv k\kappa/D_c$, and $\PeC\equiv\chi_0/M_0$ remain as defined previously. 
In this case, the interactions between type A and B particles are effectively nonreciprocal: Type A particles, by consuming the chemical, deplete it locally, creating a gradient that repels B particles away from A. Meanwhile, B particles also consume the chemical, but this depletion attracts A particles toward B. Thus, B is attracted to A (via chemotaxis toward depleted regions) while A is repelled by B (via chemotaxis away from depleted regions).

Fig.~\ref{fig::CMIPS2} and Supplementary Movie S11 show simulations of this A-B mixture in the $\PeC-\alpha$ phase space, starting from a uniform mixture with $\phi_\A(\mathbf{x}) = \phi_{\text{A}0}=\phi_\B(\mathbf{x})=\phi_{\text{B}0}=0.3$ and $\PeR=10^{-3}$, $\Da=0.5$, and $S/k=1$. In the non-chemotactic limit ($\PeC=0$), conventional MIPS occurs and A and B particles are well-mixed in the dense phase---they are indistinguishable. As $\PeC$ increases, chemotaxis drives segregation of A and B into distinct phases, with the degree of segregation increasing with $\PeC$. This segregation arises from the effective nonreciprocal interactions: A particles move away from B-rich regions while B particles are drawn toward A-rich regions. At large $\alpha$ (slow chemical diffusion), oscillatory patterns emerge as in the single-component case. In these oscillatory patterns, the A-rich phase forms waves while the B-rich phase does not exhibit patterns. When the two phases are separated into two parallel domains, traveling waves emerge in the A-rich phase propagating in the direction normal to the A-B interface, consistent with the traveling waves that are found in consumers of chemoattractant discussed in \S\ref{sec::oscillatory_pattern}. When the A-rich phase is encapsulated within the B-rich phase and forms a circular domain, irregular waves that travel in different directions emerge within the A-rich phase, while the A-rich phase as a whole also moves in time-varying directions. At the upper left corner of the phase diagram, while there are no oscillatory patterns within either phase, the interface moves in time-varying directions. 

In summary, the effective nonreciprocal interactions between particles with different chemotactic properties give rise to spatial segregation and selective oscillations, creating complex dynamic patterns not observed in either single-component systems or equilibrium mixtures. These findings provide a foundation for understanding pattern formation in diverse multi-component active matter systems, from bacterial ecosystems to engineered colloidal suspensions.

\section{Discussion}
\label{sec::discussion}
In this work, we have established a theoretical framework for understanding how collective chemotaxis fundamentally alters motility-induced phase separation. By combining linear stability analysis, amplitude equation theory, and numerical simulations, we demonstrated that the competition between MIPS clustering and chemotactic dispersal/aggregation generates a rich landscape of pattern-forming instabilities---from stationary dots and stripes to traveling waves and spirals. This landscape is governed by three new dimensionless parameters beyond those controlling standard MIPS ($\PeR,\phi$): the chemotactic P\'{e}clet number $\PeC$ comparing chemotactic to diffusive transport, the Damk{\"o}hler number $\Da$ comparing chemical reaction to diffusion, and the ratio $\alpha$ comparing particle to chemical diffusivity.

Our central finding is that chemotaxis can either suppress or enhance MIPS depending on the sign of the product $\chi_0 g$. When particles consume chemoattractant or produce chemorepellent ($\chi_0 g>0$), chemotaxis drives dispersal, opposing the clustering tendency of MIPS. In this regime, we find three distinct outcomes depending on the relative rates of chemical diffusion and uptake: (i) complete suppression of phase separation when chemotaxis is sufficiently fast (large $\PeC$), (ii) arrested coarsening leading to stationary patterns of finite wavelength when chemical diffusion is fast (small $\alpha$), or (iii) oscillatory instabilities generating traveling waves and spirals when chemical diffusion is slow (large $\alpha$). Conversely, when particles produce chemoattractant or consume chemorepellent ($\chi_0 g<0$), chemotaxis reinforces MIPS, expanding the phase separation region. This behavior is qualitatively universal: producer and consumer particles exhibit the same phenomenology, differing only in the sign of $\PeC$.

A key advance of our work is the quantitative prediction of the patterns that emerge. Using weakly nonlinear analysis, we derived analytical expressions for the amplitude, wavelength, and velocity of both stationary and oscillatory patterns. These predictions remain accurate even far from the linear stability boundary---remarkably, plane wave solutions quantitatively describe traveling dot arrays despite their fundamentally different topology. We identified four distinct bifurcation types governing transitions between different patterns: pitchfork bifurcations at the onset of stationary stripes, saddle-node bifurcations producing hysteresis between dots and uniform states, infinite-period bifurcations where traveling wave velocity grows continuously from zero, and supercritical Hopf bifurcations where velocity jumps discontinuously. The critical conditions for these bifurcations collapse onto universal scaling laws when expressed in terms of the reduced parameters $\alpha'$, $\text{Da}'$, and $\PeC'$.

Beyond single-component systems, we showed that mixtures of particles with different chemotactic properties exhibit effective nonreciprocal interactions that drive spatial segregation and selective pattern formation. These results reveal how chemical signaling can create complex dynamic patterns in multi-component active matter, with potential implications for understanding spatial organization in microbial consortia and designing programmable interactions in synthetic active colloids. 

Our work also provides testable predictions for experiments. Using typical parameter values for motile bacteria---$\text{Pe}_\text{R}\sim10^{-2}$~\cite{Liu2019}, $\alpha \lesssim 1$~\cite{Alert2022,Liu2019,Fu2018}, $\text{Da}\sim 10^2$ (with a typical chemical depletion length $\sqrt{D_c/k}\sim1~\upmu\text{m}$~\cite{Alert2022} and persistence length $\sim20~\upmu\text{m}$~\cite{Cates2012})---we estimate that chemotactic bacteria likely satisfy $\alpha'<\alpha'_\text{crit}$, placing them in the regime where chemotaxis suppresses MIPS (left of the dotted boundary in Fig.~\ref{fig::PeC_alpha_phase_diagram}). Combined with substantial chemotactic response ($\text{Pe}_\text{C}\sim10$~\cite{Alert2022,Fu2018,Bhattacharjee2021}), our results offer a possible explanation for why MIPS is rarely observed in bacterial collectives~\cite{Cates2015,Liu2019} despite cells satisfying the motility criteria for phase separation. However, synthetic active colloids offer a versatile experimental platform to explore a broader range of the phase space mapped out in this paper. For example, by tuning swimming speed $U_0$, catalyst concentration (controlling $k$), particle surface chemistry (controlling $\chi_0$), and external chemical supply, one can systematically explore the full phase diagram of chemotactic MIPS. For example, in light-activated Janus colloids~\cite{Palacci2013}, spatiotemporal control of photocatalytic activity would enable dynamic tuning across the stationary-to-oscillatory transition, providing direct experimental access to the traveling waves, spirals, and transient dynamics predicted here.

Our study has several limitations that motivate future work. First, our continuum model provides a coarse-grained description of chemotactic ABPs over timescales much longer than the reorientation time $\tau_R$ and length scales much larger than the persistence length $U_0\tau_R$. At shorter time and length scales or sufficiently large $\PeC$, particle orientation dynamics may couple to chemotaxis in ways not captured by our model. Particle-based simulations could validate the continuum predictions in these regimes and identify when they break down, requiring one to model the dynamics of both ABP density and orientation fields. Second, we assumed Keller-Segel-type chemotaxis arising from translational drift and orientational bias. Real cells and synthetic colloids employ many other strategies to perform chemotaxis, such as by modulating run-and-tumble frequency~\cite{Keller1970}, directional persistence~\cite{Bhattacharjee2022}, and sensing of higher-order gradients of chemical concentration~\cite{Mahdisoltani2021}. Each of these could potentially couple differently to MIPS. Third, our Cahn-Hilliard-based free energy captures conventional MIPS but not additional phenomena like bubble formation observed in some active systems. Extending our analysis to more general models of ABPs (e.g., active model B+)~\cite{Wittkowski2014,Tjhung2018} could reveal new pattern types and bifurcation sequences.  Our model chemotactic MIPS can also be extended to account for higher-order effects such as the anisotropic chemical reaction due to the asymmetry of Janus particles~\cite{Liebchen2015,Liebchen2017} and the screening of the chemical field due to the volume occupied by the particles, especially in high-density clusters~\cite{Pohl2014,Stark2018}. These effects may also lead to novel coupling dynamics between particles and hence affect the pattern formation. Finally, real living systems also involve proliferation and death~\cite{Hallatschek2023}. Incorporating these processes represents an important direction for understanding self-organization in living systems.

Nevertheless, our approach can be applied to other nonreciprocal active systems where pattern formation emerges from the interplay between phase separation and directional fluxes~\cite{Roca-Cusachs2013,Shellard2020,SenGupta2021,Sunyer2016,Alert2019a,Cohen2014,Mijalkov2016,Palagi2018,Saha2020,You2020Marchetti,Wysocki2016,Zhang2021,Yin2023,Bois2011,Radszuweit2013,Banerjee2015,Weber2018,Brauns2024,Demarchi2023}. Examples include enzyme-enriched condensates exhibiting self-propulsion, epithelial tissues with mechanochemical feedback, and mixtures of active and passive particles with differential motility. In each case, the amplitude equation method could provide quantitative predictions for pattern onset, morphology selection, and dynamical transitions.

Our work thus establishes chemotactic MIPS as a paradigmatic example of pattern formation in nonreciprocal active matter. The interplay between self-propulsion, phase separation, and chemical signaling yields a rich array of new nonequilibrium phenomena. For bacterial collectives, our findings suggest that chemical communication profoundly shapes spatial structure, potentially regulating functions from nutrient access to collective migration. For synthetic microswimmers, the ability to program pattern formation through controlled chemical gradients opens possibilities for self-assembling materials with tunable structure and dynamics. By providing quantitative principles connecting microscopic parameters to emergent patterns, our framework bridges the gap between theory and experiment, enabling rational design of active materials and deeper understandings of self-organization in living systems.\\

\noindent\textbf{Acknowledgements.} 
We thank Alejandro Martínez-Calvo for valuable feedback on the manuscript.
This work was supported by NSF Grants CBET-1941716, DMR-2011750, and EF-2124863, the Camille Dreyfus Teacher-Scholar Program, the Pew Biomedical Scholars Program, a Princeton Bioengineering Initiative (PBI$^2$) Postdoctoral Fellowship, and a Harold W. Dodds Fellowship.

\clearpage
\onecolumngrid
\setcounter{equation}{0}
\setcounter{figure}{0}
\setcounter{table}{0}
\setcounter{section}{0}
\renewcommand{\theequation}{S\arabic{equation}}
\renewcommand{\thefigure}{S\arabic{figure}}
\renewcommand{\thetable}{S\arabic{table}}
\renewcommand{\thesection}{S\arabic{section}}
\renewcommand{\thefootnote}{\alph{footnote}}

\begin{center}
{\large\bfseries Supplementary Information}
\end{center}

\tableofcontents

\section{Linear stability analysis}
\subsection{Connecting the general linear stability analysis to specific models of the chemotactic ABPs}
\label{sec::SI_LSA}
The general stability analysis derived in the main text applies to arbitrary constitutive relations for chemical kinetics and chemotactic sensing. Here, we connect these general results to the specific models used throughout the main text. In particular, we summarize the functional forms of chemical reaction kinetics ($g(\tilde{c})$, $S(\tilde{c})$) for consumer and producer particles used in the main text in Table~\ref{table::consumer_producer} as well as the relationship between dimensionless parameters ($\alpha'$ and $\alpha$, $\text{Da}'$ and $\text{Da}$, $\PeC'$ and $\PeC$). The chemotactic sensing function is $f(\tilde{c})=\tilde{c}$.
With these relationships, we can readily convert the stability conditions $\text{Da}' \geq \text{Da}'_\text{crit}$ and $\text{Pe}'_\text{C} \geq \text{Pe}'_{\text{C,crit}}$ from the linear stability analysis to an expression written explicitly in terms of $\partial_\phi\tilde{\mu}_h$ as also shown in Table~\ref{table::consumer_producer}.
Because $\tilde{\mu}_h$ depends linearly on $\PeC$ [Eqs.~\eqref{eqn::ABP_2D}, \eqref{eqn::ABP_3D}], we can then express the stability condition explicitly in terms of $\PeR$ and then study how the stability boundary shifts in the $\PeR-\phi_0$ phase diagram as $\alpha$, $\Da$ and $\PeC$ varies. For example, based on the 2D chemical potential Eq.~\eqref{eqn::ABP_2D}, we find
\begin{equation} \label{eqn::PeR}
  \PeR = \frac{\pi}{4} \frac{(\phi-\phi_m)^2}{\phi_m (\phi - 2\phi_m)} \left( \frac{1}{\phi} - 2 - 0.6 \phi - \partial_\phi \tilde{\mu}_h \right).
\end{equation}
Substituting the expressions for the stability conditions in terms of $\partial_\phi\tilde{\mu}_h$ into Eq.~\eqref{eqn::PeR}, we obtain the expression for $\PeR-\phi_0$ relation at the stability boundary.

From the expression of $\text{Da}'\geq \text{Da}'_\text{crit}$ we see that 
\begin{enumerate}
  \item $\text{Da}'<\text{Da}'_\text{crit}$ corresponds to the region below the boundary of $\text{Da}'=\text{Da}'_\text{crit}$ in the $\PeR-\phi_0$ phase diagram because $\text{Da}'<\text{Da}'_\text{crit}$ is equivalent to 
  \begin{equation}
    \PeR < \frac{\pi}{4} \frac{(\phi-\phi_m)^2}{\phi_m (\phi - 2\phi_m)} \left( \frac{1}{\phi} - 2 - 0.6 \phi + \frac{2\phi_0\sqrt{\alpha \text{Da}} + 1}{\alpha \phi_0} \right).
  \end{equation}
  \item The boundary of $\text{Da}'=\text{Da}'_\text{crit}$ always lies within the spinodal region because $-\partial_\phi\tilde{\mu}_h = \frac{2\phi_0\sqrt{\alpha \text{Da}} + 1}{\alpha \phi_0}>0$.
  \item The boundary of $\text{Da}'=\text{Da}'_\text{crit}$ is independent of $\PeC$.
  \item At a given $\text{Da}$ and $\phi_0$, the boundary of $\text{Da}'=\text{Da}'_\text{crit}$ shifts upward with increasing $\alpha$, because $\partial_\phi \tilde{\mu}_h$ (and hence $\PeR$) increases.
  \item At a given $\alpha$ and $\phi_0$, the boundary of $\text{Da}'=\text{Da}'_\text{crit}$ shifts downward with increasing $\text{Da}$, because $\partial_\phi \tilde{\mu}_h$ (and hence $\PeR$) decreases.
\end{enumerate}

From the expressions of $\text{Pe}'_\text{C} \geq \text{Pe}'_{\text{C,crit}}$ for consumers and producers, we see that 
\begin{enumerate}
  \item $\text{Pe}'_\text{C} < \text{Pe}'_{\text{C,crit}}$ corresponds to the region below the boundary of $\text{Pe}'_\text{C} = \text{Pe}'_{\text{C,crit}}$ in the $\PeR-\phi_0$ phase diagram for the same reason as above.
  \item The boundary of $\text{Pe}'_\text{C} = \text{Pe}'_{\text{C,crit}}$ always lies within the spinodal region because at the boundary $-\partial_\phi\tilde{\mu}_h>0$.
  \item The boundary of $\text{Pe}'_\text{C} = \text{Pe}'_{\text{C,crit}}$ is independent of $\alpha$.
  \item At a given $\text{Da}$ and $\phi_0$, the boundary of $\text{Pe}'_\text{C} = \text{Pe}'_{\text{C,crit}}$ shifts downward with increasing $\PeC$ for consumers, because $\partial_\phi \tilde{\mu}_h$ (and hence $\PeR$) decreases, while it shifts upward with increasing $\PeC$ for producers.
  \item At a given $\Da$ and $\phi_0$, the boundary of $\text{Pe}'_\text{C} = \text{Pe}'_{\text{C,crit}}$ is independent of $\text{Da}$ at small chemotactic dispersal rate or when chemotaxis leads to aggregation, specifically when $\PeC<kS_0^{-1}\text{Da}\phi_0^3$ for consumers and $-\PeC<k g_0^{-1} \text{Da}$ for producers. Otherwise at higher chemotactic dispersal rate, the boundary of $\PeC'=\text{Pe}'_{C\text{crit}}$ shifts downward with increasing $\text{Da}$ because $\partial_\phi \tilde{\mu}_h$ (and hence $\PeR$) decreases.
\end{enumerate}

\begin{table}
  \caption{Functional forms of $g(\tilde{c})$, $S(\tilde{c})$, the relationship between $\alpha'$ and $\alpha$, $\text{Da}'$ and $\text{Da}$, $\PeC'$ and $\PeC$, and the equivalent expressions for Eq.~\eqref{eqn::Da_crit} and Eq.~\eqref{eqn::PeC_crit} written explicitly in terms of $\partial_\phi\tilde{\mu}_h$ for both consumers and producers. $h(x)=x$ when $0<x < 1$, and $h(x)=2\sqrt{x}-1$ when $x > 1$.}
  \label{table::consumer_producer}
  \centering
  \begin{ruledtabular}
    \begin{tabular}{c|c}
      Consumer & Producer \\ \hline
      {$\!\begin{aligned}
      g(\tilde{c}) &= k\tilde{c} \\
      S(\tilde{c}) &= -S_0 \\
      \alpha' &= \alpha \phi_0 |\partial_\phi \tilde{\mu}_h| \\
      \text{Da}' &= \text{Da} \frac{\phi_0}{|\partial_\phi \tilde{\mu}_h|} \\
      \PeC' &= \PeC \frac{1}{\phi_0^2 |\partial_\phi \tilde{\mu}_h|}\frac{S_0}{k}
      \end{aligned}$} &
      {$\!\begin{aligned}
      g(\tilde{c}) &= -g_0 \\
      S(\tilde{c}) &= k\tilde{c} \\
      \alpha' &= \alpha \phi_0 |\partial_\phi \tilde{\mu}_h| \\
      \text{Da}' &= \text{Da} \frac{1}{|\partial_\phi \tilde{\mu}_h|} \\
      \PeC' &= -\PeC \frac{1}{|\partial_\phi \tilde{\mu}_h|}\frac{g_0}{k}
      \end{aligned}$} \\ \hline 
      \multicolumn{2}{c}{$\text{Da}' \geq \text{Da}'_\text{crit}$ [Eq.~\eqref{eqn::Da_crit}] is equivalent to}\\
      $-\partial_\phi\tilde{\mu}_h \leq \frac{2\phi_0\sqrt{\alpha \text{Da}} + 1}{\alpha \phi_0}$ &
      $-\partial_\phi\tilde{\mu}_h \leq \frac{2\sqrt{\phi_0 \alpha \text{Da}} + 1}{\alpha \phi_0}$ \\ \hline
      \multicolumn{2}{c}{$\text{Pe}'_\text{C} \geq \text{Pe}'_{\text{C,crit}}$ [Eq.~\eqref{eqn::PeC_crit}] is equivalent to}\\
      $-\partial_\phi\tilde{\mu}_h \leq \phi_0 \text{Da}\cdot h\left(\frac{S_0 \PeC}{k \text{Da}\phi_0^3}\right)$ &
      $-\partial_\phi\tilde{\mu}_h \leq \text{Da} \cdot h\left(-\frac{g_0 \PeC}{k\text{Da}}\right)$
  \end{tabular}
  \end{ruledtabular}
\end{table}

\subsection{Derivation of the oscillatory instability condition from linear stability analysis}
\label{sec::LSA_oscillatory}
As described in the linear stability analysis in Sec.~\ref{sec::LSA}, the condition for oscillatory instability is that there exists $\tilde{q}$ for which $\trJ>0$ and $\Delta<0$.
From Eq.~\eqref{eqn::tr}, the roots of $\trJ=0$ are 
\begin{equation}
  \tilde{q}_\pm^2 = \frac{1}{2\alpha'} \left(\alpha'-1 \pm \sqrt{(\alpha'-1)^2-4\alpha'\text{Da}'} \right).
\end{equation}
$\trJ>0$ when $\tilde{q}_-<\tilde{q}<\tilde{q}_+$.
Therefore, the critical $\text{Pe}_{\text{C},*}'$ above which unstable oscillatory mode emerges corresponds to when there exists a critical $\tilde{q}^* \in [\tilde{q}_-,\tilde{q}_+]$ for which $\Delta=0$ and for all $\tilde{q} \in [\tilde{q}_-,\tilde{q}_+]$, $\Delta\geq 0$.
As show in ref.~\cite{Zhao2023}, the critical wavenumber $\tilde{q}^*$ can be either $\tilde{q}_\pm$, or $\tilde{q}_{lr}'$, which are the roots of $\Delta$ of multiplicity 2 ($\Delta (\tilde{q}_{lr})=0$, and $d\Delta / d\tilde{q} (\tilde{q}_{lr})=0$).
$\tilde{q}_{lr}'$ exists when $(\alpha'+1)^2-12\alpha'\text{Da}'\geq 0$, and their expressions are given by
\begin{equation}
  \tilde{q}_{lr}^{\prime 2} = \frac{1}{6\alpha'} \left(\alpha'+1 \pm \sqrt{(\alpha'+1)^2-12\alpha'\text{Da}'} \right).
\end{equation}
Furthermore, if $\tilde{q}_-<\tilde{q}_l<\tilde{q}_+$, $\tilde{q}^*$ is $\tilde{q}_+$, $\tilde{q}_-$, or $\tilde{q}_l$, whichever makes $\Delta(\tilde{q}^*)=0$ at the smallest $\PeC'$, otherwise, $\tilde{q}^*$ is $\tilde{q}_+$ or $\tilde{q}_-$.

The minimum $\PeC'$ at which $\Delta(\tilde{q})=0$ is given by
\begin{equation}
  \text{Pe}_{\text{C},\Delta=0}(\tilde{q}) = \frac{\alpha'}{4\tilde{q}^{2}\da'} \left[ \tilde{q}^{4} + \left(\sigma-\frac{1}{\alpha'}\right) \tilde{q}^{2} - \frac{\text{Da}'}{\alpha'} \right]^2.
\end{equation}

Defining $\text{Pe}'_{\text{C},\pm} =  \text{Pe}_{\text{C},\Delta=0}(\tilde{q}_\pm)$ and $\text{Pe}'_{\text{C},l} =  \text{Pe}_{\text{C},\Delta=0}(\tilde{q}_l)$, then the critical $\text{Pe}_{\text{C},*}'$ of unstable oscillation is
\begin{equation}
  \label{eqn::critical_condition_oscillatory}
  \text{Pe}_{\text{C},*}' =
  \begin{cases}
    \text{min}\{\text{Pe}_{C,+}', \text{Pe}_{C,-}', \text{Pe}_{C,l}'\},  & \text{when }\tilde{q}_-<\tilde{q}_l<\tilde{q}_+ \text{ or equivalently }  \text{Da}'<\frac{(\alpha'+1)^2}{12\alpha'}, \, \text{Da}'>1-\frac{2}{\alpha'}, \, \text{and } \alpha'>5 \\
    \text{min}\{\text{Pe}_{C,+}', \text{Pe}_{C,-}'\},  & \text{otherwise}
  \end{cases}
\end{equation}

\section{Amplitude equation: Theory}
\label{sec::SI_amplitude}

In this section, we present details on deriving the morphologies of both stationary and oscillatory patterns (Secs.~V and VI in the main text).

\subsection{Stationary patterns}
\label{sec::SI_stationary}
We start with the amplitude equations [Eq.~\eqref{Eq:amplitude} in the main text] with $N=3$, which govern the dynamics of the complex amplitudes $A_c$ and $A_\phi$.
It will be convenient to rescale space by $l_0=\sqrt{\kappa}$ and rescale time by $\tau_0=\kappa/D_c$ (which should not be confused with $t_0$ or $\tau_R$ in the main text). For the sake of simplicity, we drop the tilde notation for the rescaled variables.
We express the amplitudes in the polar form $A_{\phi,n}=R_{\phi,n}e^{i\psi_n}$ and $A_{c,n}=R_{c,n}e^{i\theta_n}$ and introduce new phase variables $\Theta_i=\theta_i-\psi_i$ and $\Psi = \sum_{i=1}^3\psi_i$. 
The dynamical equations for the moduli and phases of the amplitudes are
\begin{align}
    \dv{R_{c,i}}{t}    & = -\qty(k_c^2 + \da \phi_0) R_{c,i} - \da c_0 R_{\phi,i}\cos\Theta_i - \da \qty[R_{c,i+1} R_{\phi,i+2}\cos(\Psi+\Theta_i+\Theta_{i+1}) + R_{c,i+2}R_{\phi,i+1}\cos(\Psi+\Theta_i+\Theta_{i+2})], \label{eqn::SI_complex_1}      \\
    \dv{R_{\phi,i}}{t} & = \alpha \phi_0 k_c^2\qty[ \qty(a-k_c^2)R_{\phi,i} + \pec R_{c,i}\cos\Theta_i - 2\gamma R_{\phi,i+1}R_{\phi, 3}\cos\Psi - b \qty(3R_{\phi,i}^2+6R_{\phi,i+1}^2+6R_{\phi,i+2}^2)R_{\phi, 1}], \label{eqn::SI_complex_2}  \\
    \dv{\psi_i}{t}     & = \alpha \phi_0 k_c^2\qty(
    \pec \frac{R_{c,i}}{R_{\phi,i}}\sin\Theta_i + 2\gamma\frac{R_{\phi,i+1}R_{\phi,i+2}}{R_{\phi,i}}\sin\Psi
    ), \label{eqn::SI_complex_3}                                                                                                                                                                                                           \\
    \dv{\theta_i}{t}   & = \da \qty(
    c_0 \frac{R_{\phi,i}}{R_{c,i}}\sin\Theta_i
    +\frac{R_{c,i+1}R_{\phi,i+2}}{R_{c,i}}\sin\qty(\Psi+\Theta_i+\Theta_{i+1})+\frac{R_{c,i+2}R_{\phi,i+1}}{R_{c,i}}\sin\qty(\Psi+\Theta_i+\Theta_{i+2})), \label{eqn::SI_complex_4} 
\end{align}
where the index $i$ cycles through $i=1,2,3$ and the dimensionless parameters are defined by $\da = \kappa k/D_c$, $\alpha=M_0/D_c$, and $\pec = \chi_0/M_0$.
In the following, we consider the fixed points corresponding to the uniform, stripe, and dot solutions. The stability of these solutions is determined from the Jacobian of the amplitude equations evaluated at the fixed points.

\subsubsection{Uniform solution}
\label{sec::SI_uniform}
The uniform solution is represented by $R_{c,i}=R_{\phi,i}=0$, which is always a fixed point. To determine its stability, we consider the Jacobian of the $R$ equations, which can be decomposed into three identical 2-by-2 matrices:
\begin{align}
    J_1= \pdv{(\dot{R}_{c,1},\dot{R}_{\phi,1})}{(R_{c,1},R_{\phi,1})}
     & =
    \begin{pmatrix}
        -k_c^2-\da\phi_0       & \da c_0                        \\
        -\alpha\phi_0k_c^2\pec & \alpha\phi_0k_c^2\qty(a-k_c^2) \\
    \end{pmatrix}.
\end{align}
Hence, the stability condition is
\begin{align}
    \tr J_1  & = -k_c^2-\da\phi_0 +  \alpha\phi_0k_c^2\qty(a-k_c^2)<0, \\
    \det J_1 & = \alpha\phi_0k_c^2\qty[
        -(a-k_c^2)(k_c^2+\da\phi_0)+\pec\da c_0
    ]>0,
\end{align}
which simplifies to
\begin{align}
  a<\min\qty(a_c, 2\sqrt{\frac{\da}{\alpha}}+\frac{1}{\alpha\phi_0})
  =\min\qty(2\sqrt{\pec\da c_0}-\da\phi_0, 2\sqrt{\frac{\da}{\alpha}}+\frac{1}{\alpha\phi_0}).
\end{align}
This condition is equivalent to the linear stability condition obtained for the general case in Sec.~III of the main text. Indeed, mode coupling does not affect the stability of the uniform solution.

\subsubsection{Stripes}
\label{sec::SI_stripes}
For the stripe solution, we have $R_{\phi,1}=R_\phi$,  $R_{c,1}=R_c$, and $R_{\phi,2}=R_{\phi,3}=R_{c,2}=R_{c,3}=0$. The resulting amplitude equations are
\begin{align}
  \dv{R_{c}}{t}    & = -\qty(k_c^2 + \da \phi_0) R_{c} - \da c_0 R_{\phi}\cos\Theta,                           \\
  \dv{R_{\phi}}{t} & = \alpha \phi_0 k_c^2\qty[ \qty(a-k_c^2)R_{\phi} + \pec R_{c}\cos\Theta  - 3bR_{\phi}^3], \\
  \dv{\psi}{t}     & = \alpha \phi_0 k_c^2
  \pec \frac{R_{c}}{R_{\phi}}\sin\Theta
  ,                                                                                                            \\
  \dv{\theta}{t}   & = \da   c_0 \frac{R_{\phi}}{R_{c}}\sin\Theta.
\end{align}
From the last two equations, we have $\sin\Theta=0$ and thus $\Theta=0$ or $\pi$. However, only $\Theta=\pi$ is the physical solution as the first equation leads to 
\begin{align}
    \frac{R_c}{R_\phi} = \frac{\da c_0}{k_c^2 + \da\phi_0}\qty(-\cos\Theta)>0 \Rightarrow \cos\Theta=-1.
\end{align}
Therefore, the $\phi$ and $c$ fields are in antiphase. 
Substituting this into the second equation gives both amplitudes:
\begin{align}
   R_\phi & = \frac{1}{\sqrt{3b}}\sqrt{
      a-k_c^2-  \frac{\da\pec c_0}{k_c^2 + \da\phi_0}
  },\quad
  R_c = \frac{\da c_0}{k_c^2 + \da\phi_0} R_\phi.\label{Eq:Rphi stripe}
\end{align}
Therefore, the maximum amplitude is reached when $k_c^2=\sqrt{\Da\PeC c_0} - \Da\phi_0$. The stripe solution only exists when $a>a_c\equiv 2\sqrt{\Da\PeC {c}_0} - \Da\phi_0$. 

The stability is governed by the 6-by-6 Jacobian $J=\pdv{\dot{R}}{R}$, which decomposes into a 2-by-2 matrix $J_1$ (describing stability along the wavevector) and a 4-by-4 matrix $J_2$ (describing stability in the other two directions):
\begin{align}
  J_1 & = \pdv{(\dot{R}_{c,1},\dot{R}_{\phi,1})}{(R_{c,1},R_{\phi,1})}
  =
  \begin{pmatrix}
      -k_c^2-\da\phi_0       & \da c_0                                   \\
      -\alpha\phi_0k_c^2\pec & \alpha\phi_0k_c^2\qty(a-k_c^2-9bR_\phi^2) \\
  \end{pmatrix},                                                                                        \\
  J_2 & = \pdv{(\dot{R}_{c,2},\dot{R}_{c,3},\dot{R}_{\phi,2},\dot{R}_{\phi,3})}{(R_{c,2},R_{c,3},R_{\phi,2},R_{\phi,3})}
  =\begin{pmatrix}
      -k_c^2-\da\phi_0       & \da R_\phi             & \da c_0                                   & \da R_c                                   \\
      \da R_\phi             & -k_c^2-\da\phi_0       & \da R_c                                   & \da c_0                                   \\
      -\alpha\phi_0k_c^2\pec & 0                      & \alpha\phi_0k_c^2\qty(a-k_c^2-6bR_\phi^2) & \alpha\phi_0k_c^2\cdot 2\gamma R_\phi     \\
      0                      & -\alpha\phi_0k_c^2\pec & \alpha\phi_0k_c^2\cdot 2\gamma R_\phi     & \alpha\phi_0k_c^2\qty(a-k_c^2-6bR_\phi^2) \\
  \end{pmatrix}.
\end{align}
The eigenvalues of the Jacobian are computed numerically to determine the stability boundary of the stripes, which are shown in Fig.~\ref{fig::stationary_pattern} in the main text. 

The above analysis is valid for $\alpha < \alpha_c =  
\max{\left\{ \frac{4}{\PeC \tilde{c}_0}, \frac{1}{a^2}\qty(\sqrt{\Da} + \sqrt{\Da+a\phi_0^{-1}})^2 \right\}}$ [Eq.~\eqref{eqn::alpha_c} in the main text], where the pattern remains stationary. Oscillatory patterns emerge at large $\alpha$, which will be discussed in the next section. 

\subsubsection{Dots}
\label{sec::SI_dots}
The dots pattern is represented by $R_{c,1}=R_{c,2}=R_{c,3}=R_c$ and $R_{\phi,1}=R_{\phi,2}=R_{\phi,3}=R_\phi$. The equations for the phases are
\begin{align}
    \dv{\psi_1}{t}   & = \alpha \phi_0 k_c^2\qty(
    \pec \frac{R_{c}}{R_{\phi}}\sin\Theta_1 + 2\gamma R_\phi \sin\Psi
    ),                                            \\
    \dv{\theta_1}{t} & = \da \qty(
    c_0 \frac{R_{\phi}}{R_{c}}\sin\Theta_1
    +R_\phi\qty(\sin\qty(\Psi+\Theta_1+\Theta_2)+\sin\qty(\Psi+\Theta_1+\Theta_3))).
\end{align}
Since our analysis is in the vicinity of the bifurcation point, the amplitudes are small, but their ratios are finite. Thus, setting the leading order of the right-hand-side to zero leads to $\sin\Theta_1=0$ and therefore $\Theta_1 = 0$ or $\pi$. Similar to the stripe case, it can be shown that only the antiphase solution is physical $\Theta_{1,2,3}=\pi$, while the in-phase solution leads to negative $R$. The equations for $R$ reduce to
\begin{align}
    \dv{R_{c,i}}{t}    & = -\qty(k_c^2 + \da \phi_0) R_{c,i} + \da c_0 R_{\phi,i} - \da \cos\Psi \qty(R_{c,i+1} R_{\phi,i+2} + R_{c,i+2}R_{\phi,i+1}),       \\
    \dv{R_{\phi,i}}{t} & = \alpha \phi_0 k_c^2\qty[ \qty(a-k_c^2)R_{\phi,i} - \pec R_{c,i}- 2\gamma R_{\phi,i+1}R_{\phi, 3}\cos\Psi - b \qty(3R_{\phi,i}^2+6R_{\phi,i+1}^2+6R_{\phi,i+2}^2)R_{\phi, 1}].
\end{align}
From the first equation, we have
\begin{align}
    R_c = \frac{\da c_0 R_\phi}{k_c^2+\da \phi_0 + 2\da R_\phi \cos\Psi}.
\end{align}
Substituting this into the second equation leads to
\begin{align}
    15bR_\phi^2  + 2\gamma R_{\phi}\cos\Psi -a + k_c^2 +  \frac{\pec \da c_0 }{k_c^2+\da \phi_0 + 2\da R_\phi \cos\Psi} =0.
\end{align}
Expanding to the quadratic order in $R_\phi$, we find
\begin{align}
  15bR_\phi^2  + 2\gamma R_{\phi}\cos\Psi +  \frac{\pec \da c_0 }{k_c^2+\da \phi_0}\qty( - \frac{2\da \cos\Psi}{k_c^2+\da \phi_0}R_\phi +\qty(\frac{2\da \cos\Psi}{k_c^2+\da \phi_0})^2R_\phi^2) = a-k_c^2 -\frac{\pec \da c_0 }{k_c^2+\da \phi_0 }.
\end{align}
The right hand side is maximized at $k_c^2=\sqrt{\Da\PeC c_0} - \Da\phi_0$, which simplifies the equation to
\begin{align}
  \qty(15b+\frac{4\da^2\cos^2\Psi}{\sqrt{\pec \da c_0}}) R_\phi^2  + 2(\gamma-\da)\cos\Psi  R_{\phi} = a-a_c.
\end{align}
In practice, $15 b\gg \frac{4\da^2}{\sqrt{\pec \da c_0}}$, so the condition for the discriminant to be positive is
\begin{align}
   \qty[(\gamma-\da)\cos\Psi]^2 +15b(a-a_c)>0 \Rightarrow  a>a_c - \frac{(\gamma-\da)^2\cos^2\Psi}{15b}.
\end{align}
Therefore, the minimum $a$ that can sustain the dots pattern is $a_\mathrm{dots} =a_c - \frac{(\gamma-\da)^2}{15b}$, which is reached when $\cos\Psi = \mathrm{sgn}\qty(\da-\gamma)$ (the sign guarantees a positive amplitude). The corresponding $R_\phi$ is
\begin{align}
  R_\phi = \frac{1}{15b}\qty(
  \abs{\gamma-\da} + \sqrt{\qty(\gamma-\da)^2 + 15b(a-a_c)}
  )= \frac{1}{15b}\qty(
      \abs{\gamma-\da} + \sqrt{15b(a-a_\mathrm{dots})}
      ).
\end{align}
Therefore, at $a=a_\mathrm{dots}$, the dots pattern emerges through a saddle-node bifurcation, where $R_\phi$ jumps from 0 to a finite value ($\frac{ \abs{\gamma-\da}}{15b}$). 

The stability of the fixed point is determined by numerically computing the eigenvalues of the Jacobian:
\begin{align}
    J & = \pdv{(\dot{R}_c,\dot{R}_\phi)}{(R_c,R_\phi)}\nonumber\\
    &=  \begin{pmatrix}
        -k_c^2-\da\phi_0    & \da R_\phi          & \da R_\phi          & \da c_0                                         & \da R_c                                         & \da R_c                                         \\
        \da R_\phi          & -k_c^2-\da\phi_0    & \da R_\phi          & \da R_c                                         & \da c_0                                         & \da R_c                                         \\
        \da R_\phi          & \da R_\phi          & -k_c^2-\da\phi_0    & \da R_c                                         & \da R_c                                         & \da c_0                                         \\
        -\tilde{\alpha}\pec & 0                   & 0                   & \tilde{\alpha}\qty(a-k_c^2-21bR_\phi^2)         & \tilde{\alpha}\qty(2\gamma R_\phi-12b R_\phi^2) & \tilde{\alpha}\qty(2\gamma R_\phi-12b R_\phi^2) \\
        0                   & -\tilde{\alpha}\pec & 0                   & \tilde{\alpha}\qty(2\gamma R_\phi-12b R_\phi^2) & \tilde{\alpha}\qty(a-k_c^2-21bR_\phi^2)         & \tilde{\alpha}\qty(2\gamma R_\phi-12b R_\phi^2) \\
        0                   & 0                   & -\tilde{\alpha}\pec & \tilde{\alpha}\qty(2\gamma R_\phi-12b R_\phi^2) & \tilde{\alpha}\qty(2\gamma R_\phi-12b R_\phi^2) & \tilde{\alpha}\qty(a-k_c^2-21bR_\phi^2)         \\
    \end{pmatrix}.
\end{align}
where $\tilde{\alpha} = \alpha\phi_0k_c^2$.

The stability of stationary pattern morphologies is summarized in the phase diagram in Fig.~\ref{fig::stationary_pattern} in the main text. 

\subsection{Oscillatory patterns}
\label{sec::SI_oscillatory}
As discussed in the main text, we consider the traveling wave solution $\phi(\mathbf{x},t) - \phi_0 = R_\phi e^{i (k_c x + \omega t)}+\mathrm{c.c.},$ and $\tilde{c} - \tilde{c}_0 =  R_c e^{i (k_c x + \omega t + \Theta) }+\mathrm{c.c.}$. The equations for the phases are
\begin{align}
    \dv{\psi}{t}     & = \omega =  \alpha \phi_0 k_c^2    \pec \frac{R_{c}}{R_{\phi}}\sin\Theta \label{Eq:Psi},                       \\
    \dv{\theta}{t}   & = \omega + \dot{\Theta} = \da
    c_0 \frac{R_{\phi}}{R_{c}}\sin\Theta.
\end{align}
Setting $\dot{\Theta}=0$ and multiplying the two equations lead to $\omega^2 = \alpha \phi_0 k_c^2    \pec \da c_0 \sin^2\Theta$. Hence, the (rescaled) phase velocity is
\begin{align}
  \tilde v = \frac{\omega}{k_c} = \sqrt{\da c_0 \phi_0\alpha \pec} \sin\Theta.
\end{align}
The (dimensional) phase velocity is
\begin{align}
  v = \frac{l_0}{\tau_0} \frac{\omega}{k_c} = \frac{D_c}{\sqrt{\kappa}} \sqrt{\da c_0 \phi_0\alpha \pec} \sin\Theta = v_0 \sqrt{\da c_0 \phi_0\alpha^{-1} \pec} \sin\Theta,
\end{align}
where $v_0 = M_0/\sqrt{\kappa} = \alpha D_c/\sqrt{\kappa}$. This is the result in Eq.~\eqref{eqn::traveling_wave_velocity} of the main text. 

Next, we consider the equations for $R_{c,\phi}$:
\begin{align}
    \dv{R_{c}}{t}    & = -\qty(k_c^2 + \da \phi_0) R_{c} - \da c_0 R_{\phi}\cos\Theta=0\label{Eq:OsciRc},                            \\
    \dv{R_{\phi}}{t} & = \alpha \phi_0 k_c^2\qty[ \qty(a-k_c^2)R_{\phi} + \pec R_{c}\cos\Theta  - 3b R_{\phi}^3]=0.\label{Eq:OsciRphi}
\end{align}
Combining Eq.~\eqref{Eq:Psi} and Eq.~\eqref{Eq:OsciRc}, we have
\begin{align}
  \frac{R_{c}}{R_{\phi}} = \frac{\omega}{ \alpha \phi_0 k_c^2    \pec\sin\Theta} = -\frac{\da c_0 \cos\Theta}{k_c^2 + \da \phi_0} \Rightarrow \cos\Theta = - \frac{k_c^2 + \da \phi_0}{k_c\sqrt{\da c_0 \alpha \phi_0 \pec}},
\end{align}
which is Eq.~\eqref{eqn::traveling_wave_phase_difference} in the main text. A necessary condition is $\cos\Theta>-1$, which requires
\begin{align}
  k_c + \frac{\da\cdot\phi_0}{k_c} < \sqrt{\da c_0\alpha \phi_0\pec},
\end{align}
which is Eq.~\eqref{eqn::traveling_wave_existence} in the main text. 
Substituting $\frac{R_{c}}{R_{\phi}}$ into Eq.~\eqref{Eq:OsciRphi} leads to 
\begin{align}
  R_\phi = \frac{1}{\sqrt{3b}}\sqrt{ a-k_c^2 -  \frac{\pec\da c_0}{k_c^2+\da\phi_0}\cos^2\Theta},
\end{align}
which is Eq.~\eqref{eqn::R_phi_traveling_wave} in the main text.

The stability of the fixed points is determined from the Jacobian:
\begin{align} \label{eqn::SI_oscillatory_J}
  J = \pdv{(\dot{R}_c,\dot{R}_\phi)}{(R_c,R_\phi)} =
  \begin{pmatrix}
      -(k_c^2+ \da\phi_0)                 & -\da c_0 \cos\Theta                            \\
      \alpha \phi_0 k_c^2 \pec \cos\Theta & \alpha \phi_0 k_c^2\qty(a-k_c^2 - 9b R_\phi^2)
  \end{pmatrix}.
\end{align}
Stability requires $\det J>0$ and $\tr J<0$:
It can be shown that $\det J = 6b R_\phi^2(k_c^2+ \da\phi_0)\cdot \alpha \phi_0 k_c^2 >0$, while $\tr J = 2\alpha \phi_0 k_c^2 \qty(k_c^2 + \frac{\da}{k_c^2 \alpha} + \frac{1}{\alpha\phi_0}-a)$. Therefore, the stability condition is
\begin{align} \label{eqn::SI_oscillatory_J_stability}
  k_c^2 + \frac{\da}{k_c^2 \alpha} + \frac{1}{\alpha\phi_0}<a,
\end{align}
which is Eq.~\eqref{eqn::traveling_wave_stability} in the main text. 

To summarize, the condition for a stable traveling wave is
\begin{itemize}
  \item Existence: $\cos\Theta>-1 \Rightarrow k_c + \frac{\da\phi_0}{k_c} < \sqrt{\da c_0\alpha \phi_0\pec}$. 
  \item Stability: $\tr J<0 \Rightarrow k_c^2 + \frac{\da}{k_c^2 \alpha} + \frac{1}{\alpha\phi_0}<a$.
\end{itemize}
The traveling waves can be destabilized by violating either of these conditions, which correspond to two different types of bifurcations: a Hopf bifurcation (when the stability condition is more restrictive) and an infinite-period bifurcation (when the existence condition is more restrictive). They are discussed in detail in the main text (Section~\ref{sec::oscillatory_pattern}).

\section{Amplitude equation: Comparison with simulation}
In this section, we provide further details on the comparison between the theoretical results of the amplitude equation and numerical simulations to supplement the main text.

\subsection{Stationary patterns}
\label{sec::SI_stationary_patterns}

\subsubsection{Pattern hysteresis by ramping parameters}
\begin{figure*}
    \centering
    \includegraphics[width=0.49\textwidth]{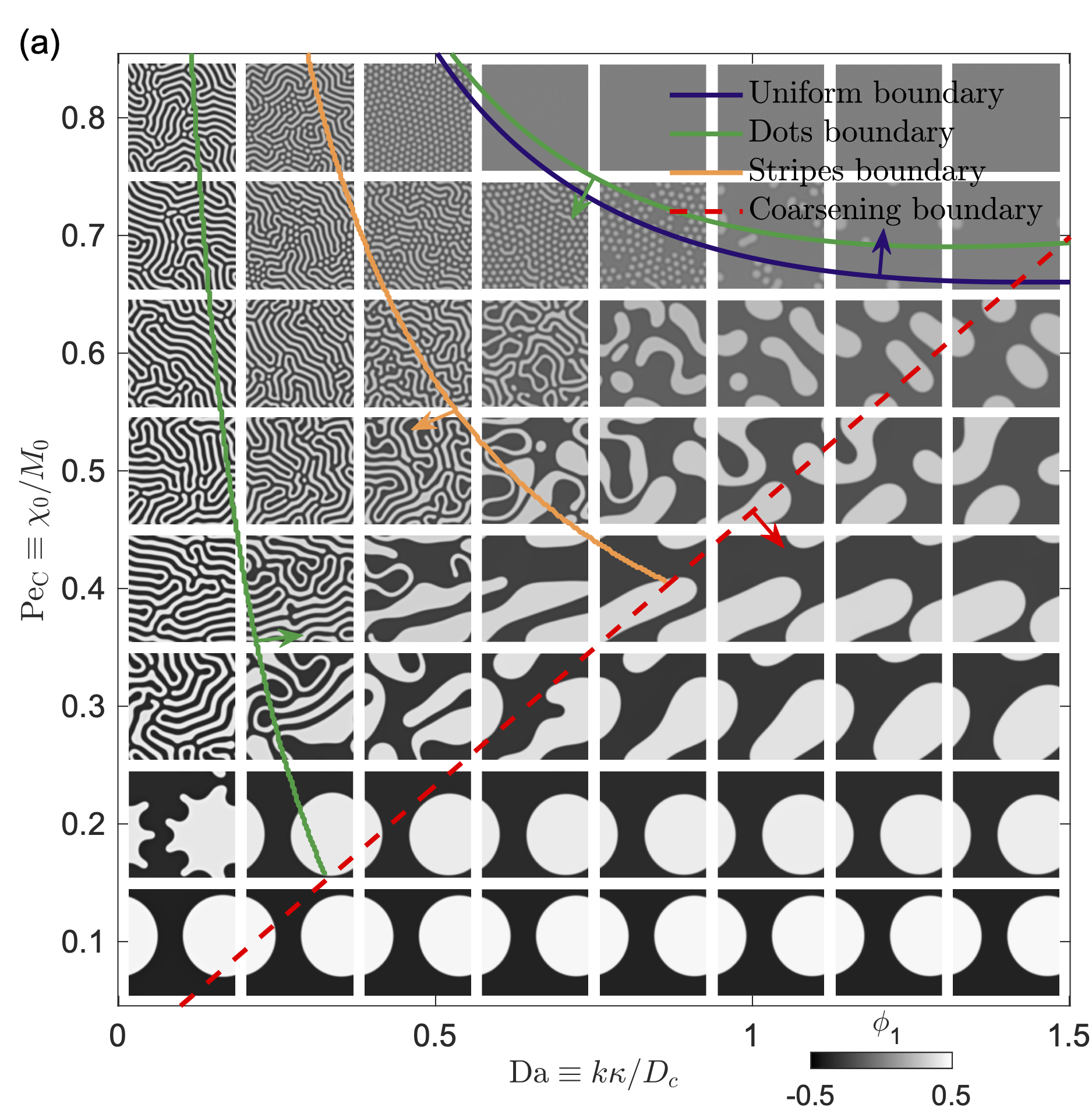}
    \includegraphics[width=0.49\textwidth]{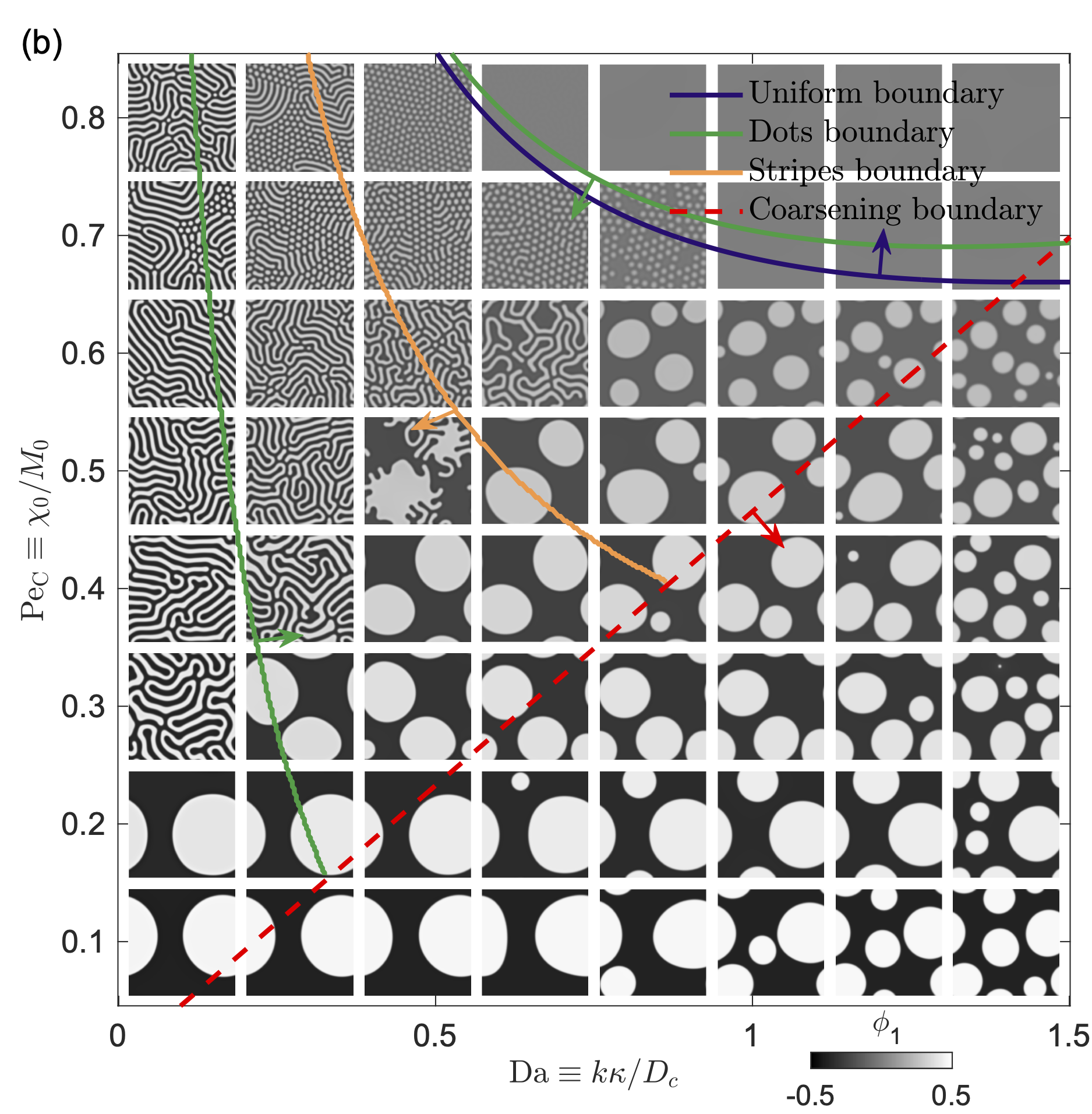}
    \caption{Comparison of simulations of the weakly nonlinear model and the amplitude equation results. The simulations are performed by (a) ramping up and (b) ramping down the value of $\da$ for each $\PeC$, using the final state displayed in the phase diagram at the previous $\da$ as the initial condition for the next $\da$ value. All parameter settings are the same as in Fig.~\ref{fig::stationary_pattern} in the main text. A duration of $t=5\times 10^4 t_0$ is used for each simulation.}
    \label{fig::stationary_pattern_hysteresis}
\end{figure*}

Fig. \ref{fig::stationary_pattern} in the main text shows simulations of the weakly nonlinear model, all of which start from the uniform state with the same random perturbation as the initial condition. Here, we ramp up and down the value of $\da$ for each $\PeC$, using the final state at the previous $\da$ as the initial condition for the next $\da$ value. The results are shown in Fig.~\ref{fig::stationary_pattern_hysteresis}. Comparing results of ramping up the value of $\da$ [Fig.~\ref{fig::stationary_pattern_hysteresis}(a)] and ramping down the value of $\da$ [Fig.~\ref{fig::stationary_pattern_hysteresis}(b)] In the region where both the dots and stripes solutions are stable, and in particular

From the rows of $\PeC = 0.7$ and $\PeC = 0.8$, we see that by ramping up the value of $\da$ [Fig.~\ref{fig::stationary_pattern_hysteresis}(a)], at low $\da$, the pattern transitions from stripes to dots. In the region where both stripes and dots are stable, the pattern is dominated by stripes with a small number of dots in between. 
At higher $\da$, the system transitions from dot pattern to the uniform state. In the region where both dots and uniform state are stable, we observe a coexistence of dots and uniform state.
In the opposite direction, we start from a uniform state at high $\da$ and ramp down the value of $\da$ [Fig.~\ref{fig::stationary_pattern_hysteresis}(b)]. The uniform state persists through the region where both dots and uniform state are stable until the uniform state is no longer stable and dots emerge. At lower $\da$, in the region where both stripes and dots are stable, the pattern is dominated by dots with a small number of stripes.
Therefore, the results of ramping up and ramping down the value of $\da$ show that in the region where multiple states are stable, the coexistence or dominance of one state over the other depends on the initial condition, and hysteresis is observed in the transition between the two states.

\subsubsection{Comparison with the full-model simulation}
\begin{figure*}
    \centering
    \includegraphics[width=0.6\textwidth]{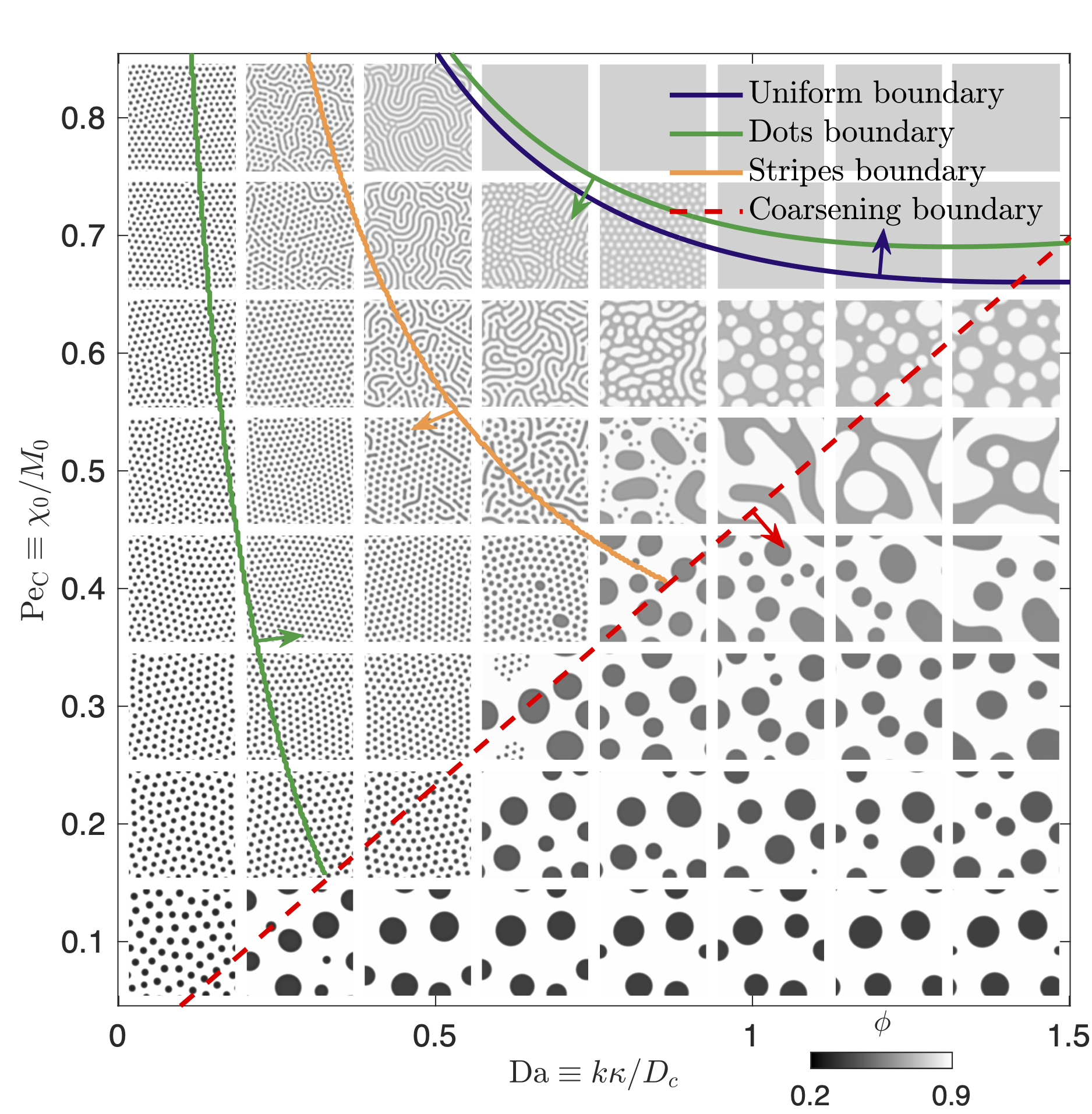}
    \caption{Comparison of simulations of the full model and the amplitude equation results. All parameter settings are the same as in Fig.~\ref{fig::stationary_pattern} in the main text. All simulations start from the uniform state with the same random perturbation as the initial condition.} 
    \label{fig::stationary_pattern_full_model}
\end{figure*}

Next, we repeat the simulations using the full models while all other parameters are identical to Fig.~\ref{fig::stationary_pattern}. The results are shown in Fig.~\ref{fig::stationary_pattern_full_model}. We see that the prediction of the dots region is only valid near the boundary of the uniform region, since away from the uniform region, the amplitude of the deviation from the uniform state becomes higher and the model enters the deeply nonlinear regime, which the weakly nonlinear model cannot capture. Nevertheless, $\PeC > \da \phi_0^2 \tilde{c}_0^{-1}$ (below the red dashed line) remains a good estimate for when coarsening occurs.

\subsection{Oscillatory patterns}
\subsubsection{Traveling stripes and dots in 2D}
\label{sec::SI_oscillatory_patterns}
\begin{figure*}
    \centering
    \includegraphics[width=\textwidth]{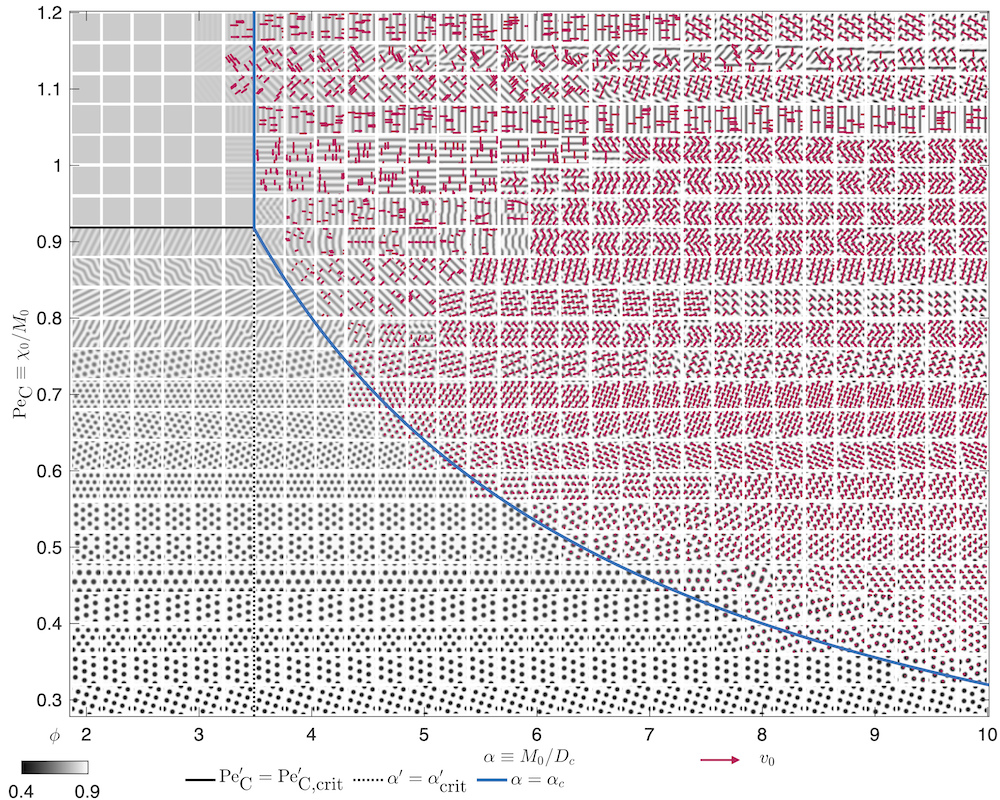}
    \caption{Simulations used to generate Fig.~\ref{fig::alpha_ramp} in the main text. Simulations are performed on a $[50l_0,50l_0]$ periodic domain and solved on a $128\times 128$ grid. The snapshots are taken at $t=5\times 10^3 t_0$. Arrows indicated the velocity of the patterns. Speed smaller than $0.05v_0$ is not shown. For stripe patterns, only regions where $|\tilde{\nabla} \phi|>0.1$ are included for calculating the level set velocities, of all eligible pixels, at most 10 pixels are randomly chosen to display the velocities (red arrows).  The blue solid line is the critical $\alpha$ for the onset of oscillatory patterns predicted by the amplitude equation.}
    \label{fig::alpha_ramp_phase_diagram}
\end{figure*}

Fig.~\ref{fig::alpha_ramp_phase_diagram} shows the snapshots of the simulations used to generate Fig.~\ref{fig::alpha_ramp} in the main text. We first ramp up the value of $\alpha$ for each $\PeC$, using the final state at the smaller $\alpha$ as the initial condition for the next $\alpha$ value. Then we ramp down the value of $\alpha$ in the opposite direction. 
The results during the ramp-down are displayed and used in Fig.~\ref{fig::alpha_ramp} in the main text. This method of continuation enhances the smoothness of the velocity and amplitude versus $\alpha$ curves, although discontinuity may still occur as the system picks another pattern with a jump in its characteristic wavevector, often associated with the transition between traveling band and traveling dots. 
For patterns whose aspect ratio is smaller than 3.7, the velocity, represented by the red arrows, is obtained by tracking the center of mass of the pattern; otherwise, the velocity is defined by the level set velocity as described in the main text. 
We find excellent agreement between the predicted boundary of oscillatory patterns $\alpha_c = 4/(\PeC \tilde{c})$, denoted by the blue dashed curve, and the simulation.
With this particular set of parameters, traveling band patterns tend to occur at high $\PeC$ and low $\alpha$, while traveling dot or short stripes patterns occur at low $\PeC$ and high $\alpha$.

\subsubsection{Traveling wave in 1D}
\label{sec::SI_oscillatory_patterns_1D}
\begin{figure}
  \centering
  \includegraphics[width=0.7\textwidth]{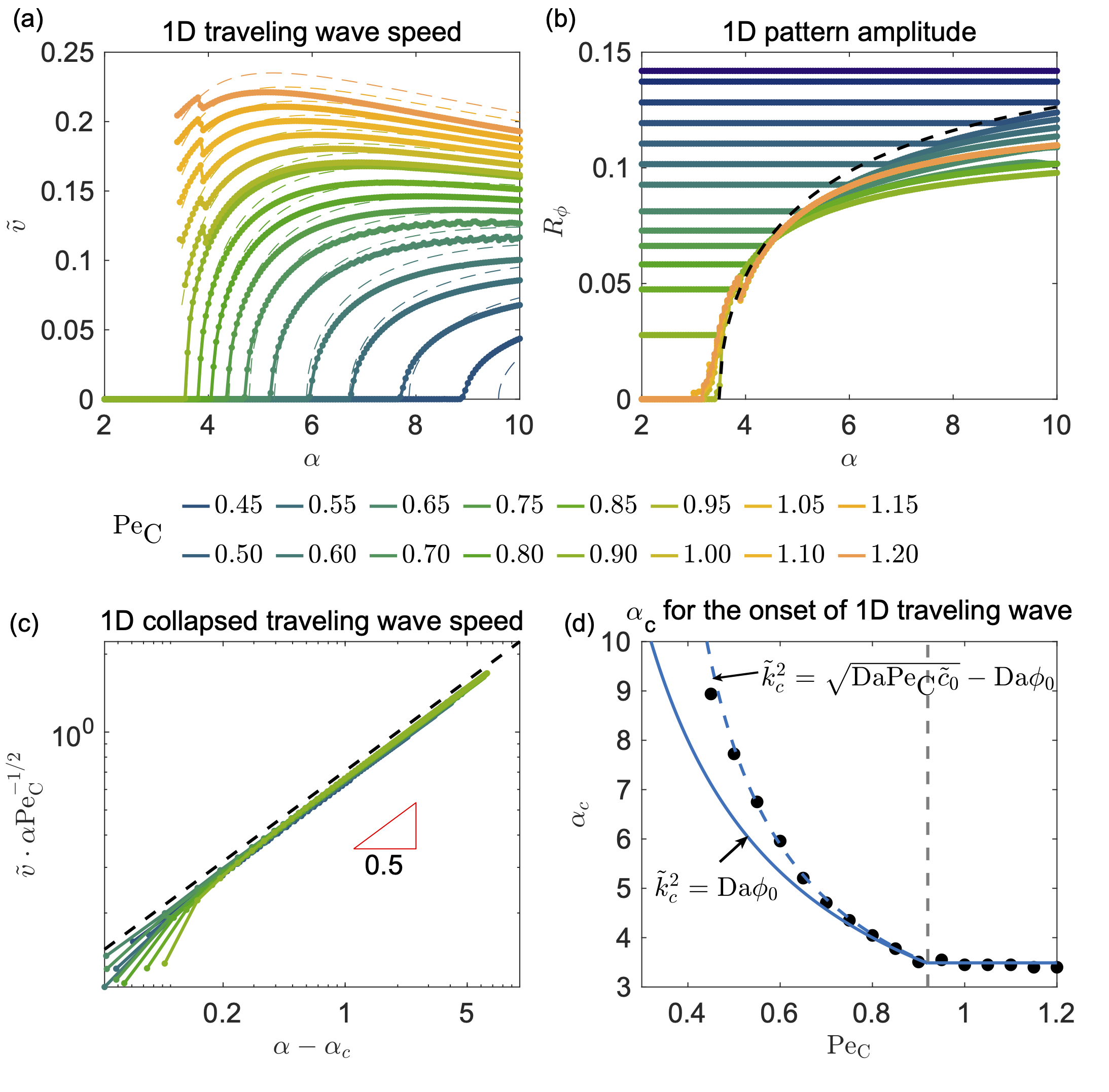}
  \caption{ Properties of the traveling waves in 1D. $\text{Da}=0.5$, $\PeR=10^{-3}$, $\phi_0=0.8$, and $S_0/k=1$. Simulations are performed in a 1D periodic domain whose length is $200 l_0$. (a) The average normalized speed $\tilde{v}=v/v_0$ as a function of $\alpha$ at different $\PeC$. The speed is obtained by tracking the peaks. The dashed curves of the same color are the analytical prediction from the amplitude equation [Eqs.~\eqref{eqn::traveling_wave_velocity_low_PeC} and \eqref{eqn::traveling_wave_velocity_high_PeC}]. 
  (c) The amplitude of the pattern as a function of $\alpha$ that corresponds to (b). The black dashed curve is the analytical prediction from the amplitude equation at low $\PeC$ [Eq.~\eqref{eqn::R_phi_traveling_wave_high_Pe}].
  (d) The critical $\alpha$ as a function of $\PeC$ for the onset of oscillatory patterns. Dots are from simulations (first $\alpha$ when $\tilde{v}>0.01$). The blue solid curve is Eq.~\eqref{eqn::alpha_c} $\alpha_c = 4/(\PeC \tilde{c}_0)$ when $\tilde{k}_c = \sqrt{\da \phi_0}$. The blue dashed curve is given by Eq.~\eqref{eqn::alpha_c_1D} when $k_c^2 = \sqrt{\da \PeC \tilde{c}_0} - \da \phi_0$ to the left of the vertical dashed line, which indicates infinite period bifurcation to the left and supercritical Hopf bifurcation to the right.}
  \label{fig::alpha_ramp_1D}
\end{figure}

Next, we repeat the above simulations in 1D, thus eliminating the dot patterns. Similarly, we ramp up and then down the value of $\alpha$ for each $\PeC$ and ensure that the system chooses either stationary stripes or traveling bands. 
As a result, compared to Fig.~\ref{fig::alpha_ramp}, smoother $\tilde{v}-\alpha$ and $R_\phi-\alpha$ curves are obtained (Fig.~\ref{fig::alpha_ramp_1D}). 
Again, in Fig.~\ref{fig::alpha_ramp_1D}(c), we find that $\tilde{v}-\alpha$ curves in the low $\PeC$ regime where infinite-period bifurcation occurs collapse well to the analytical prediction  $\tilde{v} \cdot \alpha \PeC^{-1/2} = \sqrt{\da \tilde{c}_0 \phi_0 (\alpha-\alpha_c)}$ denoted by the dashed line. However, the critical $\alpha_c$ in the low $\PeC$ regime differs from that in Fig.~\ref{fig::alpha_ramp} in 2D.
In 2D, the critical $\alpha_c$ obtained from simulation agrees with $\alpha_c = 4/(\PeC \tilde{c}_0)$ (blue dashed curve), which is the minimum $\alpha$ above which traveling wave exists, according to Eq.~\eqref{eqn::traveling_wave_existence}, which corresponds to a wavenumber of $\tilde{k}_c = \sqrt{\da \phi_0}$. At other wavenumbers, the critical $\alpha_c$ is larger. We find that in 1D simulations, at low $\PeC$ when the pattern transitions from stationary to traveling wave, the wavelength of the traveling wave near the critical $\alpha_c$ transitions continuously from the wavelength of the stationary wave pattern at lower $\alpha$, which suggests that $\tilde{k}_c$ is equal to that of the stationary wave whose analytical expression is obtained in Sec.~\ref{sec::stationary_pattern}: $k_c^2 = \sqrt{\da \PeC \tilde{c}_0} - \da \phi_0$. Hence, the critical $\alpha_c$ predicted by the existence condition Eq.~\eqref{eqn::traveling_wave_existence} that corresponds to the wavenumber of the stationary pattern is
\begin{equation} \label{eqn::alpha_c_1D}
  \alpha_c = \eval{
      \frac{\qty(k_c^2+\da\phi_0)^2}{k_c^2 {\da \tilde{c}_0 \phi_0\PeC}}
  }_{k_c^2=\sqrt{\PeC \da \tilde{c}_0} - \da  \phi_0} = \frac{1}{\phi_0\qty(\sqrt{\PeC \da \tilde{c}_0} - \da  \phi_0)}. 
\end{equation}
Recall that at high $\PeC$, the critical $\alpha_c$ is determined by the stability condition, hence overall
\begin{equation}
  \alpha_c = \max{\left\{ \frac{1}{\phi_0\qty(\sqrt{\PeC \da \tilde{c}_0} - \da  \phi_0)}, \frac{1}{a^2}\qty(\sqrt{\da} + \sqrt{\da+a\phi_0^{-1}})^2 \right\}}.
\end{equation}
We see that the critical $\alpha_c$ agrees well with this prediction as shown by the black dashed curve in Fig.~\ref{fig::alpha_ramp_1D}(d) and is larger than the blue dashed curve $\alpha_c = 4/(\PeC \tilde{c}_0)$. The predicted traveling wave velocities at various $\PeC$ are plotted as the dashed curves. In the low $\PeC$ regime, it is given by Eq.~\eqref{eqn::traveling_wave_velocity} and \eqref{eqn::traveling_wave_phase_difference} using $k_c^2 = \sqrt{\da \PeC \tilde{c}_0} - \da \phi_0$, while in the high $\PeC$ regime, it is given by Eq.~\eqref{eqn::traveling_wave_velocity_high_PeC}. 

\subsubsection{Spiral waves}
\label{sec::spiral_waves}
\begin{figure}
  \centering
  \includegraphics[width=\textwidth]{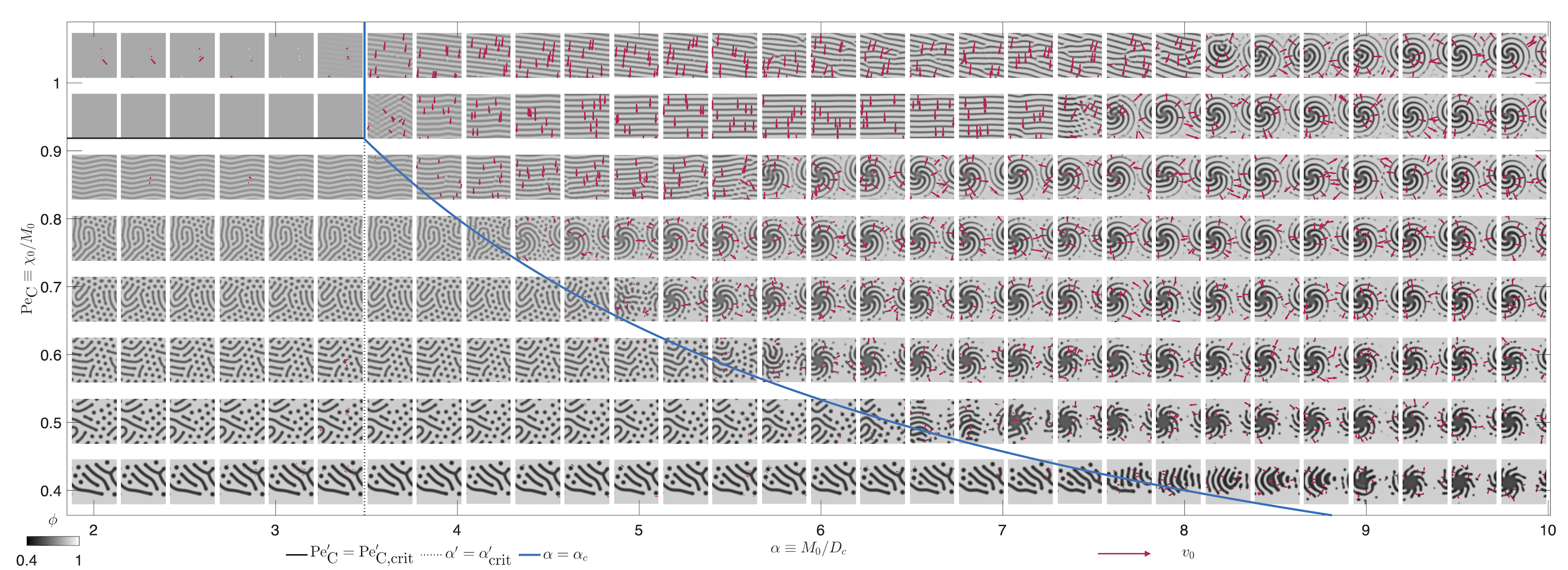}
  \caption{Phase diagram of the spiral waves in the $\PeC-\da$ plane. For each $\PeC$, the value of $\alpha$ is ramped down from $\alpha=10$, using the final state of the previous simulation at a higher $\alpha$ as the initial condition. Simulations are performed on a $[50l_0,50l_0]$ periodic domain and solved on a $256\times 256$ grid. The snapshots are taken at $t=5\times 10^3 t_0$. Arrows indicated the level set velocity of the patterns. Speed smaller than $0.05v_0$ is not shown. Only regions where $|\tilde{\nabla} \phi|>0.1$ are included for calculating the level set velocities.  The blue solid line is the critical $\alpha$ for the onset of oscillatory patterns predicted by the amplitude equation.}
  \label{fig::spiral_phase_diagram}
\end{figure}

Starting from spiral waves in Fig.~\ref{fig::spiral} at different $\PeC$ as the initial condition, we ramp down $\alpha$ and plot the simulations in the $\PeC-\da$ phase diagram in Fig.~\ref{fig::spiral_phase_diagram}. We observe that, at low $\PeC$, the spiral wave solution remains stable as $\alpha$ decreases all the way until the pattern becomes stationary. On the other hand, at high $\PeC$, the spiral wave becomes unstable below a certain $\alpha$ and becomes traveling plane waves.

\begin{figure}
  \centering
  \includegraphics[width=0.75\textwidth]{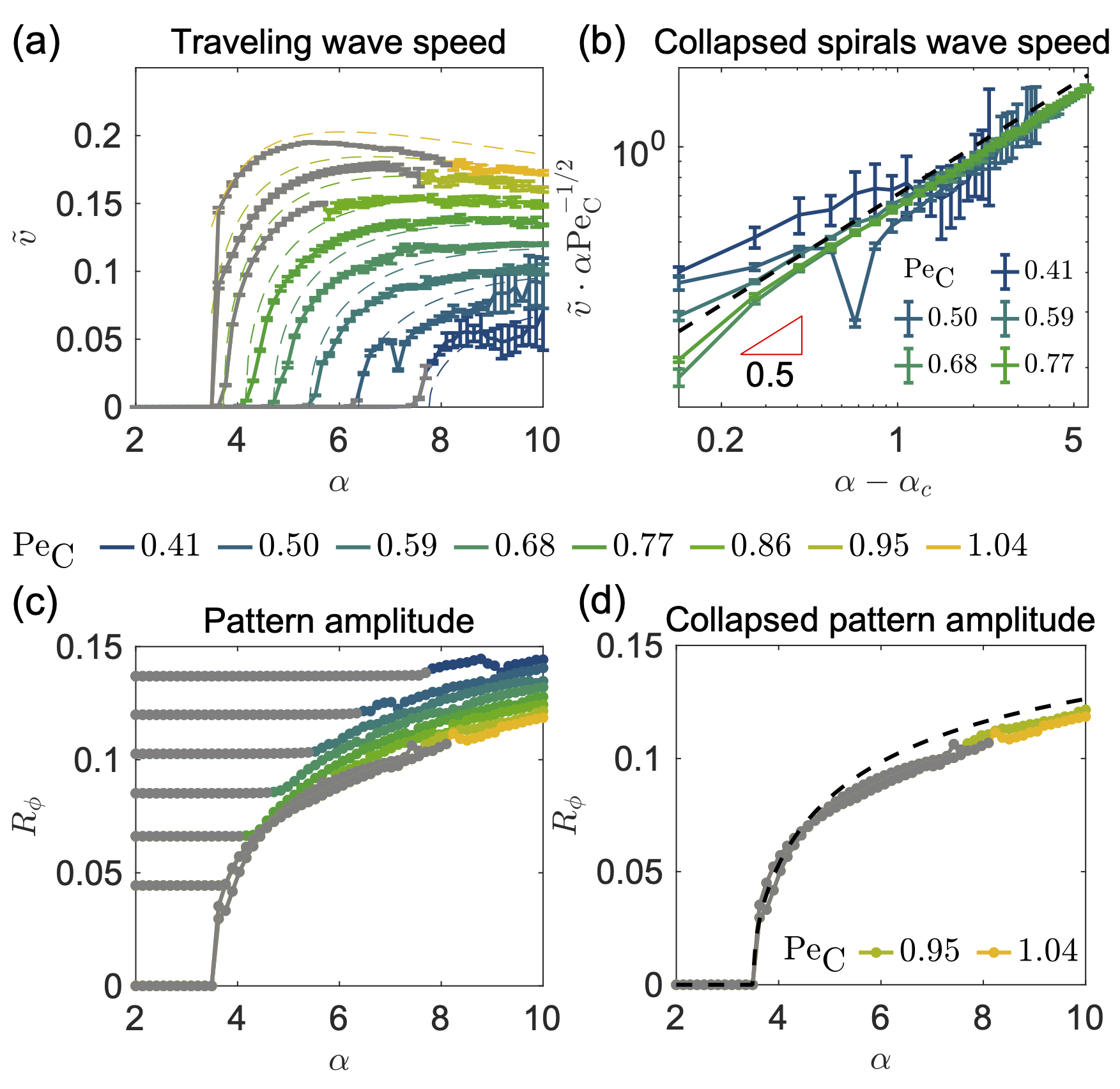}
  \caption{Properties of the spiral waves extracted from Fig.~\ref{fig::spiral_phase_diagram}. (a) The average normalized speed $\tilde{v}=v/v_0$ as a function of $\alpha$ and $\PeC$. The speed is obtained by averaging the level set velocity in space and time. The dashed curves of the same color are the analytical prediction from the amplitude equation [Eq.~\eqref{eqn::traveling_wave_velocity_low_PeC} and \eqref{eqn::traveling_wave_velocity_high_PeC}]. Non-spiral wave patterns (including stationary and traveling waves) are colored in gray.
  (b) The collapse of the spiral wave speed according to $\tilde{v} \cdot \alpha \PeC^{-1/2} = \sqrt{\da \tilde{c}_0 \phi_0 (\alpha-\alpha_c)}$ (dashed line). (c) The amplitude of the pattern as a function of $\alpha$ and $\PeC$. 
  (d) The amplitude of the pattern in the high $\PeC$ regime. The black dashed curve is the analytical prediction from the amplitude equation at high $\PeC$ [Eq.~\eqref{eqn::R_phi_traveling_wave_high_Pe}].}
  \label{fig::spiral_properties}
\end{figure}

Fig.~\ref{fig::spiral_properties}(a) shows the level set velocity averaged in space and time as a function of $\PeC$ and $\alpha$. The non-spiral-wave patterns are colored in gray. We see that the trend of the speed obtained from simulations, including the spiral waves, agrees with the prediction of the amplitude equation, which is shown in dashed curves of the same color. Similar to traveling dots and stripes, the spiral wave speed collapses according to $\tilde{v} \cdot \alpha \PeC^{-1/2} = \sqrt{\da \tilde{c}_0 \phi_0 (\alpha-\alpha_c)}$ as shown in Fig.~\ref{fig::spiral_properties}(b).
Fig.~\ref{fig::spiral_properties}(c-d) shows the amplitude $R_\phi$ as a function of $\PeC$ and $\alpha$. Similar to traveling dots and stripes, the amplitude of the spiral wave increases with increasing $\alpha$ and decreasing $\PeC$. The amplitude of the spiral wave in the regime of supercritical Hopf bifurcation agrees well with the prediction of the amplitude equation [Eq.~\eqref{eqn::R_phi_traveling_wave_high_Pe}] denoted by the dashed curve.

\section{Supplemental movies}
All movies are available at this 
\href{https://drive.google.com/drive/folders/1FGNizLdX5susRQzl18rI6C3ECAg6cv3v?usp=sharing}{link}.
\begin{enumerate}
    \item 2D traveling wave pattern (Fig. 8a). $\PeR=1.9\times 10^{-3}$, $\phi_0=0.70$, $\Da=0.2$, $\alpha=4$, $\PeC=1$. 
    \item 2D traveling dots pattern (Fig. 8b). $\PeR=1.2\times 10^{-3}$, $\phi_0=0.81$, $\Da=0.2$, $\alpha=4$, $\PeC=1$.
    \item 2D short traveling stripes pattern (Fig. 8c). $\PeR=1.4\times 10^{-3}$, $\phi_0=0.79$, $\Da=0.56$, $\alpha=15$, $\PeC=0.44$.
    \item 2D dot pattern that exhibits intermittent bursts (Fig. 8d). $\PeR=2.4\times 10^{-3}$, $\phi_0=0.80$, $\Da=0.5$, $\alpha=10$, $\PeC=0.34$.
    \item 3D traveling plane wave pattern (Fig. 9a). $\PeR=10^{-3}$, $\text{Da}=1$, $\alpha=10$. $\phi_0=0.55$, $\PeC=0.40$.
    \item 3D traveling plane wave pattern (Fig. 9b). $\PeR=10^{-3}$, $\text{Da}=1$, $\alpha=10$. $\phi_0=0.55$, $\PeC=0.71$.
    \item 3D traveling cylinder pattern (Fig. 9c). $\PeR=10^{-3}$, $\text{Da}=1$, $\alpha=10$. $\phi_0=0.58$, $\PeC=0.42$.
    \item 3D traveling cylinder pattern (Fig. 9d). $\PeR=10^{-3}$, $\text{Da}=1$, $\alpha=10$. $\phi_0=0.58$, $\PeC=0.80$.
    \item 3D traveling spheroid pattern (Fig. 9e). $\PeR=10^{-3}$, $\text{Da}=1$, $\alpha=10$. $\phi_0=0.61$, $\PeC=0.37$.
    \item 3D traveling spheroid pattern (Fig. 9f). $\PeR=10^{-3}$, $\text{Da}=1$, $\alpha=10$. $\phi_0=0.61$, $\PeC=0.59$.
    \item Dynamics of a mixture of chemoattractant consumers (green) and chemorepellent consumers (magenta) (Fig. 12).
\end{enumerate}

\end{document}